\documentclass[preprint,12pt]{elsarticle}

\usepackage{graphicx}
\usepackage{amssymb}

\usepackage{lineno}

\biboptions{comma,round}

\usepackage[hidelinks]{hyperref}

\journal{International Journal of Heat and Mass Transfer}
\usepackage[version=4]{mhchem}
\usepackage{upgreek}
\usepackage{interval}
\usepackage[rgb]{xcolor}
\usepackage{tikz}
\usepackage{caption}
\usepackage{float}
\usepackage{empheq}
\usetikzlibrary{arrows.meta, patterns, angles, chains, shapes, spy, quotes, arrows, positioning,shapes.geometric}
\makeatletter
\tikzset{every loop/.style={}}
\tikzset{edge/.style={->,> = latex'}}
\usepackage{relsize}
\usepackage{siunitx}
\usepackage{hyperref}
\usepackage{arydshln}
\usepackage{etoolbox}
\usepackage[figurename=Fig.]{caption}
\usepackage{setspace}
\tikzset{
    quote/.style={{|[width=2.5mm]}-{|[width=2.5mm]}}
}
\tikzset{
    quote2/.style={{|[width=1.25mm]}-{|[width=1.25mm]}}
}
\usepackage{subcaption}
\usepackage{booktabs}
\definecolor{blue2}{RGB}{0,72,153}
\definecolor{red2}{RGB}{185,0,0}
\definecolor{green2}{RGB}{0,150,0}
\definecolor{yellow2}{RGB}{255,234,0}

\usepackage[margin=2.25cm]{geometry}
\usepackage[nameinlink,capitalize]{cleveref}
\crefname{appendix}{}{}
\DeclareMathOperator\erfc{erfc}
\usepackage{multirow}

\usepackage{bbm}
  \newcommand*\filtered[1]{\overline{#1}}

\usepackage{bm}
\renewcommand\vec{\bm}
\usepackage{bm}
\newcommand{\uveci}{{\bm{\hat{\textnormal{\bfseries\i}}}}}

\usepackage{pifont}

\usepackage[symbol]{footmisc}
\setcitestyle{square}

\usetikzlibrary{calc,spy,shapes}

\begin{document}

\begin{frontmatter}

\title{Numerical and experimental study of open-cell foams for the characterization of heat exchangers}


\date{February 16, 2023}

\author[CERN,UPM]{Aitor Amatriain$^*$}
\author[CERN]{Corrado Gargiulo}
\author[UPM,CSC]{Gonzalo Rubio}

\address[CERN]{ALICE Collaboration, CERN, Geneva 23, 1211, Switzerland}
\address[UPM]{E.T.S. Ingeniería Aeronáutica y del Espacio, Universidad Politécnica de Madrid, Plaza Cardenal Cisneros 3, 28040, Madrid, Spain}
\address[CSC]{Center for Computational Simulation, Universidad Politécnica de Madrid, Campus de Montegancedo, Boadilla del Monte, 28660, Madrid, Spain}

\begin{abstract}
    A multiscale model of open-cell foams is developed for the characterization of heat exchangers. The model is applicable to a wide range of materials, cell sizes, and porosities. The microscopic geometry is based on a periodic model that is defined by the porosity and the specific surface area of the foam considered. The representative geometrical scales of the model are validated with microscope images and computed tomography scans. The outputs of the microscopic model are the coefficients of the parabolic pressure loss curve, the thermal conductivity, and the Nusselt number. These values are used as inputs of the macroscopic model that determines the thermal performance of a macroscopic system. The results given by the models are compared with experimental data obtained from the literature, and from an experimental setup built at CERN. It is concluded that the multiscale model provides accurate results in all open-cell foams considered. 
\end{abstract}


\end{frontmatter}


\section{Introduction}
\label{S:1}

Foam materials are cellular structures that consist of a solid material that contains a high number of pores. Two different variants exist depending on the internal structure: open-cell foams and closed-cell foams. The former have a network of ligaments but no cell walls, while the cells of the latter are surrounded by thin cell walls and are sealed off from neighboring cells. Owing to its fluid permeability, the open-cell structure is adequate for applications where fluid transport is demanded, while the isolated pores in the closed-cell structure offer potential advantages for thermal insulation purposes \cite{WYPYCH2022}. Both structure types have excellent specific properties that can be tuned by varying the precursor material---usually metals, ceramics, or carbon---and/or the production process.

The possibility of producing open-cell foams with low cell sizes (down to $\ell \sim 10^{-4}$  \si{\meter}) and base materials of high thermal conductivity leads to their use as heat exchangers. When a foam is placed in contact with a material that dissipates heat, the heat can be transferred to the foam by conduction, to then be removed from the foam with a fluid flow that can be a gas, a liquid, or a two-phase mixture. Systems such as modern electronic devices \cite{LI2011} and power plants \cite{KURUNERU2020} benefit from the combination of both heat transfer mechanisms. In the last years, foams have been studied as potential candidates for improving the performance of heating, ventilation, and air conditioning (HVAC) systems \cite{DONGELLINI2022}, as well as enhanced catalyst carriers to replace random packed beds of pellets for tubular reactors \cite{TRONCONI2014}.

In high-energy physics, an important parameter that is characteristic of each material is the radiation length $\tilde{X}_0$. This value is representative of the energy loss of particles when passing through matter \cite{KOLANOSKI2020}. The radiation length of a particle detector layer of thickness $h$ is usually given as a ratio with respect to the radiation length of the material: $X/X_0 = h/(\tilde{X}_0 \rho$), with $\rho$ referring to the material density. Lower values of $X/X_0$ lead to a higher accuracy of the measurement of the momentum of some particles. Thus, for a fixed detector layer thickness, the radiation length and the density of the material should be minimized. The radiation length of carbon (42.7 g/cm$^2$) is higher than other materials such as aluminum (24.01 g/cm$^2$) and copper (12.86 g/cm$^2$), which motivates its use over the traditional metallic structures that were the basis of previous particle detectors. In addition, since $X/X_0 \sim 1/\rho$ and the foam density is one order of magnitude lower than the density of the base material, carbon foams are widely used at CERN.

Currently, in three large experiments of the Large Hadron Collider (ATLAS, CMS, and LHCb), \ce{CO2} cooling systems are used. The heat dissipated by the silicon sensors of the particle detectors is transferred by conduction to open-cell foams and to titanium pipes, to then be removed by convection with a two-phase \ce{CO2} flow \cite{ATLAS2017S,CMS2017,MOUNTAIN2019}. In the ALICE experiment, open-cell carbon foams are planned to be used for the first time as heat exchangers in contact with the silicon sensors of the inner detectors. The heat dissipated by the sensors will be transferred by conduction to the foam, and then removed by convection with a forced air flow \cite{ITS32019}. Additional (lighter) open-cell foams are also required for structural purposes. In what follows, closed-cell foams will not be considered, so this work is focused on open-cell foams.

The multiple applications of foams have motivated the realization of numerous experimental studies for the characterization of foam properties such as the thermal conductivity \cite{CALMIDI1999,SADEGHI2011,PAEK2000}, the pressure drop \cite{LACROIX2007,BONNET2008}, and the heat transfer coefficient in natural convection \cite{PHANIKUMAR2002} and forced convection \cite{YOUNIS1993,MANCIN2013}. The experimental works have lead to analytical models based on correlations for different parameters such as the thermal conductivity \cite{BOOMSMA2001,KUMAR2016}, the pressure drop \cite{DUKHAN2006,INAYAT2016}, and the Nusselt number \cite{CALMIDI2000}. Simulations have been performed to compute the heat transfer coefficient in aluminum foams with geometries extracted from computed tomography scans \cite{IASIELLO2020}, although reduced domains are simulated, and thus the fully developed flow regimes are not considered. Moreover, the possibility of producing foams with a wide ranges of cell sizes (from $250$ \si{\micro\meter} up to $10$ mm) and porosities (from $\Phi=0.8$ to $\Phi=0.97$) limit the accuracy of the correlations and the analytical models that rely on experimental data. The correlations consider different geometrical parameters as building blocks \cite{G2022}. Since the repeatability of the foam production process is not close to 100 \% and different measurement techniques are used to obtain them, the comparison between correlations of different foams is not straightforward. In addition, the thermal resistance of the joint between carbon foams and solid surfaces---and, in particular, the effect of the amount of glue that penetrates into the foam--- which is expected to play a major role in the thermal performance of foams, has not been studied in depth. In this line, an initial study applied to the previously mentioned \ce{CO2} cooling systems has been performed at CERN \citep{MUSSGILLER2016}, where a titanium-foam interface is considered. However, the applicability of this study to other configurations with different interfaces such as silicon-foam is unclear, since the direct contact between these materials can damage the silicon sensors. A different configuration of the interface is required such as the one of \cref{interface-ts} that is explained in detail in \cref{methodology}.

This work is focused on foam characterization with the goal of selecting adequate foams to be used in the High-Energy Physics (HEP) particle detectors; in particular, in terms of the pressure loss, the thermal conductivity, and the Nusselt number. To the authors knowledge, there is no analytical and/or numerical methodology to compute these properties with a reasonable accuracy in a general case. The studies cited provide great accuracy, but for a limited range of validity that is usually unknown. The present work tackles this problem, and proposes a multiscale foam model that considers two geometric scales: microscopic and macroscopic, which are represented by the foam length and the length of the system where the foam is used, respectively. The model can be applied to a wide range of porosities, cell sizes, and materials. The microscopic model is based on the representation of the microscopic geometry of open-cell foams, and the outputs of the model are used as inputs of a macroscopic model. This multiscale methodology reduces the computational cost in cases of practical application. Moreover, the effect of the glue penetration in the temperature of the foam-silicon interface is studied in a case of practical application with numerical and experimental results. This work is divided as follows: first, the methodology consisting on the description of the multiscale model of foams and the experimental setup used for model validation is explained in \cref{methodology}. Then, the results of the simulations and the experiments are discussed in \cref{results}. Finally, the main conclusions are drawn in \cref{conclusions}.

\section{Methodology}
\label{methodology}

The foam characterization is based on the development of a multiscale model applied to two geometric scales: the microscopic scale $\ell$, which is representative of the characteristic foam cell size, and the macroscopic scale $\mathcal{L}$, which is representative of the characteristic length of the system where the foam is used. The microscopic model gives as outputs averaged properties. These properties are inputs of the macroscopic model, which is used to determine the thermal behavior of a macroscopic system. First, the microscopic and macroscopic models for the foam geometry are described in \cref{m-model}. Then, an experimental setup built at CERN for model validation is described in \cref{exp-setup}. In this work, four types of foams will be studied:

\begin{itemize}
    \item Duocel$^{\text{\textregistered}}$ Al: Aluminum foams with densities ranging from 100 \si{\kilogram\per\meter\cubed} to 325 \si{\kilogram\per\meter\cubed} and thermal conductivities of 2 - 7 \si{\watt\per\meter\per\kelvin}. These foams are not planned to be used in future HEP detectors. However, since extensive results of experimental studies are available in the literature, these foams will be used to validate the models developed for the calculation of the parabolic pressure loss curve (\cref{deltap-adj}), the thermal conductivity (\cref{k-def}), the Nusselt number (\cref{Nu-def}), and the overall heat transfer coefficient (\cref{power-h}). It should be noted that the Nusselt number is referred to the (local) heat transfer coefficient $h$ (\cref{hv-coef}) that takes into account the microscopic geometry, while the overall heat transfer coefficient is related to the performance of a system at the macroscopic scale.
    \item Duocel$^{\text{\textregistered}}$ RVC: Foams made of reticulated vitreous carbon, which is one of the morphological structures of vitreous carbon. The density $\rho =45$ \si{\kilogram\per\meter\cubed}, and different options are available for a wide of cell sizes. These foams are thermal insulators $(\kappa_f \approx 0.05$ \si{\watt\per\meter\per\kelvin}, therefore they cannot be used as heat exchangers for thermal applications. However, the low density and high stiffness motivate its use structural parts in particle detectors. Thus, in these foams the only important result to be taken into account is the pressure loss.
    \item Lockheed Martin K9: Foam made of reticulated vitreous carbon with graphite added by chemical vapor deposition to achieve thermal conductivity values $\kappa_f \approx 25$ \si{\watt\per\meter\per\kelvin}\footnote{Private communication with Lockheed Martin}. This foam of density $\rho \approx 200$ \si{\kilogram\per\meter\cubed} is currently used in HEP particle detectors at CERN. The microscopic model determines the coefficients of the parabolic pressure loss curve, the thermal conductivity, and the Nusselt number. Then, the macroscopic model will be used to assess thermal performance in a case of practical application in HEP.
    \item CFOAM$^{\text{\textregistered}}$ 35 HTC: Foam made from mesophase pitch feedstock with density $\rho_f \approx 350$ \si{\kilogram\per\meter\cubed}. This foam is anisotropic, and the thermal conductivity in the vertical direction is 20-30 \si{\watt\per\meter\per\kelvin}, which is approximately twice as the one in the planar directions\footnote{Private communication with CFOAM LLC}. It has some closed cells, and it considered as an alternative to the K9 foam to be used as a heat exchanger.
\end{itemize}

\cref{foams-appendix} provides more details about the foams mentioned, including images taken from microscopy and computed tomography scans. After defining the main geometrical parameters of foams in \cref{m-model}, the values corresponding to the foams mentioned are presented in \cref{geop-foams}.

\subsection{Multiscale model}
\label{m-model}

\begin{figure}[H]
 	\centering
        \begin{tikzpicture}[xscale=1, yscale=1, axis/.style={->,thick},line cap=rect]
        	\begin{scope}[spy using outlines={circle, magnification=4, size=1.625cm, connect spies}]
      	\node[anchor=south west,inner sep=0] at (0,0) {\includegraphics[trim={325 62.5 400 75},width=0.285\textwidth, clip]{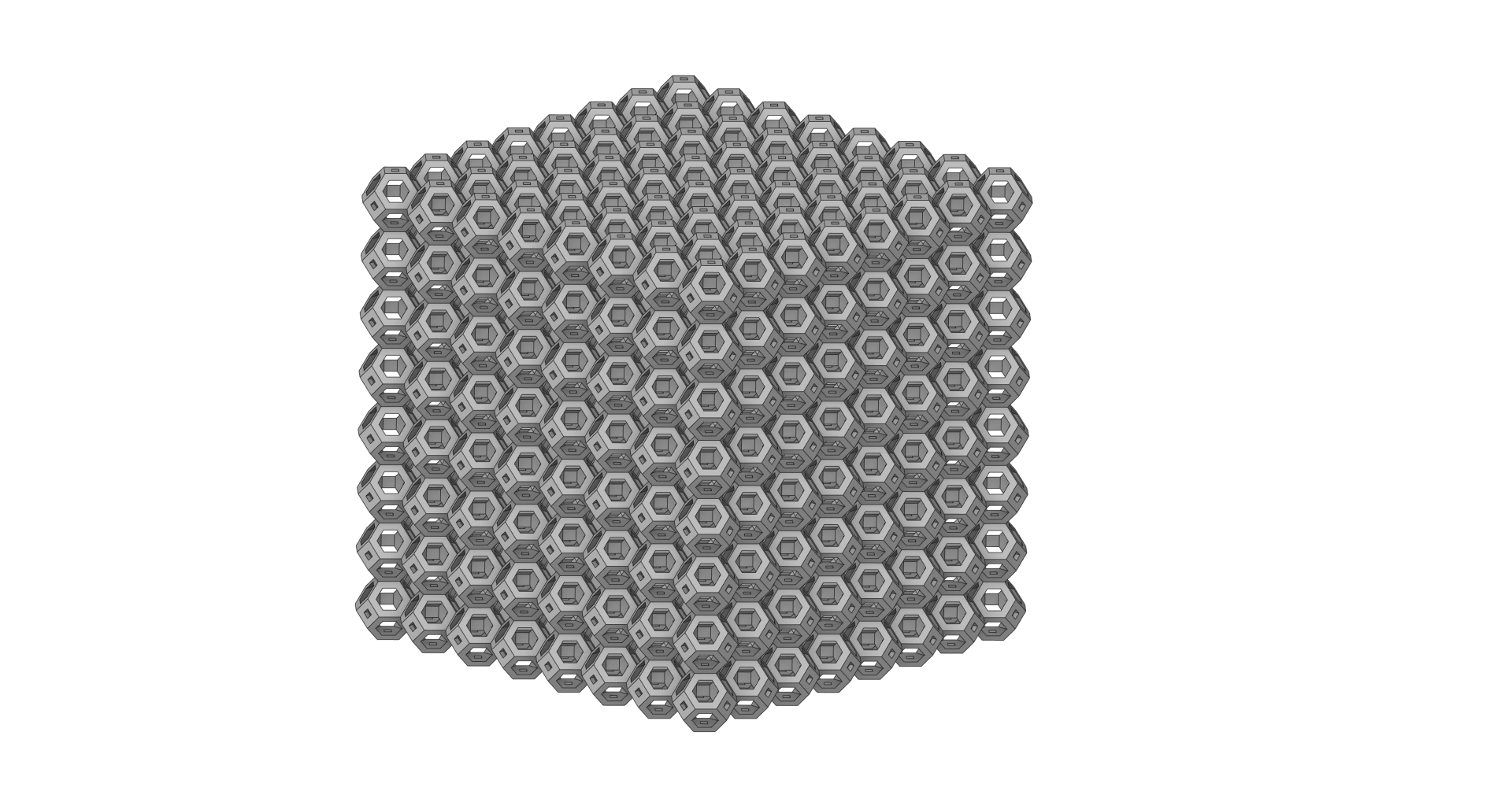}};
      	\spy on (1.5,2.31) in node (spy)  at (-2,2);
      	      	\node[anchor=south west,inner sep=0] at (7,-0.25) {\includegraphics[trim={575 150 575 100},width=0.265\textwidth, clip]{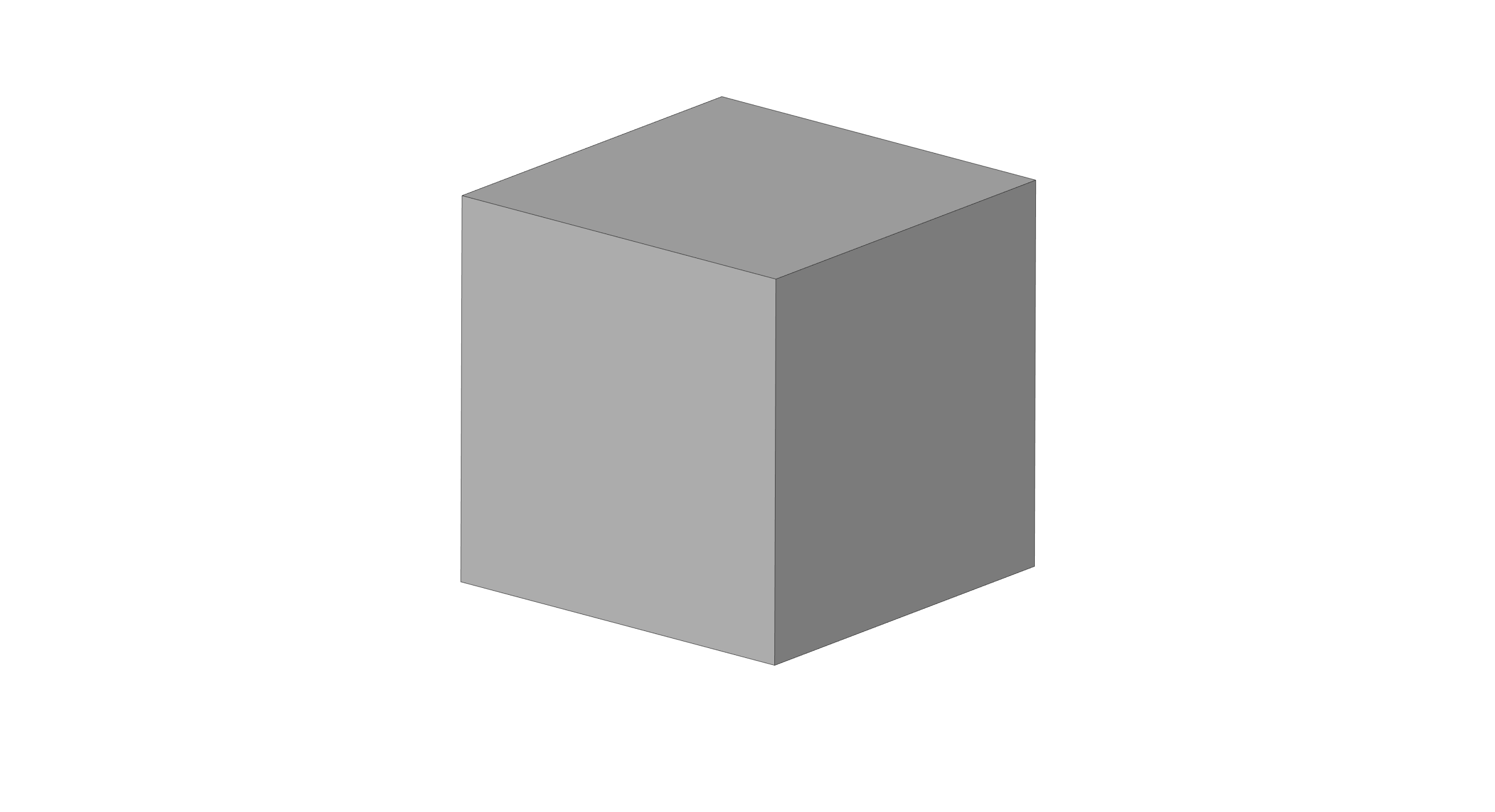}};
      	      		    \node  at (-3.25,2) {$\ell$};
      	      		       \node  at (-2,0.75) {\small Microscopic};
      	      		          \node  at (-2,0.3) {\small model};
      	      		 \node  at (0.5,4) {\small Foam};
      	    \draw[quote,line width=0.25mm] (-3,1.2)  -- (-3,2.75);
      	       	    \draw[quote,line width=0.25mm] (4.675,0.75)  -- (4.675,3.5);
      	       	      \node  at (5,2) {$\mathcal{L}$};
      	       	        	      \node  at (6.5,2) {$\mathcal{L}$};
      	       	      	\draw[quote,line width=0.25mm] (6.75,0.6)  -- (6.75,3.5);
      	       	      	      \node  at (7,0.15) {\small Macroscopic};
      	       	      	    \node  at (7,-0.3) {\small model};
      	\end{scope}   
  		\end{tikzpicture}
  		\caption{Geometric scales of the submodels}
  		\label{model-overview}
  \end{figure}
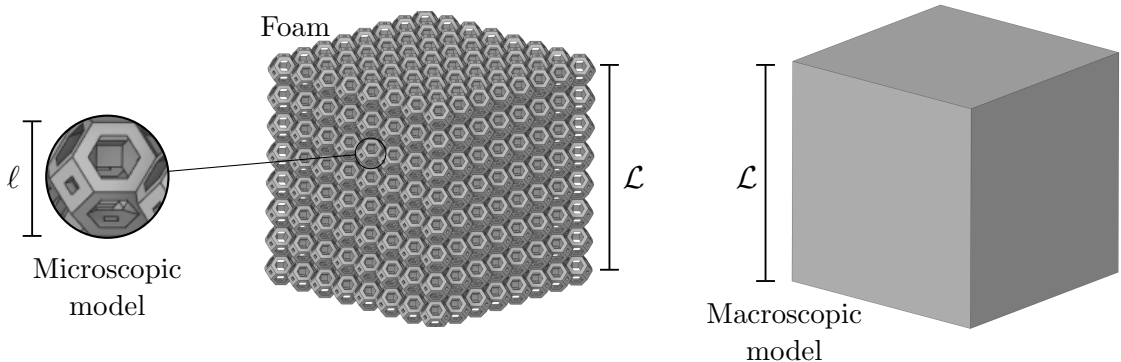

The geometry of the simulations is the main building block of the multiscale model, which is divided into two submodels (see \cref{model-overview}):

\begin{itemize}
    \item Microscopic model. In the microscopic domain $\Omega_f^{\ell}$, the geometry is based on a model of the unit cell of foams derived from the literature. After solving the Navier-Stokes equations in the microscopic domain with periodic boundary conditions, the coefficients of the parabolic pressure loss curve, the thermal conductivity, and the Nusselt number of foams are obtained.
    \item Macroscopic model: In the macroscopic domain $\Omega_f^{\mathcal{L}}$, mean variables are considered that are obtained by averaging the fluid magnitudes in volume regions $\mathcal{L}>\ell^3$ for a sufficient $\mathcal{L}/\ell$ ratio that will be mentioned in the results section. This model uses a simplified geometry of the foam---which acts as a porous medium---and solves the Navier-Stokes equations with the addition of source terms. The source terms take into account the effect of the small scales of the order of $\ell$ on the large scales of the order of $\mathcal{L}$, and contain parameters obtained from the simulations of the microscopic model.
\end{itemize}
      \vspace{-5mm}

\subsubsection{Microscopic model}
\label{geometric-model}

\subsubsection*{Geometry}

When looking for the space-filling arrangement of cells of equal volume that has minimum surface area, Lord Kelvin proposed a 14-sided truncated octahedron---named as tetrakaidecahedron by Kelvin---as a solution \cite{THOMSON1887}. In the miscoscopic model, foams are modeled as uniformly-distributed cells of tetrakaidecahedrons. The starting point of the model is the entirely solid geometry, and then volume is removed to obtain the frame structure (see \cref{foam-model}).

Taking as references microscopy images of aluminum \cite{PERROT2007} and graphite (\cref{sem-ld-zoom}) foams (see \cref{foams-appendix}), triangular cross-section is considered in the ligaments, with equal length of the sides of the rectangular and hexagonal faces. This implies that the size of the square holes decreases as the porosity decreases. When $\Phi \approx 0.82$, the square holes disappear, therefore the model proposed is assumed to be valid for $\Phi>0.82$. This result is applicable to the CFOAM, which has a lower number of holes per cell than the other foams (see \cref{ct-plane}).

 \begin{figure}[H]
 \centering
 \begin{subfigure}[t]{.23\linewidth}
 	\centering
\includegraphics[trim={650 250 825 50}, width=0.85\textwidth,clip]{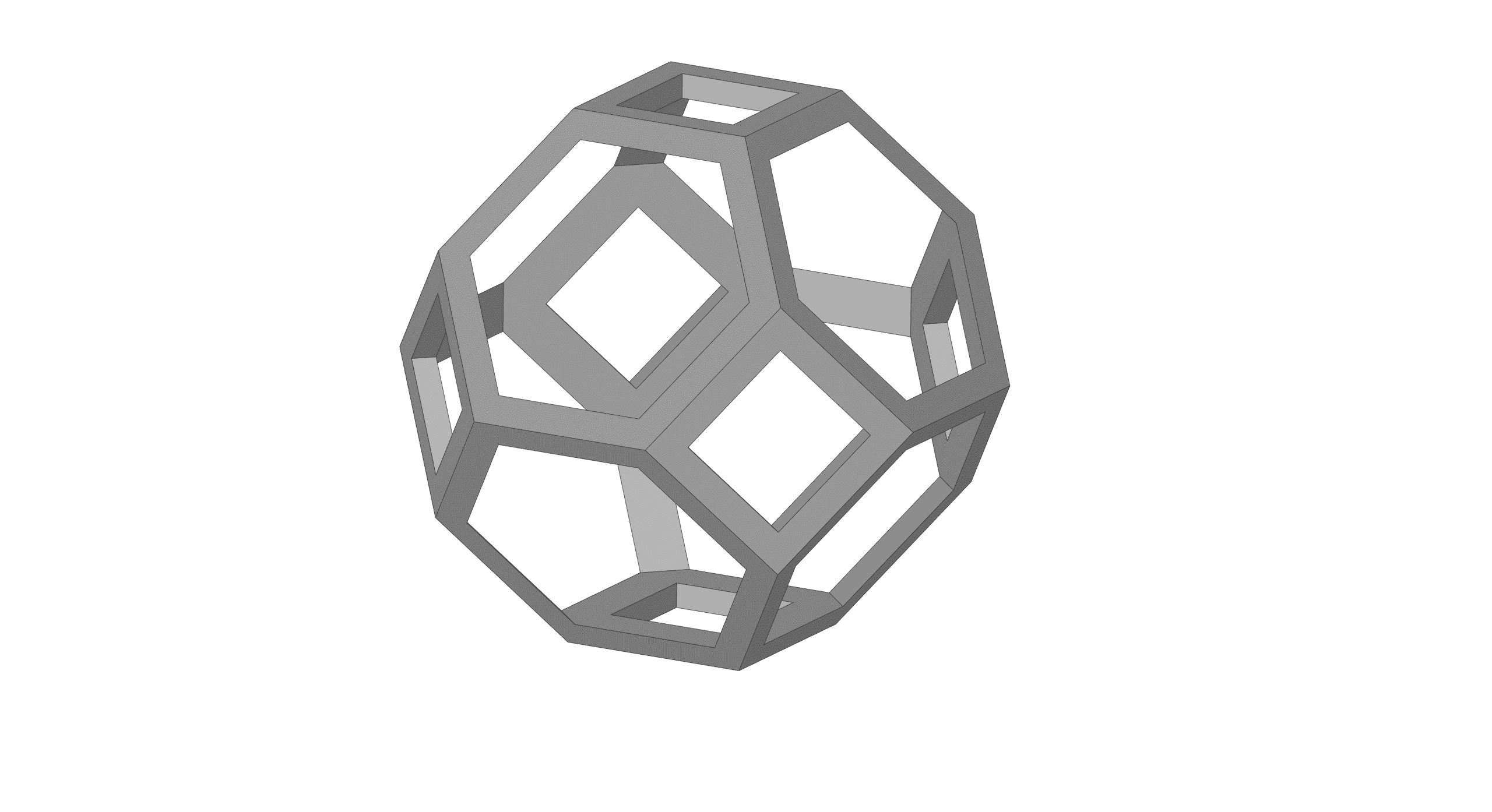}
    \caption{$\Phi=0.97$ (RVC)}
  \end{subfigure}
    \begin{subfigure}[t]{.23\linewidth}
    \centering
          \includegraphics[trim={1150 300 1150 175},width=0.85\textwidth, clip]{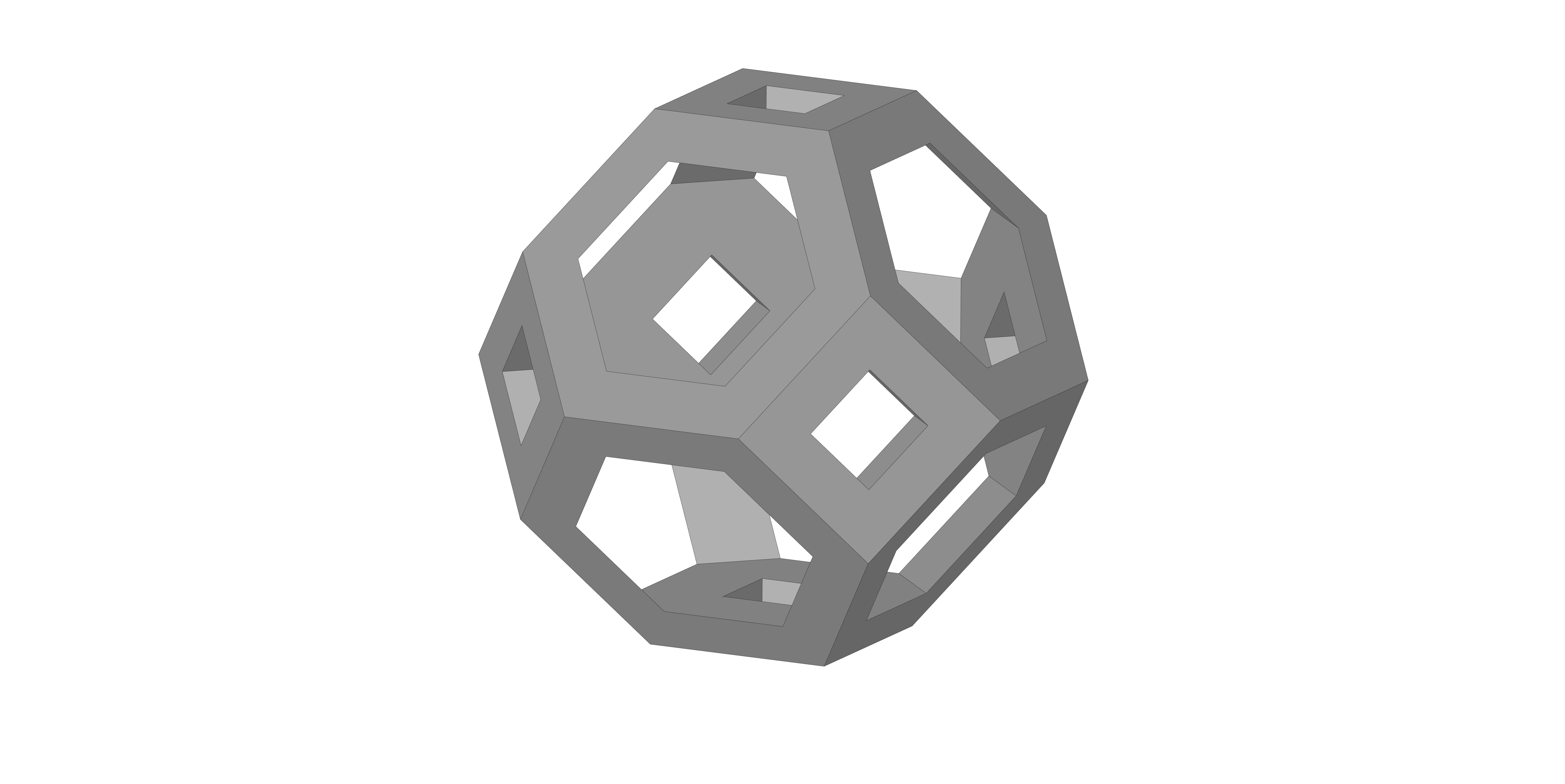}
    \caption{$\Phi=0.92$ (Al)}
  \end{subfigure}
      \begin{subfigure}[t]{.23\linewidth}
    \centering
          \includegraphics[trim={600 200 800 100},width=0.85\textwidth, clip]{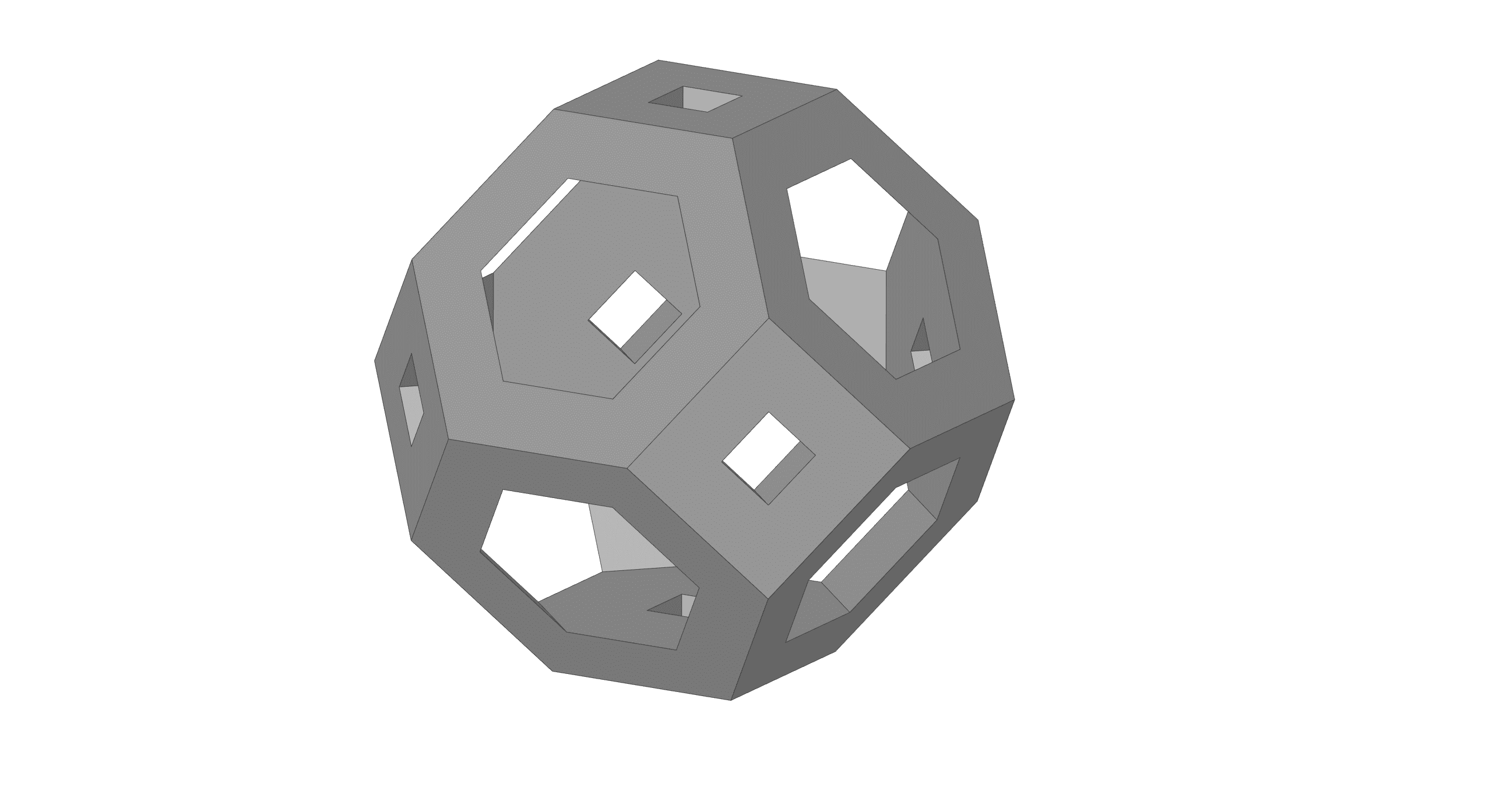}
    \caption{$\Phi=0.89$ (K9)}
  \end{subfigure}
        \begin{subfigure}[t]{.23\linewidth}
    \centering
          \includegraphics[trim={600 175 800 100},width=0.85\textwidth, clip]{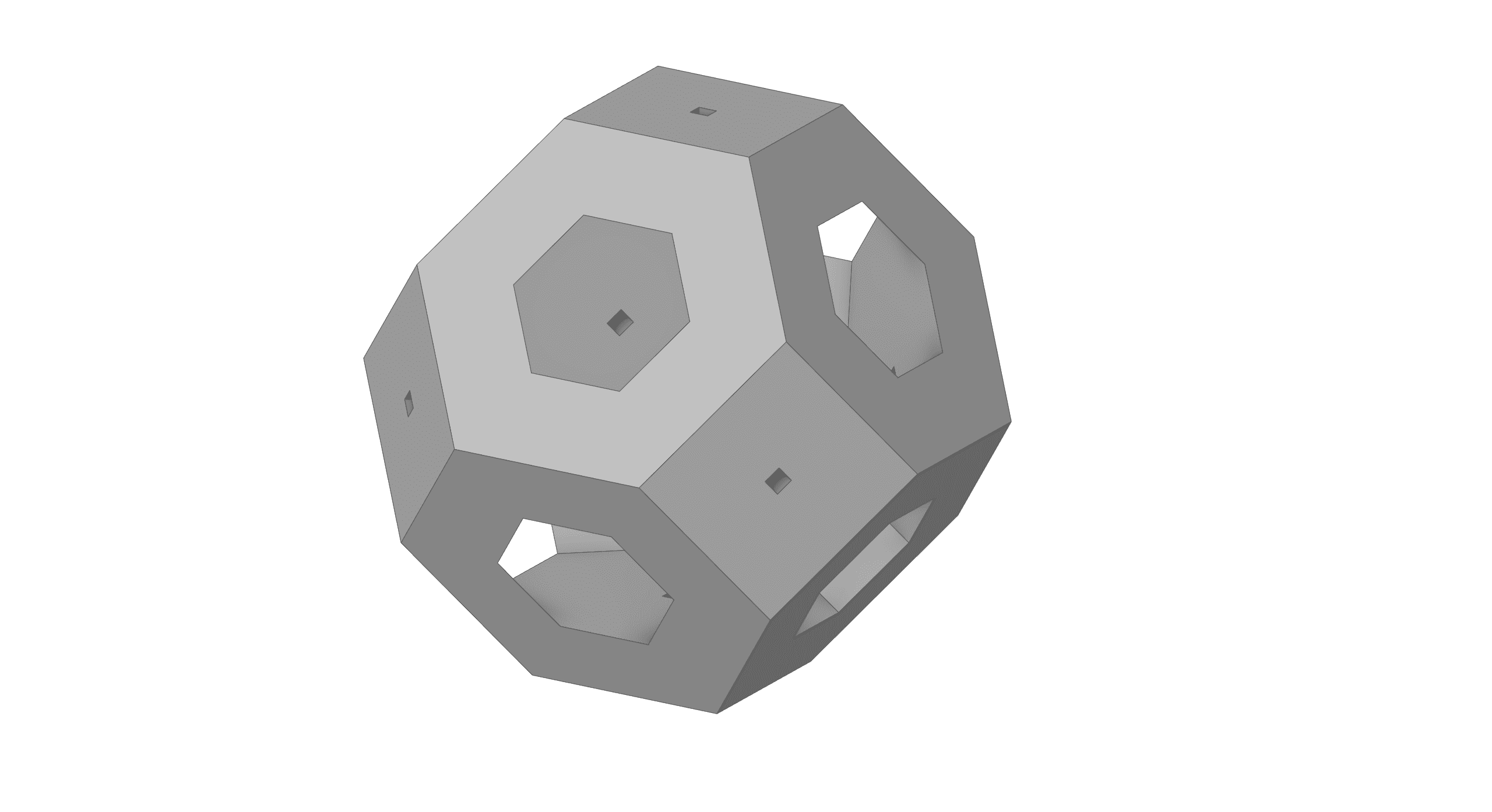}
    \caption{$\Phi=0.83$ (CFOAM)}
    \label{cfoam-t}
  \end{subfigure}
  \end{figure}
  \vspace{-5mm}
   \begin{figure}[H]\ContinuedFloat
 	\centering
  \begin{subfigure}[t]{.485\linewidth}
      \centering
            \scalebox{0.7}{
      	\begin{tikzpicture}[xscale=1, yscale=1,axis/.style={->,thick}]
      	\node[anchor=south west,inner sep=0] at (0,0) {\includegraphics[trim={750 325 750 185}, width=\textwidth, clip]{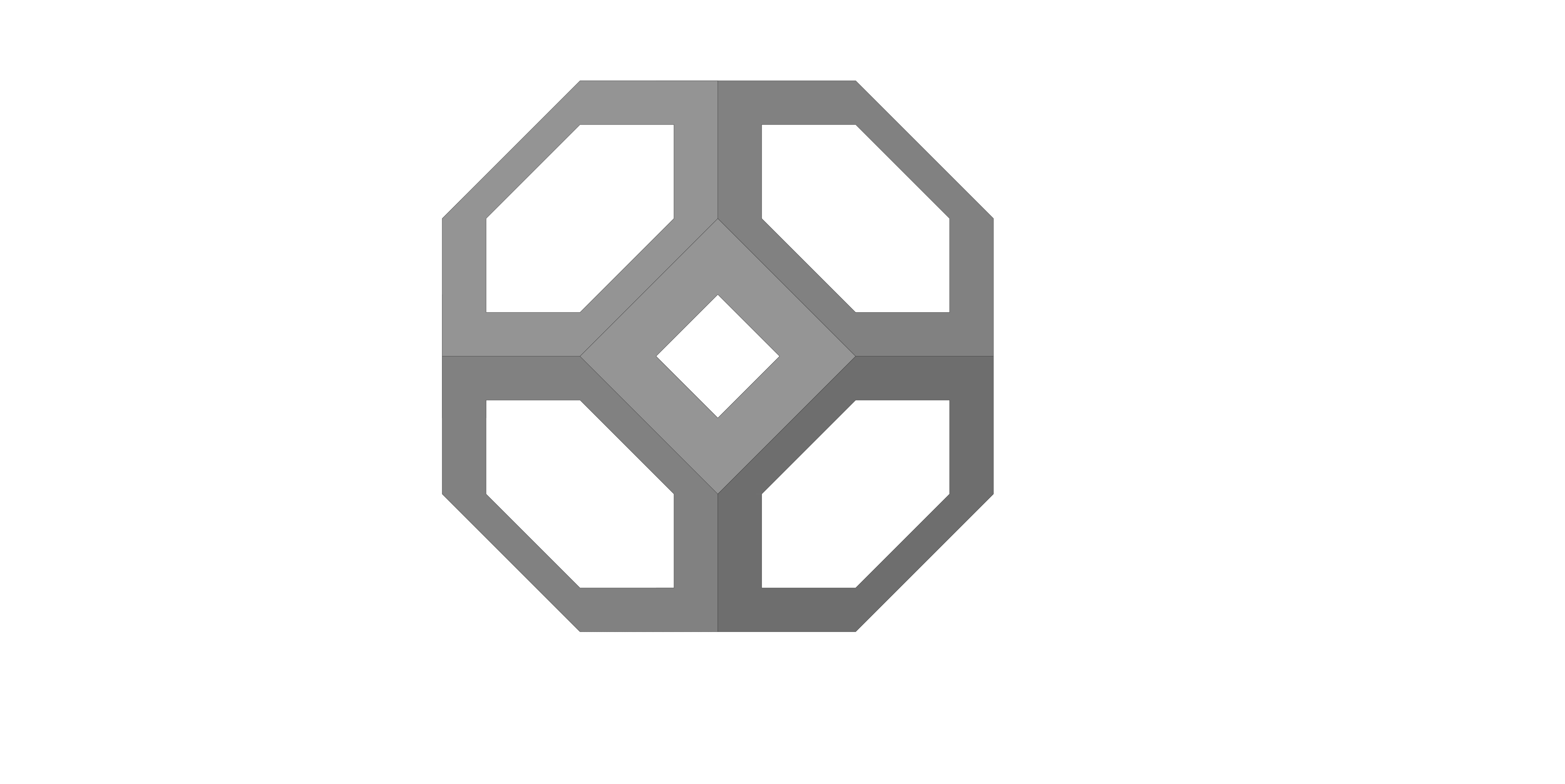}};
      		\node (A) at (4.75,0.6) {};
      		\node (B) at (3.75,0.6) {};
      		\node (C) at (5.5,1.55) {};
      		\draw[quote,line width=0.25mm] (6,0.175)  -- (6,4.975);
      		    \node (D)  at (6.5,2.65) {\Large $\ell_{cell}$};
      		\pic [draw, ->, angle eccentricity=1.5] {angle = C--A--B};
      			\draw[line width=0.125mm] (3.2,4.56)  -- (3.2,5);
      			\draw[line width=0.125mm] (3.975,4.56)  -- (3.975,5);
      		    \draw[quote,line width=0.25mm] (3.975,5.25)  -- (3.2,5.25);
      		    \node (D)  at (3.55,5.625) {\Large $\ell_{tr}$};
      		    \node  at (4.75,1.45) {\Large $120^{\circ}$};
      		    	\draw[axis] (1.2,0.15) -- (1.2,0.95) node[anchor=east]{\Large $z$};
	                \draw[axis] (1.2,0.15)-- (2,0.15) node[anchor=north]{\Large $x$};
      		\end{tikzpicture}}
    \caption{Front view for $\Phi=0.92$ (Al)}
    \label{input-def}
  \end{subfigure}
    \caption{Computational models of the foams described in \cref{foams-appendix}}
      \label{foam-model}  
  \end{figure}
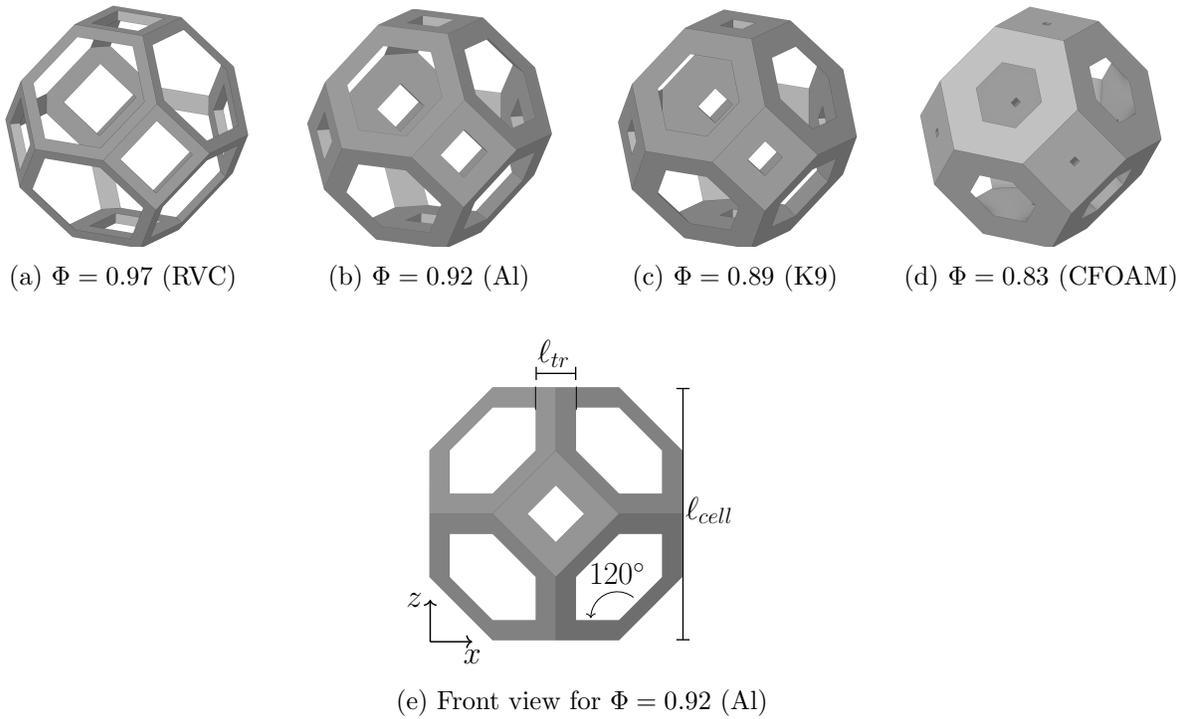

The dimensions of each cell are completely determined by specifying the values of the cell length $\ell_{cell}$ and the ligament triangle side $\ell_{tr}$, which are derived from two inputs (see \cref{input-def}):

\begin{itemize}
    \item Porosity $\Phi$: The ratio between the volume of the tetrakaidecahedron and the volume of the cube of edge length $\ell_{cell}$. It determines the amount of material that has to be removed from each tetrakaidecahedron so that the geometry is fixed up to a scaling factor.
    \item Specific surface area $\Sigma_s$: The ratio between the surface area of the tetrakaidecahedron and the volume of the cube of edge length $\ell_{cell}$. It determines the scaling factor that has to be applied to the geometrical model to obtain the completely defined geometry.
\end{itemize}

ERG Aerospace provides the values of the specific surface density of the Duocel$^{\text{\textregistered}}$ foams. The ligament thickness of the K9 foam is obtained at CERN by taking the mean value of the thicknesses of the ligaments derived from microscope images (see \cref{sem-ld-zoom}), while the values of the CFOAM are obtained from a computed tomography scan performed at CERN (see \cref{foams-appendix}). The geometric parameters of the foams considered in this work are presented in \cref{geop-foams}. The values given by the computational model provide a reasonable approximation in all cases.

 \begin{table}[H]
\centering
\begin{tabular}{c c c c c c c c}
	\toprule
	  \textbf{Foam}  &  $\boldsymbol{\Phi}$ & $\boldsymbol{\Sigma_s}$ \textbf{(1/m)}  & $\boldsymbol{\ell_{cell}}$ \textbf{($\boldsymbol{\mu}$m)} & $\boldsymbol{\ell_{tr}}$ \textbf{($\boldsymbol{\mu}$m)} &   \textbf{Experiments} &   \textbf{Reference} \\
	\midrule
	RVC  & 0.97 & 6600 & 390 & 35 & $\filtered{\ell_{cell}}\approx 400$ $\mu$m & \cref{rvc-sem}\\
	\rule{0pt}{2.5ex}
	 Al-10 &  0.92 &  750 & 5000 &  800 & $\filtered{\ell_{tr}}\approx 640$  $\mu$m& \cite{PERROT2007}\\	
	\rule{0pt}{2.5ex}
	 Al-20  &  0.92 & 1260 &  2975 &   475 & $\filtered{\ell_{tr}}\approx 450$  $\mu$m & \cite{PERROT2007} \\	
	 \rule{0pt}{2.5ex}
	 Al-40  &  0.92 & 1800 &  2125 &   340 & $\filtered{\ell_{cell}}\approx 2250$  $\mu$m& \cite{AMANI2018}\\	
	\rule{0pt}{2.5ex}
	 K9 &  0.89 &   14700 & 285 &  55  & $\filtered{\ell_{cell}}\approx 300$ $\mu$m & \cref{sem-ld-zoom} \\
	 	\rule{0pt}{2.5ex}
	 CFOAM & 0.83 & 2500 & 3500 & 900 & $\filtered{\ell_{cell}}\approx 3000$ $\mu$m & \cref{ct-plane} \\
	\bottomrule
	 \end{tabular}
	\caption{Geometrical parameters of the microscopic model of the foams studied in this work}  %
	\label{geop-foams}
	\end{table}
	
\subsubsection*{Equations}
\label{g-equations}

The incompressible filtered Navier-Stokes equations for the air are \cite{POPE2000}:

\begin{empheq}[left={\empheqlbrace}]{alignat=3}
	& \nabla \cdot \boldsymbol{}\filtered{\Vec{v}}=0 \hspace{5mm} \label{continuity}\\
	& \rho\bigg(\frac{\partial \filtered{\Vec{v}}}{\partial t}+ \filtered{\Vec{v}} \cdot \nabla \filtered{\Vec{v}}\bigg) =-\nabla \filtered{p}+ \nabla \cdot [(\mu+\mu_t)\nabla \filtered{\Vec{v}}] +\filtered{\Vec{S}} \label{momentum}\\
    	& \rho c_p \bigg(\gamma\frac{\partial \filtered{T}}{\partial t}+\filtered{\Vec{v}} \cdot  \nabla \filtered{T}\bigg) =\gamma\nabla \cdot [(\kappa+\kappa_t)\nabla \filtered{T}]+\filtered{Q}, \label{energy}
\end{empheq}

\noindent where $\filtered{\Vec{v}}$ is the filtered velocity, $t$ is time, $\filtered{p}$ filtered pressure, $\mu$ dynamic viscosity, $c_p$ specific heat capacity at constant pressure, $\filtered{T}$ filtered temperature, and $\kappa$ thermal conductivity. The \crefrange{continuity}{energy} also contain the following asumptions, terms, and parameters:

\begin{itemize}
   \item No explicit filter is applied. Thus, it is assumed that the finite support of the computational mesh together with the low-pass characteristics of the discrete differentiating operators act as an effective filter \citep{LUND2003}.
   \item The isotropic part of the \textit{subgrid stress tensor} has been included in the filtered pressure, and the turbulent viscosity $\mu_t$ derived from the Boussinesq assumption is provided by the WALE model \cite{NICOUD1999} with the wall constant $C_w=0.325$. 
   \item The \textit{filtered viscous dissipation function} $\filtered{\phi}_v=\nabla \cdot (\filtered{\tau} \cdot \filtered{\Vec{v}})-\filtered{\Vec{v}} \cdot (\nabla \cdot \filtered{\tau})$ that contains the filtered viscous stress tensor $\filtered{\tau}$ is neglected, since in terms of the Mach number $M^2/Re \ll 1$. 

\item The \textit{turbulent thermal conductivity} $\kappa_t=\mu_tc_p/Pr_t$, where $Pr_t=0.85$ is the turbulent Prandtl number \citep{KAYS1994}.
\item The \textit{filtered heat flux} $\filtered{Q}=0$ and the coefficient $\gamma=1$.
   \end{itemize}

\subsubsection*{Computation of the pressure loss, the thermal conductivity, and the heat transfer coefficient}

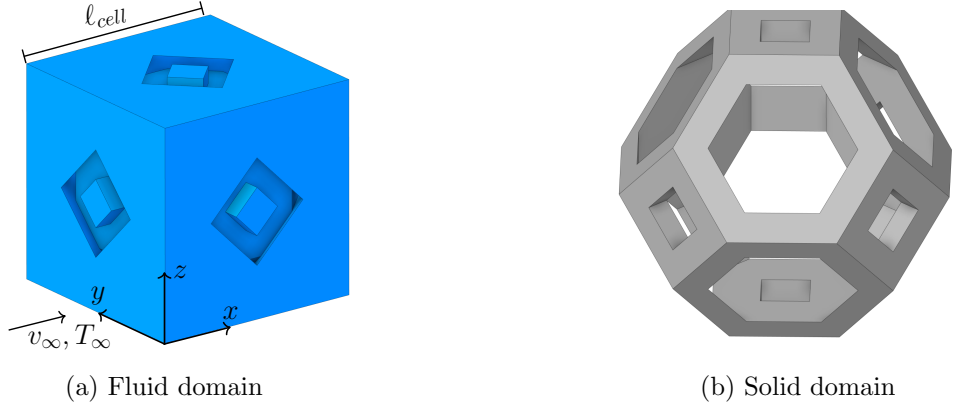
\begin{figure}[H]
  	\centering
    \begin{subfigure}[t]{.485\linewidth}
    \centering 
                \scalebox{0.75}{
        \begin{tikzpicture}[xscale=1, yscale=1, axis/.style={->,thick},line cap=rect]
        \hspace{5mm}
      	\node[anchor=south west,inner sep=0] at (0,0) {\includegraphics[trim={400 100 400 25},width=0.9\textwidth, clip]{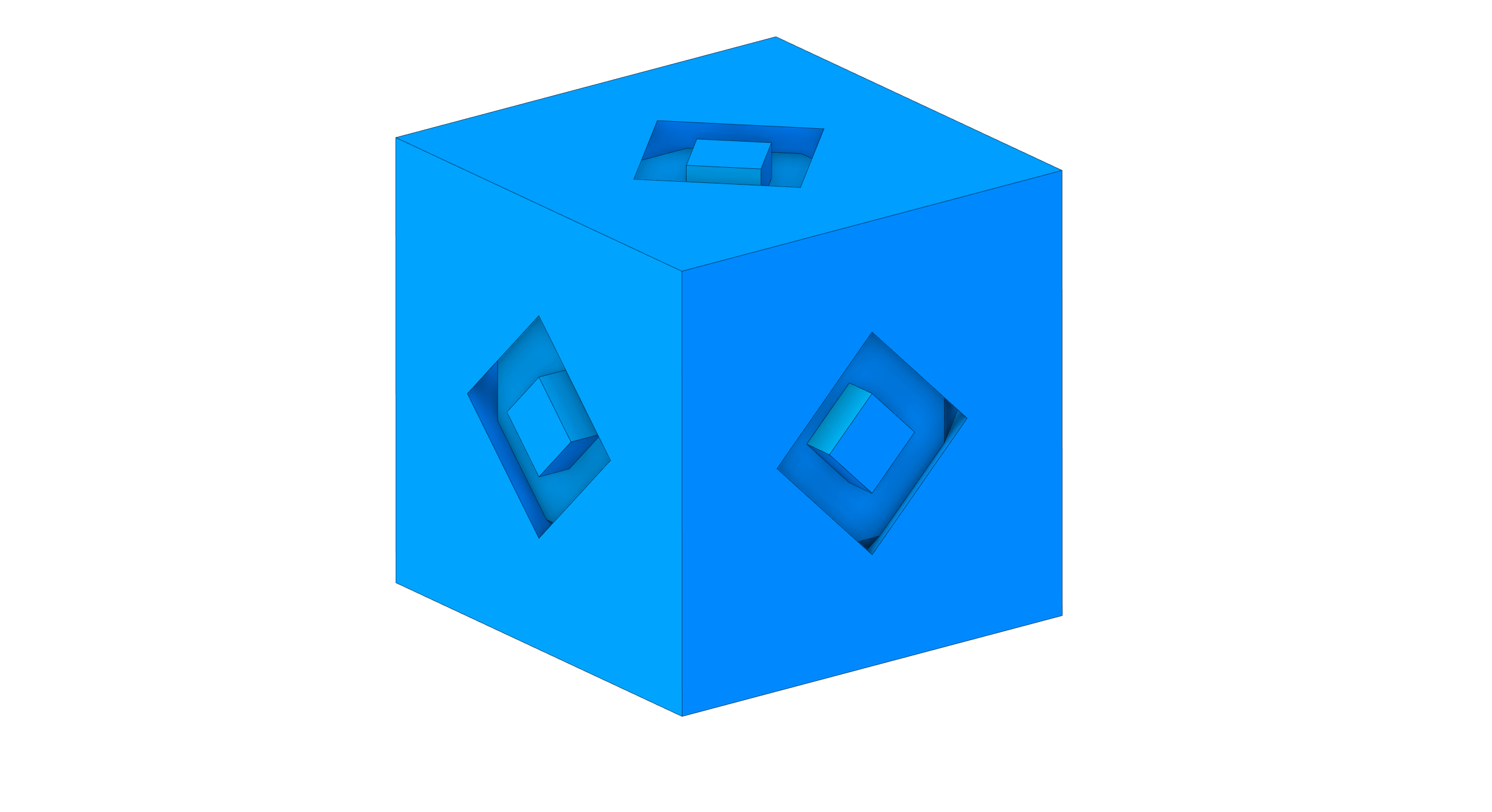}};
	\draw[axis] (3.175,0.5-0.35) -- (4.35,0.45) node[anchor=south]{\large $x$};
	\draw[axis] (3.175,0.5-0.35) -- (3.175,1.425) node[anchor=west]{\large $z$};
	\draw[axis] (3.175,0.5-0.35)-- (2,0.7) node[anchor=south]{\large $y$};
		\draw[quote,line width=0.25mm] (0.675,5.25)  -- (3.85,6.11);
      	\node [rotate=12]  (D)  at (2.125,6) {\large $\ell_{cell}$};
      \draw [->,line width=0.25mm] (0.425,0.4) -- (1+0.425,0.65);
           	\node  at (1.5,0.275) {\large $v_{\infty}, T_{\infty}$};
  		\end{tikzpicture}}
    \caption{Fluid domain}
        \label{geometry-pf}
  \end{subfigure}
   \begin{subfigure}[t]{.485\linewidth}
 	\centering
\includegraphics[trim={500 125 500 100}, width=0.625\textwidth, clip]{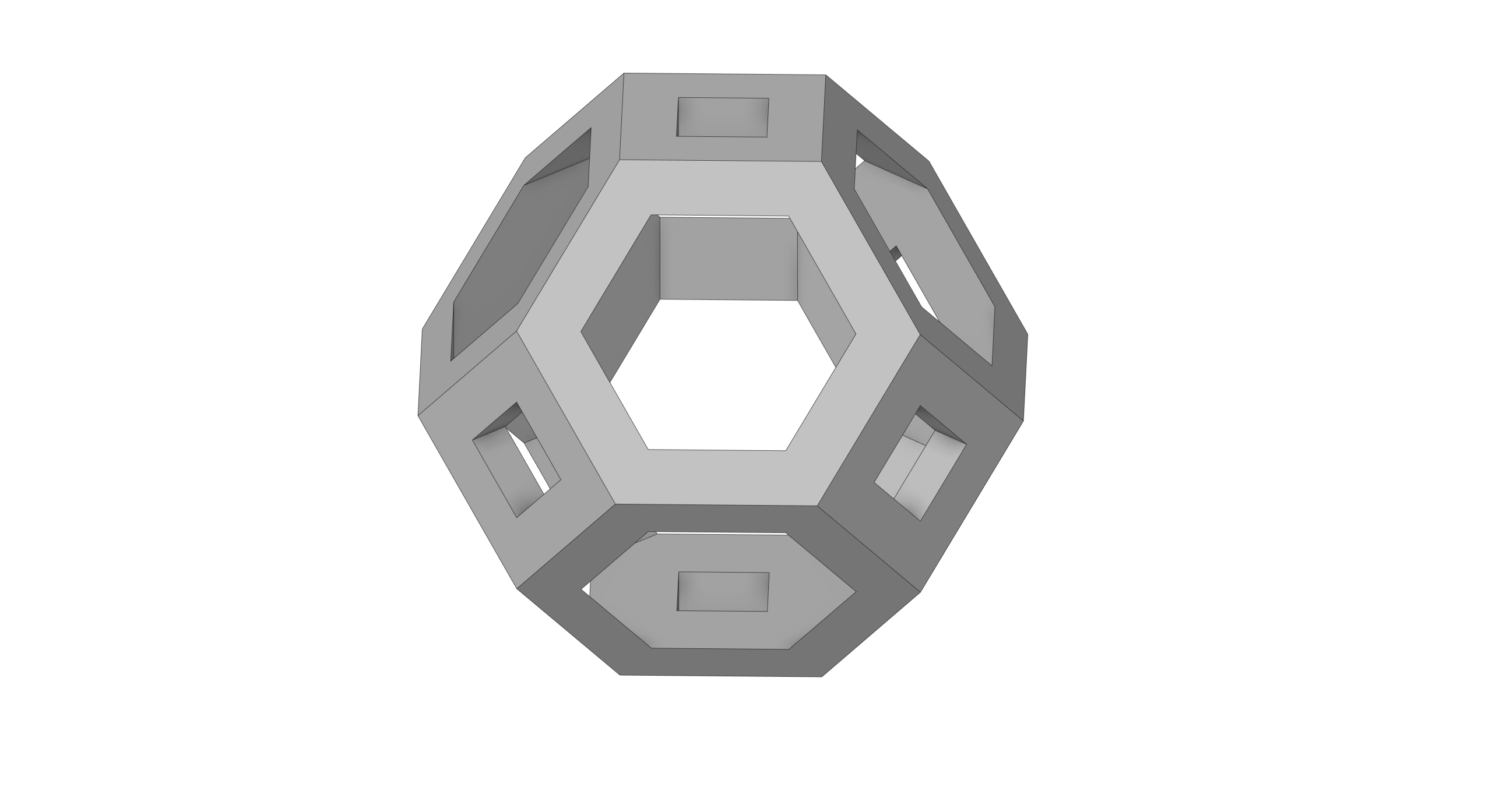}
    \caption{Solid domain}
        \label{geometry-ps}
  \end{subfigure}
  \caption{Geometry of the microscopic model}
    \label{geometry-p}
  \end{figure}

 To compute the pressure loss, a single tetrakaidecahedron cell is used with periodic boundary conditions in the three spatial directions (see \cref{geometry-pf}). The pressure difference $\Delta p=p(x=\ell_{cell})-p(x=0)$ is periodic, and the numerical  implementation is done with a body force in \cref{momentum}: $\filtered{\Vec{S}}=\beta \uveci $. The parameter $\beta$ is the pressure gradient that is updated in each iteration so that the mass flow rate is equal to $\dot{m}=\rho v_{\infty}\ell_{cell}^2$, where $v_{\infty}$ is the freestream velocity. No-slip condition is imposed at the walls of the domain.
   
The results of the simulations provide the pressure gradient that can be modeled as

\begin{equation}
\label{deltap-adj}
    \nabla p =\frac{\Delta p}{\ell_{cell}}=A v_{\infty}+Bv_{\infty}^2=\rho f \frac{v_{\infty}^2 }{2 d_h}
\end{equation}

\noindent for appropriate values of $A$ and $B$ \citep{BEAR2018}. On the right-hand side of \cref{deltap-adj}, $f$ is the friction factor, and $d_h = 4 \Phi/\Sigma_s$ is the hydraulic factor.

The thermal conductivity of foams $\kappa_f$ is obtained with a single tetrakaidecahedron (see \cref{geometry-ps}), which is in contact with two plates of negligible thermal resistance that are placed at $z_1=0$ and $z_2=\ell_{cell}$ at different temperatures. The heat equation is solved with the thermal diffusivity of the solid material $\alpha_s=\kappa_s/(\rho_s c_s)$, and adiabatic walls are considered except at the foam-plate contacts, where the equality of temperatures and heat fluxes is imposed. In regard to the radiative heat transfer, previous studies for poliurethane \citep{COQUARD2005} and metallic \citep{CONTENTO2014} foams---with cell sizes and porosities that are similar to the foams studied in this work---have shown that its contribution is $\kappa_r \sim 10^{-2}$ \si{\watt\per\meter\per\kelvin}. Since the thermal conductivities of the foams used as heat exchangers $\kappa_f >1$ \si{\watt\per\meter\per\kelvin} (see the introduction of \cref{methodology}) and $\kappa_r/\kappa_f \ll1$, then the radiative heat transfer is neglected in the model. When the simulation is finished, defining $q$ as the heat flux between the plates, the thermal conductivity of the foam is calculated as follows:

\begin{equation}
\label{k-def}
    \kappa_f = \frac{q\ell_{cell}}{(T_2-T_1)}.
\end{equation}

To calculate the heat transfer coefficient, periodic boundary conditions are applied in the momentum equation in the same way as in the calculation of the coefficients of the parabolic pressure loss curve. In this problem the Richardson number $Ri = g \beta \Delta T \mathcal{L}/v^2 \ll 1$ for the gravity acceleration $g \sim 10$ \si{\meter\per\second\squared}, the air thermal expansion coefficient $\beta \sim 10^{-3}$ K$^{-1}$, $\Delta T \sim 10$ \si{\kelvin}, $\mathcal{L} \sim $ 1 \si{\meter} and $v \sim 10$ \si{\meter\per\second}, which means that natural convection is negligible with respect to forced convection. Then, the Nusselt number is a function of the Reynolds and Prandtl numbers, so the solid temperature distribution is assumed to not have an impact on the results. Based on the previous reasoning, the solid surface temperature $T_S$ is an input of the simulations that is considered as constant. As the air flows through the porous medium, its temperature approaches the wall temperature. At sufficient distance from the inlet, the fluid-to-wall temperature differences decay exponentially to zero. In the periodic thermally developed regime, the variable $\Theta=(T(\vec{x})-T_S)/(\tilde{T}_{\infty}-T_S)$ is periodic \citep{PATANKAR1977}, where $\tilde{T}_{\infty}$ is the flux-weighted freestream temperature:

\begin{equation}
\tilde{T}_{\infty}=\dfrac{\displaystyle \int_{x=0}T |\rho \vec{v}\cdot \mathrm{d}\vec{\mathcal{S}}|}{\displaystyle \int_{x=0}|\rho \vec{v}\cdot \mathrm{d}\vec{\mathcal{S}}|},
\end{equation}

\noindent which is an input of the simulations. The heat transfer coefficient $h$ is calculated at the surfaces of the microscopic domain $\partial \Omega_f^{\ell}$ as follows:

\begin{equation}
\label{hv-coef}
    h= \frac{1}{A_p}\int_{\partial \Omega_f^{\ell}} \frac{\filtered{q}}{T_S-\tilde{T}_{\infty}} \ \mathrm{d}\mathcal{S}=\frac{1}{A_p}\int_{\partial \Omega_f^{\ell}} \filtered{h} \ \mathrm{d}\mathcal{S},
\end{equation}

\noindent where $\filtered{q}$ is the filtered heat flux, and $A_p$ is the surface area of the porous medium. It has been verified that $h$ defined in \cref{hv-coef} does not depend on the values of the inputs $T_S$ and $\tilde{T}_{\infty}$. The Nusselt number referred to the microscopic length scale $\ell$ is given by the following expression:

\begin{equation}
\label{Nu-def}
    Nu_{\ell}=\frac{h \ell}{\kappa}.
\end{equation}

\subsubsection*{Mesh}

The computational mesh consists of cubes in the bulk region, with a high-quality layered poly-prism mesh in the boundary layers. These two meshes are connected with general polyhedral elements. Two different cell sizes are considered: one for the bulk region, $\ell_{mesh}$, and other one equal to $\ell_{mesh}$/2 close to the walls. The value selected for the simulations will be deduced and justified from a mesh-independence study presented in \cref{results}. That mesh is used for the calculation of the pressure loss and the Nusselt number, while the mesh used in the calculation of the thermal conductivity has only one cell size equal to  $\ell_{mesh}$.

\subsubsection{Macroscopic model}
\label{model-macro}

\subsubsection*{Geometry}

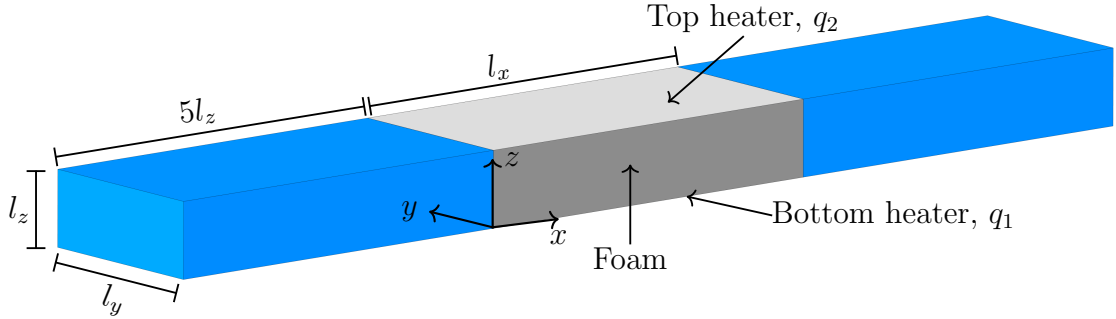
\begin{figure}[H]
    \centering 
        \begin{tikzpicture}[xscale=1, yscale=1, axis/.style={->,thick},line cap=rect]
        \hspace{3mm}
      	\node[anchor=south west,inner sep=0] at (0,0) {\includegraphics[trim={100 650 100 650},width=\textwidth, clip]{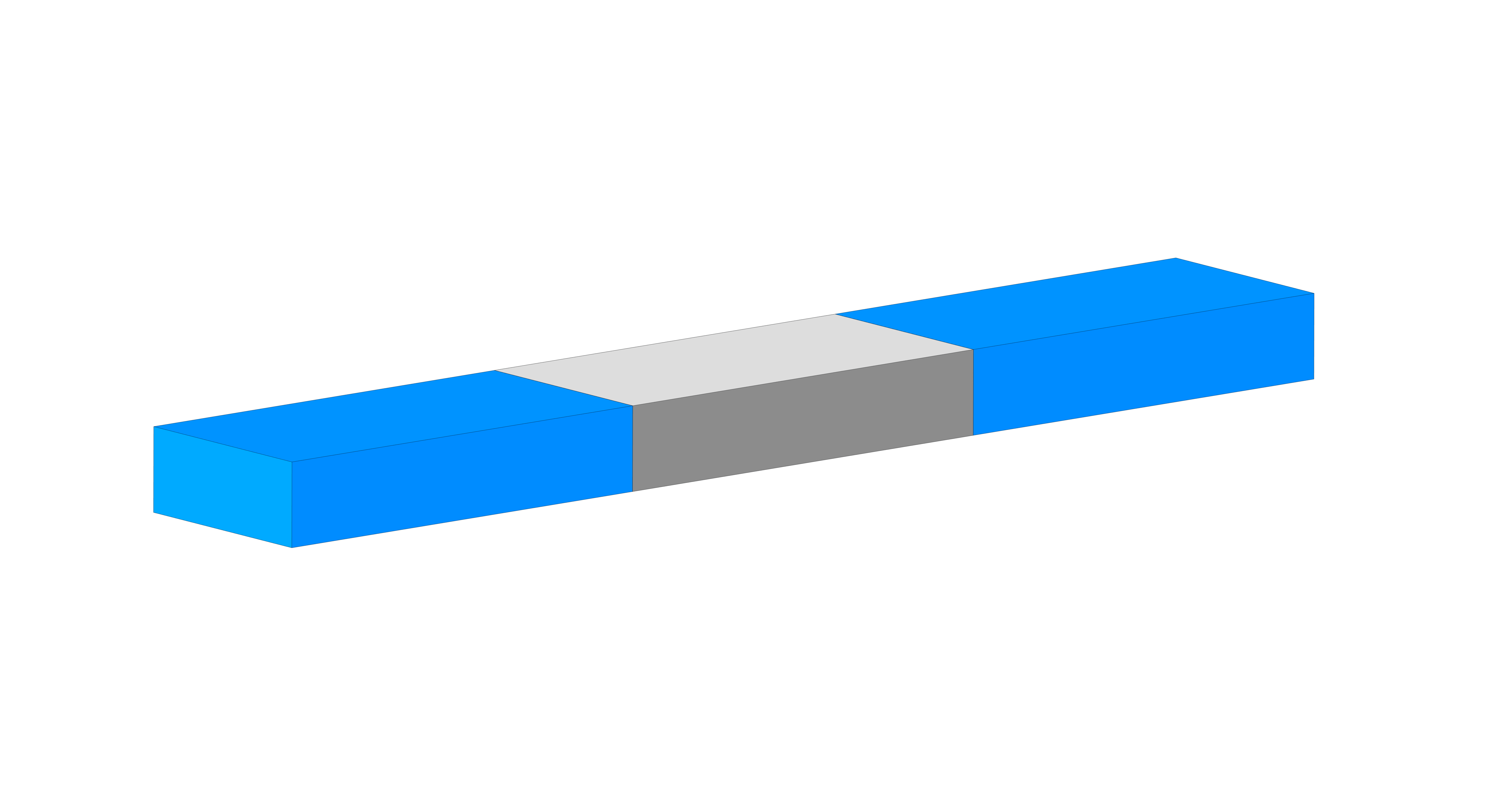}};
			    \node   (A)  at (0.875,1.75-0.675) {$l_z$};
			     \node[rotate=-8]   (B)  at (2.125,-0.125) {$l_y$};
			 	\node[rotate=8]   at (7.25,3.4-0.425) {$l_x$};
			 	\node[rotate=8]   at (3.25,2.375) {$5l_z$};
			\draw[quote,line width=0.25mm] (1.1,1-0.45)  -- (1.1,2.25-0.65);
	\draw[quote,line width=0.25mm] (1.35,0.8-0.425)  -- (2.975,-0.0375);
	\draw[quote,line width=0.25mm] (5.525,2.85-0.4)  -- (9.625,3.5-0.375);
	\draw[quote,line width=0.25mm] (1.35,1.75+0.025)  -- (5.475,2.425+0.025);
	\draw[axis] (7.17,1.4-0.56) -- (8.05,0.95) node[anchor=north]{$x$};
	\draw[axis] (7.17,1.4-0.56) -- (7.17,1.75) node[anchor=west]{$z$};
	\draw[axis] (7.17,1.4-0.56)-- (6.325,1.4-0.35) node[anchor=east]{$y$};
	 \node  at (9,0.85-0.425) {Foam};
	 \node  at (10.5,3.625) {Top heater, $q_{2}$};
	  \draw [->,line width=0.25mm] (10.5,3.375) -- (9.5,2.5);
	 \draw [->,line width=0.25mm] (9,1.05-0.425) -- (9,2.1-0.425);
	 \node  at (12.5,1) {Bottom heater, $q_{1}$};
	 \draw [->,line width=0.25mm] (10.875,1) -- (9.75,1.25);
  		\end{tikzpicture}
  \caption{Geometry of the simulations of the macroscopic model}
  \label{geo-ohtc}
  \end{figure}

The geometry of the macroscopic model is the one of the physical system to be simulated. The experiments considered in this work are rectangular channels with a foam whose inlet is placed at the origin of the coordinate system (see \cref{geo-ohtc}).

\subsubsection*{Equations}

In the regions with no foam---colored in blue in \cref{geo-ohtc}--- the equations are the same as the ones of the microscopic model. In the foam region, some modifications are done:

\begin{itemize}

\item No turbulence model is considered; that is, $\mu_t=0$ in \cref{momentum,energy}.
 \item The \textit{filtered source term} $\filtered{\Vec{S}}$ takes into account the pressure loss of the foam \cite{BEAR2018}:
    \begin{equation}
    \label{pd-coeff}
    \filtered{\Vec{S}}=\Phi \bigg(\frac{\mu}{\mathcal{P}}+\frac{\Phi\rho C_d}{\sqrt{\mathcal{P}}}|\filtered{\Vec{v}}|\bigg) \filtered{\Vec{v}}\   \mathbf {1}_{\Omega_f^{\mathcal{L}}}.
    \end{equation}
where $\mathbf {1}_{\Omega_f^{\mathcal{L}}}$ is the indicator function in the macroscopic domain $\Omega_f^{\mathcal{L}}$. The first component (Darcy's law) represents the drag of Stokes flows ($Re\ll1$), while the second component (Forchheimer's law) provides the general expression for $Re\gg1$, with $\mathcal{P}$ and $C_d$ referring to the permeability and the drag coefficients, respectively. Assuming that the flow reaches a steady state, and that viscous effects are negligible, the values are derived from \cref{deltap-adj}:
\begin{equation}
    \mathcal{P}=\frac{\mu}{A}, \hspace{5mm} C_d =\frac{B}{\rho}\sqrt{\frac{\mu}{A}}.
\end{equation}
\item The air and the solid material of the foam are considered to be an homogeneous mixture. Based on the theory of multiphase flows \cite{BRENNEN2005}, the specific enthalpy of the foam is:
\begin{equation}
    \underbrace{\rho_f h_f}_{\text{Foam}}= \underbrace{\Phi \rho c_p(T-T_{ref})}_{\text{Air}}+\underbrace{(1-\Phi)\rho_sc_s (T_s-T_{ref})}_{\text{Solid}},
\end{equation}
where $T_{ref}=298.15$ \si{\kelvin} is the reference temperature, and $T_s$ the solid temperature.
\item The air temperature is given by \cref{energy} with $\gamma=\Phi$, and the  \textit{filtered heat source} $\filtered{Q}$ models the convective heat transfer:
\begin{equation}
\label{heat-source}
\filtered{Q}= h \Sigma_s(T_s-\filtered{T}) \mathbf {1}_{\Omega_f^{\mathcal{L}}}.
\end{equation}
The heat transfer coefficient obtained from the microscopic model $h$ (\cref{hv-coef}) is used, and $T_s$ refers to the solution of the energy equation in the solid:
\begin{equation}
\label{energy-solid}
(1-\Phi)\rho_s c_s \frac{\partial T_s}{\partial t}=\kappa_f \nabla^2 T_s+h \Sigma_s(\filtered{T}-T_s)\mathbf {1}_{\Omega_f^{\mathcal{L}}}.
\end{equation}
where $\kappa_f$ is the thermal conductivity of the foam.
\end{itemize}

\subsubsection*{Computation of the overall heat transfer coefficient and the wall temperature}

In this model the freestream velocity $v_{\infty}$ and temperature $T_{\infty}$ are the boundary conditions at the inlet, and zero-gradient boundary conditions for the flow variables at the outlet. Depending on the experiment considered, there are up to two copper heaters that provide a heat flux of $q_1$ and $q_2$. This is implemented with a source term in the energy equation equal to the heat flux divided by the thickness of the heater. The plane $y=0$ is considered as a symmetry plane.

Two indicators of the performance of a system will be calculated with the model: the mean temperature of the heater $T_w$, which is the area-weighted average of the temperature of the mesh faces in contact with the heater surface, and the overall heat transfer coefficient \citep{MANCIN2013}:

\begin{equation}
\label{power-h}
U=\frac{P_h}{A_{h}\Delta T_{log}},
\end{equation}

\noindent where $A_h=l_x l_y$ is the surface area of the heater, and $P_h$ is the power supplied by the heater. Assuming that the walls of the domain are adiabatic, then the energy equation in a control volume consisting of the domain of \cref{geo-ohtc} states that all of the power is transferred to the air:

\begin{equation}
\label{p-equal0}
  P_h=P_{air}=\dot{m}(h_{0}^{out}-h_{0}^{in}),   
\end{equation}

\noindent where $\dot{m}$ is the mass flow rate and $h_0$ is the air specific stagnation enthalpy. Since the sectional areas of the inlet and outlet are the same, \cref{p-equal0} is reduced to
\begin{equation}
\label{p-equal}
 P_h=\dot{m}c_p (T^{out}-T_{\infty})=\rho v_{\infty} l_y l_z c_p (T^{out}-T_{\infty}).   
\end{equation}
The last term of \cref{power-h} to be defined is the logarithmic temperature difference  $\Delta T_{log}$:
\begin{equation}
\Delta T_{log}=\frac{(T_{w}^{in}-T_{\infty})-(T_{w}^{out}-T^{out})}{\ln \bigg(\dfrac{T_{w}^{in}-T_{\infty}}{T_{w}^{out}-T^{out}}\bigg)},
\label{log-T}
\end{equation}
\noindent where $T_w^{in}$ and $T_w^{out}$ are the inlet and outlet temperature of the wall in contact with the foam, and $T^{out}$ is the outlet air temperature.

\subsubsection*{Mesh}

The same configuration of hexahedral and polyhedral elements described in the microscopic model is used. After performing a mesh-independence study, two different cell sizes are considered taking as a reference \cref{geo-ohtc}: one for the blue domain equal to $l_z/20$, and other for the foam (grey) domain equal to $l_z/40$. The number of prism layers in the boundary layers near the walls of the domain is equal to 20. 

\subsubsection*{Analytical solution}
\label{oom}

An analytical approach will be derived to study the influence of the parameters of the problem in preliminary studies. As a first approximation, the following assumptions are made:

\begin{itemize}
    \item The air and the solid material of the porous medium are in thermal equilibrium.
    \item The flow is two-dimensional, and a steady condition is reached.
    \item The longitudinal velocity $v_{\infty}$ is the only non-zero component of the velocity vector.
\end{itemize}

In the thermal boundary layer, the energy equation and the boundary conditions are:

\begin{empheq}[left={\empheqlbrace}]{alignat=3}
	&  v_{\infty}\frac{\partial T}{\partial x}=\alpha_f \frac{\partial^2 T}{\partial z^2} & \text{in } \Omega_f^{\mathcal{L}}\\
	&  \kappa_f \frac{\partial T}{\partial z}=q & \text{in } \partial \Omega_f^{\mathcal{L}} \\
	& T \to T_c \hspace{5mm} &  \text{if } \eta \to \infty, \label{bc-inf}
\end{empheq}

\noindent where $\alpha_f=\kappa_f/(\rho c_p)$ is the foam equivalent thermal diffusivity, and $T_c$ is a characteristic temperature. Introducing the following self-similar variables

\begin{equation}
\eta = \sqrt{\dfrac{v_{\infty}}{\alpha_f}}\frac{z}{\sqrt{x}}, \hspace{5mm} \Theta=\sqrt{\dfrac{v_{\infty}}{\alpha_f}}\frac{\kappa_f}{q }\frac{T-T_{c}}{\sqrt{x}},
\end{equation}

\medskip

\noindent then the previous equation can be solved analytically to give:

\begin{equation}
\label{T-analytical}
\Theta=\frac{2}{\sqrt{\pi}}\exp\bigg(-\frac{\eta^2}{4}\bigg)-\eta \erfc \bigg(\frac{\eta}{2}\bigg),
\end{equation}

\noindent where $\erfc$ is the complementary error function. The solution at the wall is:

\begin{equation}
\label{analytical}
T_w-T_c=2q\sqrt{\frac{x}{\pi\rho c_p \kappa_f v_{\infty}}}.
\end{equation}

\subsubsection{Numerical implementation}
\label{n-implementation}

The models are solved with the finite volume method that is implemented in Ansys Fluent 2022R1. Flow variables are stored at cell centers, and simulation values at cell faces are computed by means of interpolation schemes. A central difference interpolation scheme is used for pressure and temperature. In regards to the velocity, a central difference scheme is applied in the momentum equation, while in the continuity equation a corrected momentum interpolation is proposed to avoid pressure checkerboarding \citep{MAJUMDAR1988}.  Gradients are obtained by a least-squares procedure, and the temporal discretization is done with a second order implicit scheme. The pressure-velocity coupling is performed with the SIMPLEC algorithm \cite{FERZINGER2020}, and in each iteration the resulting system of algebraic equations is linearized and solved with an algebraic multigrid (AMG) method. The simulations are performed with a time step $\Delta t$ such that the convective Courant number $Co=v_{cell}\Delta t/\ell_{cell}<1$, and are finished when the residuals of the systems of equations and the output variables reach a statistically-steady state. Two identical domains are created for the air and solid of the foam, and \cref{energy} and \cref{energy-solid} are solved in the corresponding domain. All solid layers with high aspect ratios such as the heaters (see \cref{geo-ohtc}) are implemented using the Shell Conduction model. This model does not require that the wall thickness is taken into account in the mesh generator. The solid layers mentioned are generated automatically by the solver of Ansys Fluent \citep{ANSYS2022} before the simulation starts. The shells are adiabatic on the sides where there is no physical contact with other component, and otherwise the equality of the temperatures and heat fluxes is imposed. In the shell-foam contacts, the shells are coupled with the solid domain of the porous medium.

\subsection{Experimental setup}
\label{exp-setup}

To validate the results given by the multiscale model, an experimental setup has been built at CERN (see \cref{overview-ts}). The geometry of the foams tested are rectangular cubes of $\ell_x \times \ell_y \times \ell_z=6\times60\times6$ mm that represent characteristic dimensions of the foams to be used in future HEP particle detectors. The flow meter includes flow stabilizers at the inlet and at the outlet while providing a low pressure drop (up to $275$ \si{\pascal} for $v_{\infty}=10$ m/s) and an error of $\pm 0.07 v_{\infty}$. Simulations have been performed to deduce the minimum lengths of the different parts of the system so that turbulence levels and non-uniformities do not affect the accuracy of the measurements. Since the pressure loss of the K9 foam is unknown, the fan with the highest pressure loss is selected. The pressure loss caused by the foam is measured by means of a differential pressure sensor of error $\pm 50$ \si{\pascal} and whose ends are located at a distance of 50 mm to the center of the foam. The air temperature is measured by two PT1000 sensors of error $\pm 0.15$ \si{\kelvin} that are at a distance of 100 mm to the foam. To prevent air leaks, the 3D-printed flanges include rectangular cavities to place toric joints (see \cref{assembly-ts}). The adaptors, the flanges, and the two parts of the frame are 3D printed in Accura 25 material, the circular pipe is made of CFRP, black PVC in the case of the frames, and the material of the rectangular pipes is black Plexiglas$^{\text{\textregistered}}$. The material selection---$\kappa < 0.2 $ \si{\watt\per\meter\per\kelvin} in all of the heat transfer surfaces---has been done to maximize the thermal insulation of the system. In addition, Armaflex$^{\text{\textregistered}}$ ($\kappa = 0.03$ \si{\watt\per\meter\per\kelvin}) of 20 mm of thickness is added to reduce the heat losses of the system. To avoid the spread of carbon powder in the test room, the rectangular pipe of the outlet includes a fitting to connect the system to a vacuum cleaner. The data acquisition is done using LabVIEW software.

 \begin{figure}[H]
    \centering 
      		  \begin{subfigure}[t]{\linewidth}
      		   	\centering
      \scalebox{0.85}{
        \begin{tikzpicture}[xscale=1, yscale=1, axis/.style={->,thick},line cap=rect]
      	\node[anchor=south west,inner sep=0] at (0,0) {\includegraphics[trim={0 275 25 200},width=\textwidth, clip]{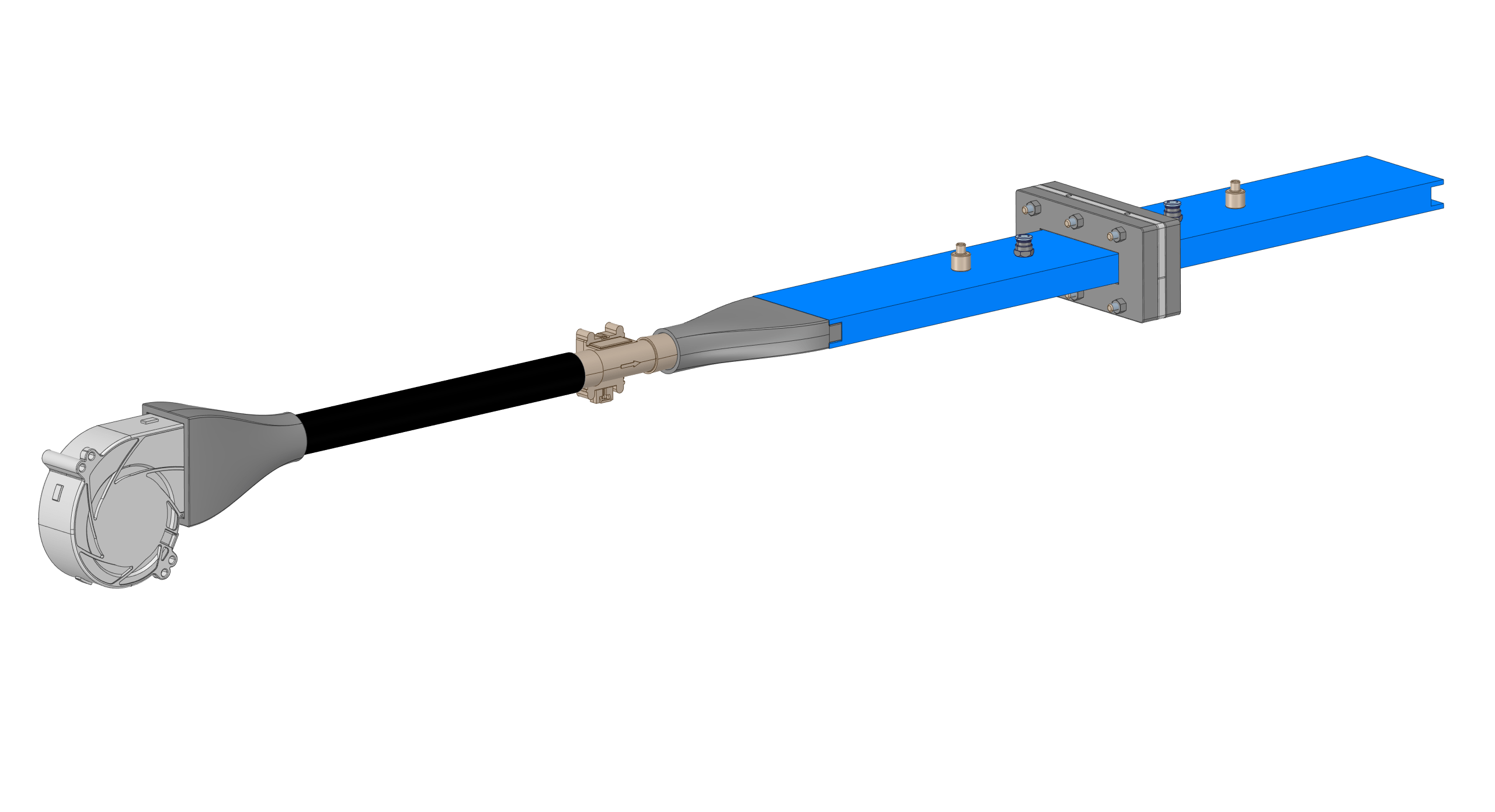}};
			 \node[rotate=11.5]    at (5.25,1.625) {\small 225 mm};
			  \node[rotate=11.5]    at (11.5,2.825) {\small 225 mm};
	\draw[quote,line width=0.25mm] (3.625,1.5)  -- (6.75,2.2);
	\draw[quote,line width=0.25mm] (9.725,2.75)  -- (12.85,3.475);
	 \node    at (0.9,2.75) {\small Fan};
	 	 \node    at (2.15,3) {\small Adaptor};
	 	 \node    at (4.5,3.1) {\small Circular pipe};
	 	  	 \node[text width=2cm]    at (7,3.825) {\small Flow meter};
	 	  	 \node    at (8.25,4.25) {\small Adaptor};
	 	  	 \node    at (12,5.25) {\small Rectangular pipes};
	 	  	 \node    at (15,3.125) {\small $p$, $T$ sensors};
	 \draw [->,line width=0.25mm] (0.9,2.5) -- (0.9,1.875);
	 \draw [->,line width=0.25mm] (2.2,2.75) -- (2.2,2.25);
	 \draw [->,line width=0.25mm] (4.45,2.85) -- (4.45,2.375);
	 \draw [->,line width=0.25mm] (7,3.575) -- (7,3.25);
	 \draw [->,line width=0.25mm] (8.25,4) -- (8.25,3.375);
	 \draw [->,line width=0.25mm] (12,5) -- (10,3.875);
	\draw [->,line width=0.25mm] (12,5) -- (15,4.75);
	\draw [->,line width=0.25mm] (15,3.5) -- (14.425,4.5);
	\draw [->,line width=0.25mm] (15,3.5) -- (13.75,4.25);
  		\end{tikzpicture}}
  \caption{General overview}
  		\label{overview-ts}
  		\end{subfigure}
  		\end{figure}
  		  \vspace{-5mm}
  		    \begin{figure}[H]\ContinuedFloat
  		     	\centering
  		 		  \begin{subfigure}[t]{.445\linewidth}
  \centering
       \scalebox{0.95}{
     \begin{tikzpicture}[xscale=1, yscale=1, axis/.style={->,thick},line cap=rect]
	  	\node[anchor=south west,inner sep=0] at (8,-1.5) {\includegraphics[trim={85 100 100 75},width=0.925\textwidth, clip]{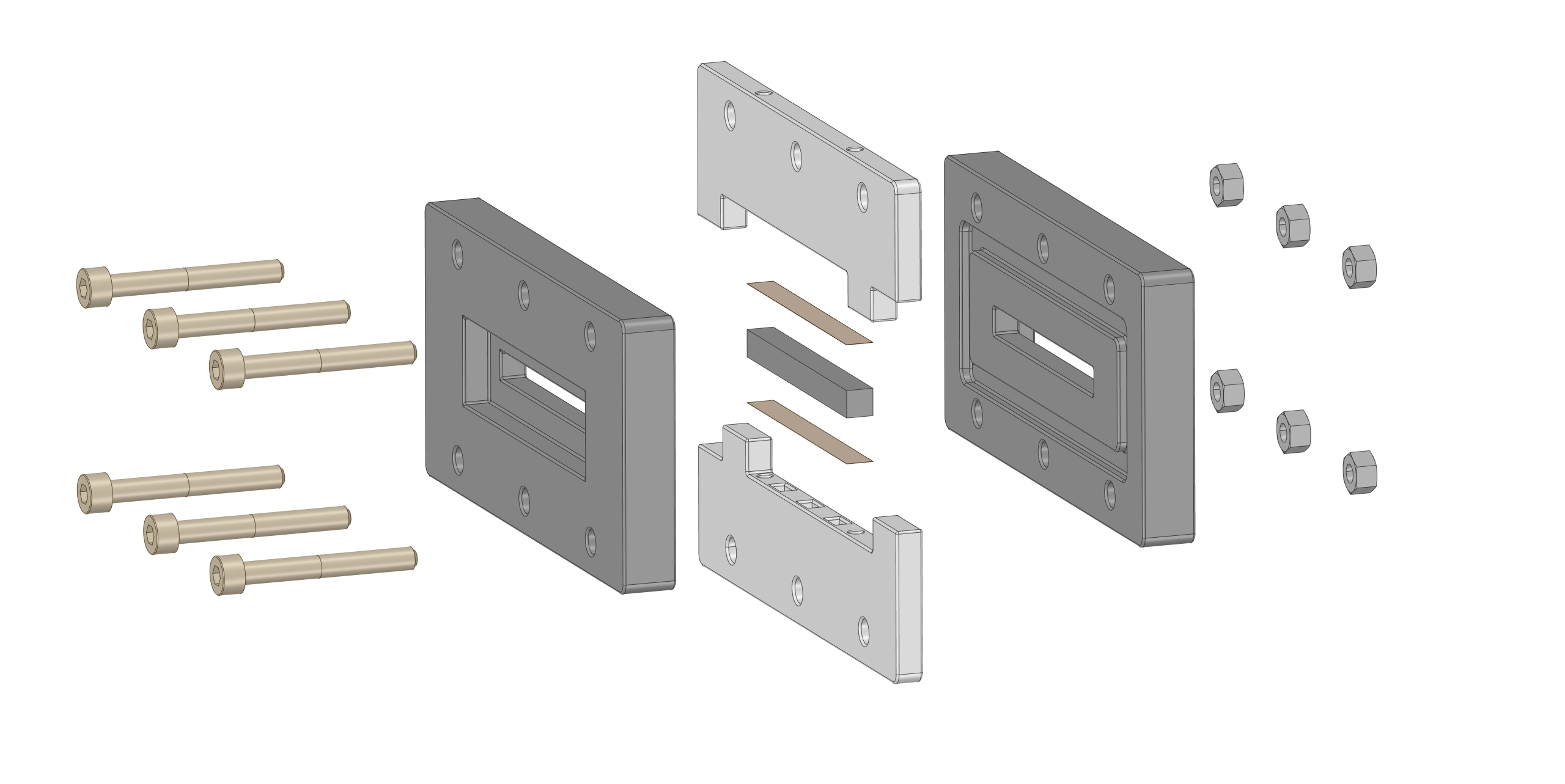}};
	    \node    at (8.5,1.25) {\footnotesize Flange};
		\draw [->,line width=0.25mm] (9.125,1.25) -- (9.875,0.9);
		\node    at (12.825,1.5) {\footnotesize Frame};
		\draw [->,line width=0.25mm] (12.25,1.5) -- (11.75,1.5);
	    \node  at (10.5,1.225) {\footnotesize Foam};
	    	\draw [->,line width=0.25mm] (10.725,1) -- (11.425,0.275);
	   \node  at (13,-1.25) {\footnotesize Heaters};
	   	\draw [->,line width=0.25mm] (13,-1.125) -- (11.95,-0.325);
	   	\draw [->,line width=0.25mm] (13,-1.125) -- (11.95,0.275);
  		\end{tikzpicture}}
  		\caption{Sample assembly}
  		\label{assembly-ts}
  \end{subfigure}
      		  \begin{subfigure}[t]{.525\linewidth}
  \centering
       \scalebox{0.95}{
     \begin{tikzpicture}[xscale=1, yscale=1, axis/.style={->,thick},line cap=rect]
	  	\node[anchor=south west,inner sep=0] at (-0.425,0) {\includegraphics[trim={200 350 1000 625},width=0.725\textwidth, clip]{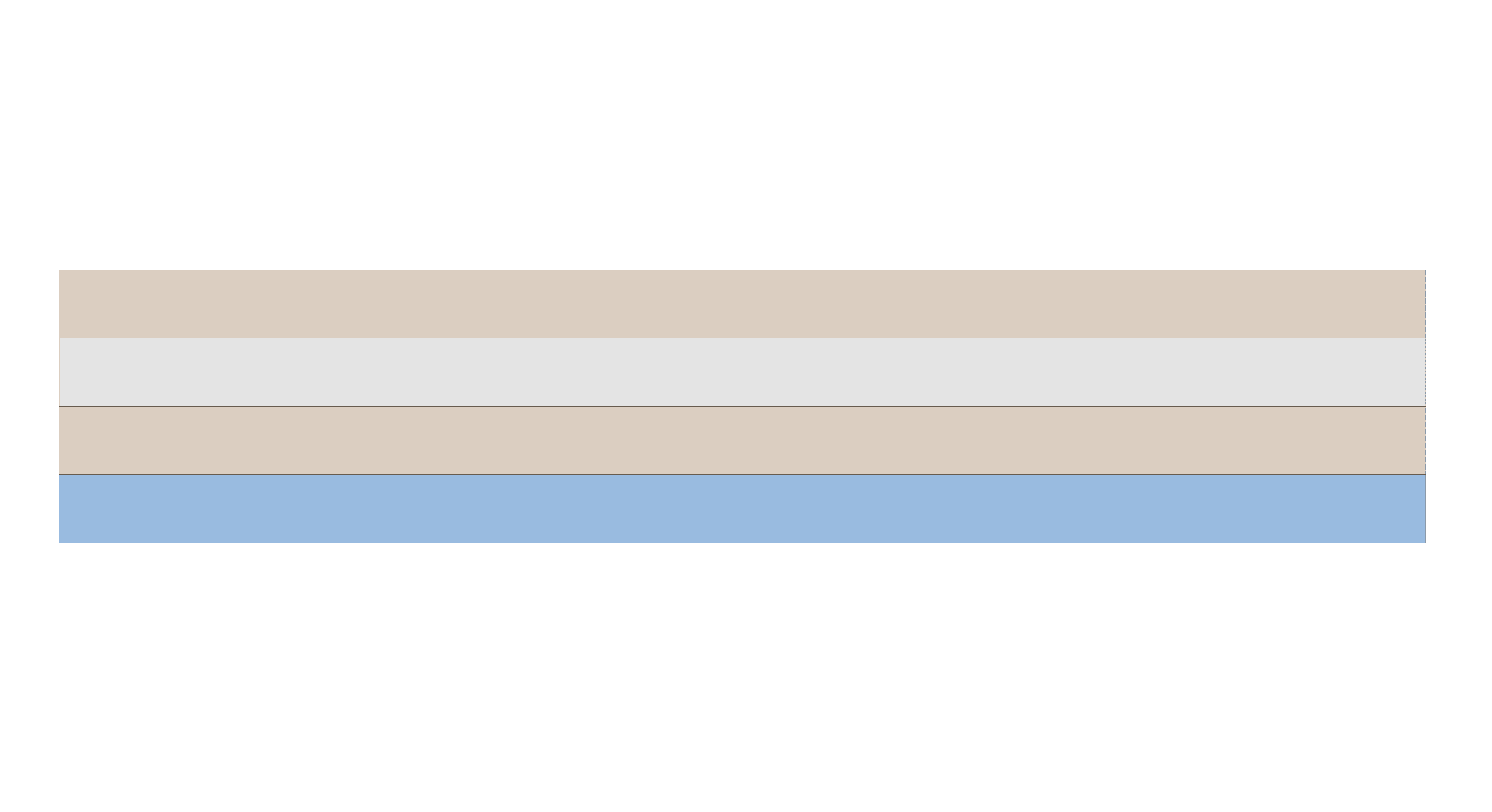}};
	  	    \node at (-0.2,0.825)[circle,fill,inner sep=1.75pt]{};
	  	    \node at (2.75,0.825)[circle,fill,inner sep=1.75pt]{};
	  	    \node    at (0.325,0.75) {\footnotesize  $T_{side}$};
	  	    \node    at (3.375,0.75) {\footnotesize  $T_{center}$};
	  		\node    at (2.75,1.175) {\footnotesize    Glue};
	  		\node    at (2.75,1.5) {\footnotesize  Polyimide};
	  		\node    at (2.75,1.9) {\footnotesize  Copper};
	  		\node    at (2.75,2.275) {\footnotesize Polyimide};
	  		   \draw[quote2,line width=0.25mm](6.175,2.125) -- (6.175,2.525);
	  		  \node    at (-1.075,1.9) {\footnotesize \ 5 $\mu$m};
	  		    \node   at (6.75,2.275) {\footnotesize \ 25 $\mu$m};
	  		   \draw[quote2,line width=0.25mm](-0.5,0.95) -- (-0.5,1.35);
	  		    \draw[quote2,line width=0.25mm](-0.5,1.75) -- (-0.5,2.13);
	  		   \node    at (-1.225,1.125) {\footnotesize \ 100 $\mu$m};
	  		   	\node   at (6.75,1.5) {\footnotesize \ 75 $\mu$m};
	  		   	\draw[quote2,line width=0.25mm](6.175,1.325) -- (6.175,1.725);  		
	  		   	\end{tikzpicture}}
  		\caption{Heater design}
  		\label{heaters}
  \end{subfigure}
  \end{figure}
  \vspace{-10mm}
    \begin{figure}[H]\ContinuedFloat
    \centering
    		  \begin{subfigure}[t]{.525\linewidth}
  \centering
        \scalebox{0.95}{
     \begin{tikzpicture}[xscale=1, yscale=1, axis/.style={->,thick},line cap=rect]
	  	\node[anchor=south west,inner sep=0] at (-0.35,0) {\includegraphics[trim={200 350 1000 625},width=0.775\textwidth, clip]{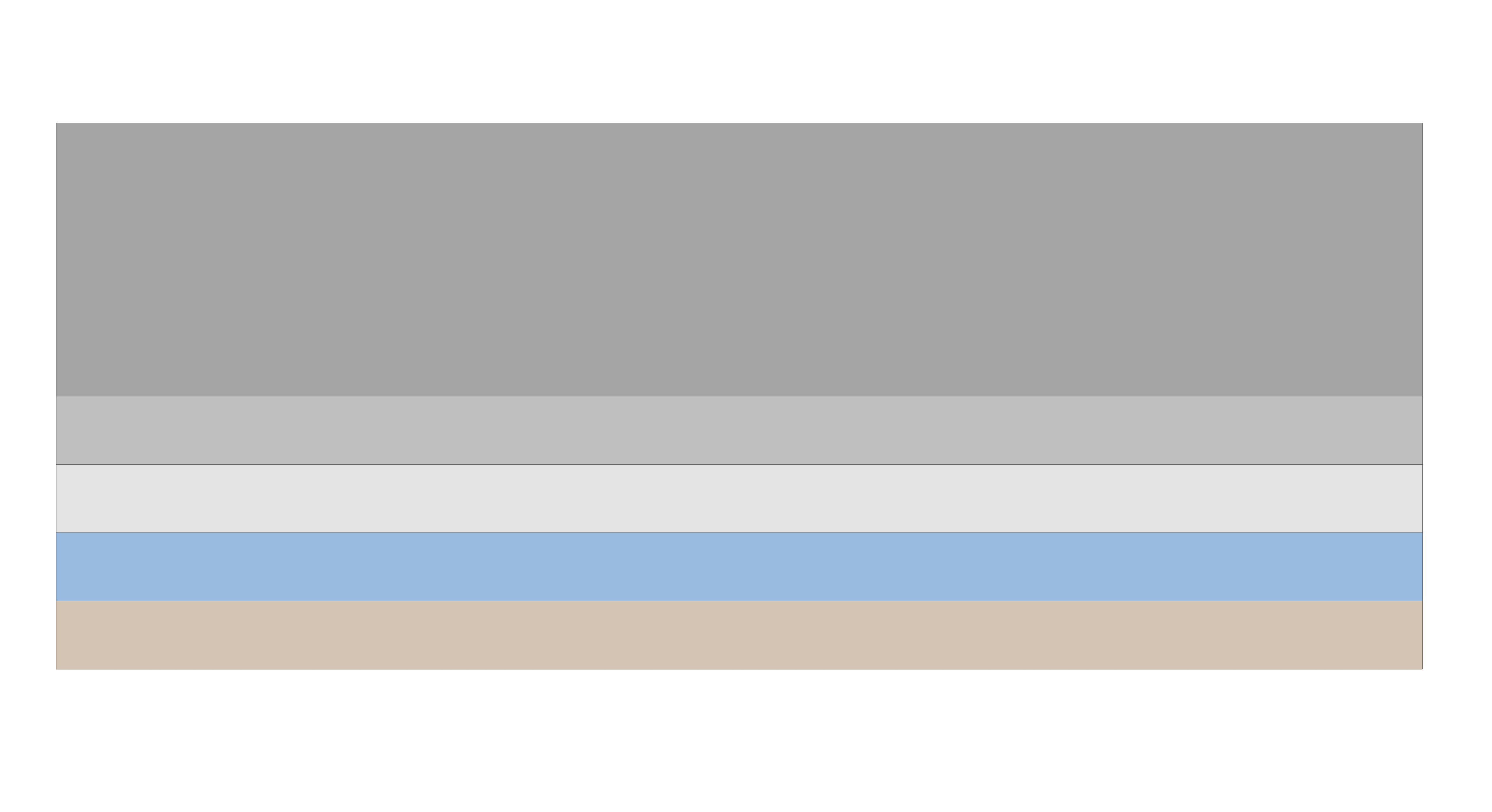}};
	  		\draw[axis] (-0.34,2.925) -- (0.44,2.925) node[anchor=north]{$y$};
	  		\draw[axis] (-0.34,2.925) -- (-0.34,2.16) node[anchor=west]{$z$};
	  		\node    at (3.125,0.275) {\footnotesize Heater};
	  		\node    at (3.125,0.725) {\footnotesize  Glue};
	  		\node    at (3.125,1.15) {\footnotesize  Fleece + glue};
	  		\node    at (3.125,1.6) {\footnotesize  Foam + glue};
	  		\node    at (3.125,2.325) {\footnotesize Foam};
	  		\node    at (3.125,3.25) {};
	  		   \draw[quote2,line width=0.25mm](-0.45,0.05) -- (-0.45,0.5);
	  		   \draw[quote2,line width=0.25mm](6.75,0.525) -- (6.75,0.975);
	  		  \node    at (-1.175,0.25) {\footnotesize \ 100 $\mu$m};
	  		    \node   at (7.375,0.74) {\footnotesize \ 100 $\mu$m};
	  		   \draw[quote2,line width=0.25mm](-0.45,0.975) -- (-0.45,1.425);
	  		   \node    at (-1.175,1.175) {\footnotesize \ 120 $\mu$m};
	  		   	\node   at (7.125,1.65) {\footnotesize \ $\langle \zeta_g\rangle $};
	  		   	\draw[quote2,line width=0.25mm](6.75,1.425) -- (6.75,1.875);  		\end{tikzpicture}}
  		\caption{Foam-heater interface}
  		\label{interface-ts}
  \end{subfigure}
  \caption{Experimental setup for foam characterization}
  \end{figure}
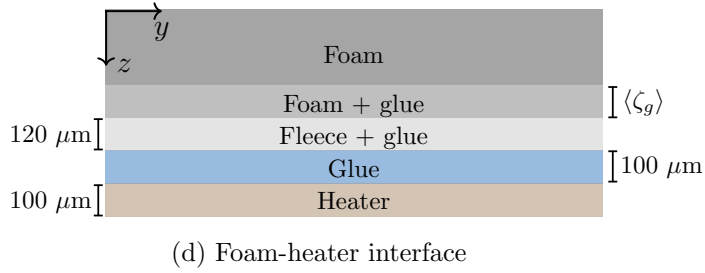

Two custom-made polyimide heaters of 100 \si{\micro\meter} of thickness--consisting of a copper layer of 5 \si{\micro\meter} surrounded by two polyimide layers ($\kappa=0.2$ \si{\watt\per\meter\per\kelvin}) of 25 and 75 \si{\micro\meter}---are placed in contact with the foam to represent the heat dissipation of the silicon sensors (see \cref{assembly-ts}), and the temperatures of the sides of the heater that are not in contact with the foam are measured by two PT1000 in each heater (see \cref{heaters}). The PT1000 have planar dimensions of $1.6 \times 1.2$ mm, and are placed in the middle of the flow direction $x$. Assuming that the walls of the frame are adiabatic, $T_{center}=T_{side}$. However, soldered wires exit from the heaters near the wall, and the effective surface decreases in the neighbourhood of the soldering point. Thus, $T_{side}$ gives additional information to determine possible gradients in the planar direction $y$.
  
The setup aims to represent the thermal interface between a foam and a silicon sensor in future HEP detectors. Previous tests performed at CERN have showed that a direct contact between the foam and the silicon sensor of 50 \si{\micro\meter} of thickness creates footprints in the sensor, which constitute a risk of deterioration of the quality of the measurements. To solve that issue, a carbon fleece of 120 \si{\micro\meter} of thickness and areal density of 8 \si{\gram\per\meter\squared} is added between the foam and the heater. The fleece glued to the foam provides a smoother contact, with additional contact points that lead to the reduction of the contact resistance and the increase of the shear strength of the joint. Moreover, the presence of the fleece helps the control of the thickness of the glue layer of mean thickness $\langle \zeta_g\rangle $ that penetrates into the foam (see \cref{interface-ts}). In the assembly, first the foam is glued to the fleece, and after the curing process the resulting part is glued to the heater. The thickness of the glue layer between the fleece and the heater is 100 \si{\micro\meter}. The glue used is the Epoxies$^{\text{\textregistered}}$ 50-3150 FR, consisting of an epoxy adhesive filled with \ce{Al2O3} powder of 20 $\si{\micro\meter}$, and with a thermal conductivity $\kappa=0.85$ \si{\watt\per\meter\per\kelvin} tested at CERN.

\section{Results and discussion}
\label{results}

In this section the accuracy of the microscopic model is assessed in \cref{micro-results} . Then, the validated outputs of the microscopic model are used in the the macroscopic model, whose results are compared with experimental data, and the analytical solution in \cref{r-macro}.

The parameters of the simulations are shown in \cref{parameters}. The physical properties are obtained from \cite{ASM1991} (Al and Cu), and by private communication with the supplier of the K9 foam, the polyimide (PI) and the glue. The density and thermal conductivity of the K9 material are the ones of the solid material (carbon). The value of the thermal conductivity of the glue tested at CERN is considered instead of the one provided by the official datasheet ($\kappa =2.16$ \si{\watt\per\meter\per\kelvin}). 

	\begin{table}[H]
	\centering
	\begin{tabular}{c c c c c}
		\toprule
		\textbf{Symbol} & \textbf{Material} & \textbf{Parameter} & \textbf{Value}  & \textbf{Units} \\
		\midrule
		$c_p$ & Air&  Specific heat at constant pressure & 1006 & \si{\joule\per\kilogram\per\kelvin}  \\
		\noalign{\vskip 1ex}
		\hdashline\noalign{\vskip 1ex}
    	$c_s$ 	& Al&  Specific heat   & 900 & \si{\joule\per\kilogram\per\kelvin} \\
		\rule{0pt}{2.5ex}
	  & Cu &    & 385 &  \\
				\rule{0pt}{2.5ex}
					 & Glue &   & 1000 &     \\
	\rule{0pt}{2.5ex}
		 & K9 &    & 710 &   \\
	\rule{0pt}{2.5ex}
			 & PI &    & 1100 &   \\
				\noalign{\vskip 1ex}
		\hdashline\noalign{\vskip 1ex}
	 	$\kappa$ & Air & Thermal conductivity  & \num{0.025} & \si{\watt\per\meter\per\kelvin}   \\
	 	\rule{0pt}{2.5ex}
	 & Al &  & \num{218} &   \\
		\rule{0pt}{2.5ex}
			& Cu &  & \num{385} & 	 \\
				\rule{0pt}{2.5ex}
					& Glue &  & \num{0.85} & 	\\
				\rule{0pt}{2.5ex}
		& K9 & 	& \num{1500} & 	 \\
					\rule{0pt}{2.5ex}
			& PI & & \num{0.2} & 	 \\
				\noalign{\vskip 1ex}
		\hdashline\noalign{\vskip 1ex}
		$\mu$ & Air& Dynamic viscosity & \num{1.79e-5}& \si{\pascal\second}  \\
				\noalign{\vskip 1ex}
		\hdashline\noalign{\vskip 1ex}
	$\rho$	& Air & 	Density  & 1.225 &  \si{\kilogram\per\meter\cubed}  \\
				\rule{0pt}{2.5ex}
	& Al & 	 & 2700 & 	 \\
				\rule{0pt}{2.5ex}
	 & Cu &	& 8930 & 	\\
				\rule{0pt}{2.5ex}
						& Glue & & 1600& 	\\
				\rule{0pt}{2.5ex}
		 & K9 & 	& 1800 &  \\
			\rule{0pt}{2.5ex}
	& PI & 	& 1420 & 	  \\
		\bottomrule
		\end{tabular}
		 \caption{Simulation parameters}
		 \label{parameters}
	\end{table}

The material properties of \cref{parameters} are taken at 288 \si{\kelvin}. In the simulations the air and solid temperatures vary between 288 \si{\kelvin} and 388 \si{\kelvin}, with the maximum values achieved in the cases of lowest Reynolds numbers considered in \cref{o-htc-r}. In this temperature range, the material properties whose variations are greater than 10 \% are the air thermal conductivity ($\approx$ 25 \% higher at 388 \si{\kelvin} in both cases) and the dynamic viscosity. This temperature-dependency is neglected in the present model, since it is assumed that the associated error is of the same order as other errors; for example, the simplifications of the multiscale model, the uncertainty on the foam production process, and the accuracy of the experimental devices.
	
\subsection{Microscopic model}
\label{micro-results}

The accuracy of the microscopic model will be tested for the pressure loss in \cref{ploss-r}, the thermal conductivity in \cref{s-k}, and the Nusselt number in \cref{s-h}.

\subsubsection{Pressure loss}
\label{ploss-r}

Four different mesh sizes are studied to obtain a reference for the numerical values. The mesh element length at the walls is two times smaller than in the interior. Other different ratios have been tested such as 1/1, 1/4, and 1/8, and the 1/2 ratio provides the optimum balance between accuracy and computational cost. For error estimation, the filtered mean pressure loss $\langle\filtered{\nabla p}\rangle$ is approximated with a second-order series expansion as a function of the interior mesh element length. The mesh of \num{2.5e5} elements is chosen for subsequent simulations, and from the series expansion its associated error is estimated to be around 8 \%. It is verified that the results obtained with the WALE model and without modeled turbulence are similar for laminar or weakly-turbulent flows as the one considered here (see \cref{mi-study}). The reason why LES is considered in all foams is that for Duocel$^{\text{\textregistered}}$ Al foams, the Reynolds number increases up to $Re_{\ell_{cell}}=3500$ in the cases that will be considered in this section. For $Re_{\ell_{cell}}>250$ the flow in foams is expected to be turbulent \cite{HALL1996,DELLATORRE2014}, with a Kolmogorov length scale that satisfies  $\ell_{cell}/\eta \sim Re_{\ell_{cell}}^{3/4}\sim 10^2$, therefore it is one order of magnitude greater than the mesh element length. With the mesh element sizes selected, simulations have been performed with cubes consisting of multiple cells in each direction and without periodic boundary conditions. The results have lead to the conclusion that the error of the periodic model is lower than 5 \% when the real geometry of the foam has more than eight cells in each direction; that is, with the geometric scales illustrated in the model in Fig. 1, the entrance and exit effects are considered to be negligible for $\mathcal{L}/\ell \geq 8$.

 \begin{table}[H]
\centering
\begin{tabular}{c c c c c}
	\toprule
	\multicolumn{2}{c}{\textbf{Mesh element length}} &\multirow{2}{*}{\begin{tabular}{@{}c@{}} \textbf{Mesh element}\\  \textbf{number} \end{tabular} }  &\multirow{2}{*}{\begin{tabular}{@{}c@{}} $\boldsymbol{\langle\filtered{\nabla p}\rangle}$  \textbf{(Pa/mm)} \\\rule{0pt}{2.5ex}  \textbf{(LES)} \end{tabular} }&\multirow{2}{*}{\begin{tabular}{@{}c@{}} $\boldsymbol{\langle\nabla p}\rangle$  \textbf{(Pa/mm)} \\\rule{0pt}{2.5ex}  \textbf{(No model)} \end{tabular} }\\\cmidrule{1-2}
	\  \textbf{Walls} & \textbf{Interior} & & \\
		\midrule
	\  $\ell_{cell}/50$& $\ell_{cell}/25$ &   \num{5.5e4} &  174.2 & 173.6 \\
 	 \rule{0pt}{2.5ex}
 	 $\ell_{cell}/100$& $\ell_{cell}/50$ &     \num{2.5e5} &  181.2 & 181.2\\
 	 	 \rule{0pt}{2.5ex}
     $\ell_{cell}/200$& $\ell_{cell}/100$ &    \num{1.4e6} &  187.9  & 187.9\\	
 	 	\rule{0pt}{2.5ex}
 	 	 $\ell_{cell}/400$& $\ell_{cell}/200$ & \num{9.5e6} &  191.7 & 191.7 \\	
	\bottomrule
	 \end{tabular}
	\caption{Results of the mesh independence study for RVC foam at $Re_{\ell_{cell}}=275$} %
	\label{mi-study}
	\end{table}

Experiments performed at CERN in the setup described in \cref{exp-setup} and data provided in the official website of ERG are taken as a reference for result comparison. \cref{pressure-loss} compares the numerical results given by the macroscopic model and from experimental data. It is shown that the results of the tests and the simulations agree notably at all values of the freestream velocity for the Duocel$^{\text{\textregistered}}$ and K9 foams, although the simulations overpredict the pressure loss in the Al-20 foam for $Re_{\ell_{cell}}>1250$. However, the porosities of RVC and K9 samples have been measured at CERN with precision balances, but the exact value of Duocel$^{\text{\textregistered}}$ Al foams tested by the manufacturer (ERG) is unknown. Thus, small deviations from the nominal value can lead to differences as the ones shown in \cref{pressure-loss-Al}. The maximum pressure loss that the fan can provide is $3400$ \si{\pascal}, while the pressure loss of a sample of the K9 foam of $6 \times 60 \times 6$ mm is $750 \times 6=4500$ \si{\pascal} (see \cref{k9-deltap}). To obtain the value of the pressure loss of the K9 foam at $Re_{\ell_{cell}} =175$, a sample of $3 \times 60 \times 6$ mm is tested. Since the cell size of the K9 foam is 285 $\si{\micro\meter}$, in 3 mm of length there are $3/0.285 \approx 10$ cells. This value is higher than the number of 8 cells from which the error of the periodic assumption is considered to be negligible. 

 \begin{figure}[H]
 	\centering
 \begin{subfigure}[t]{.485\linewidth}
 	\centering
\includegraphics[trim={0 0 0 0}, width=0.85\textwidth,clip]{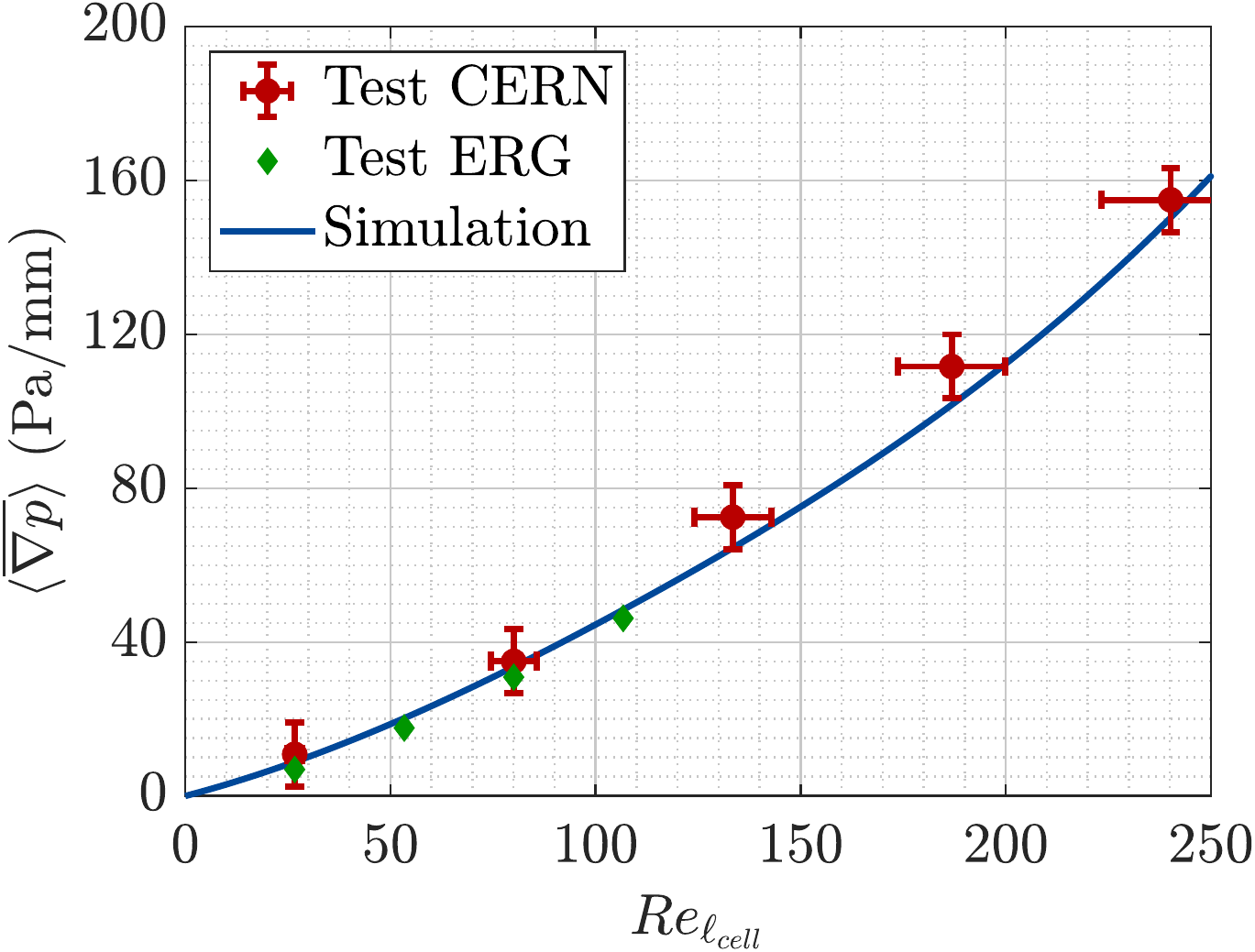}
    \caption{Duocel$^{\text{\textregistered}}$ RVC}
  \end{subfigure}
    \begin{subfigure}[t]{.485\linewidth}
    \centering
    \includegraphics[trim={0 0 0 0},width=0.85\textwidth, clip]{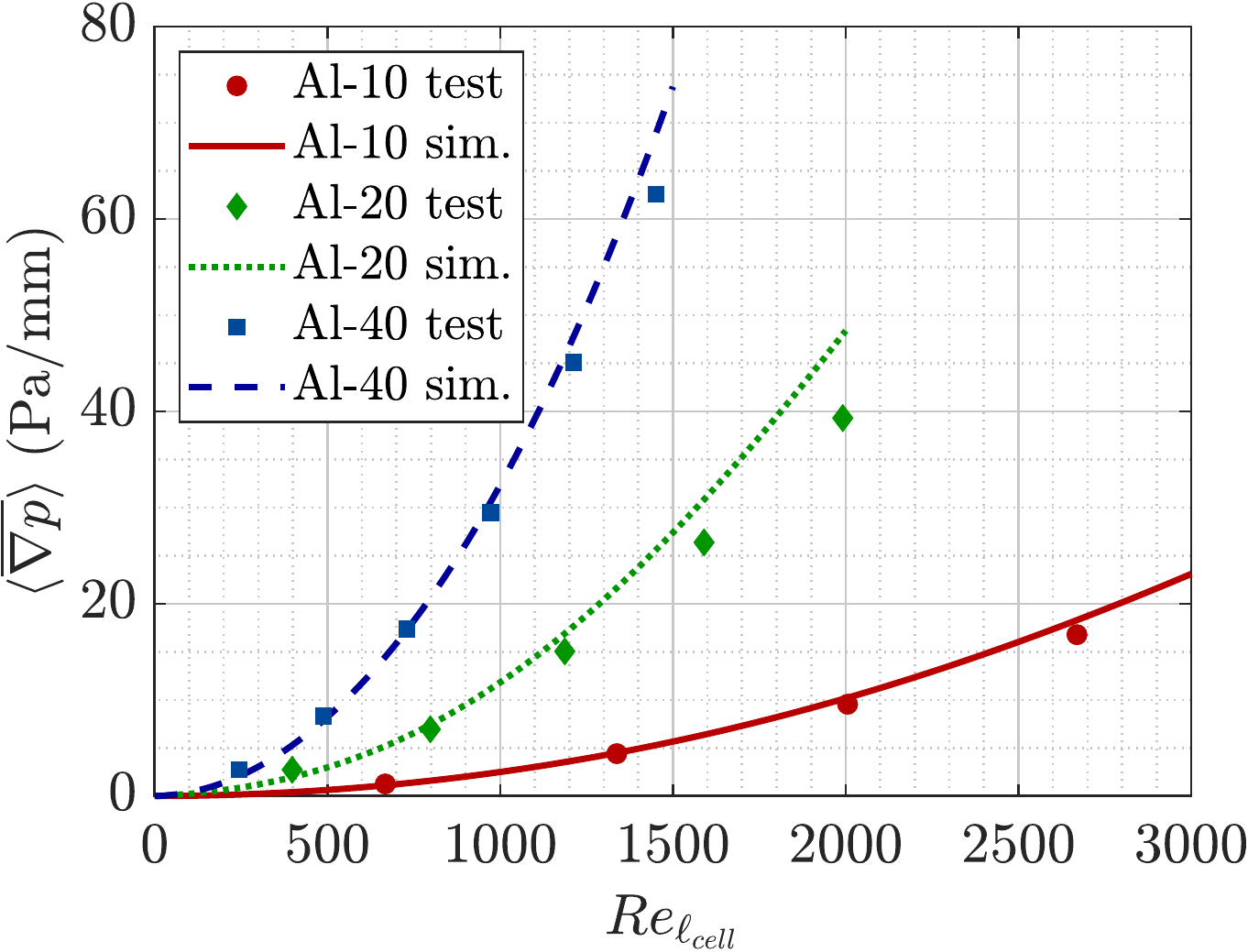}
    \caption{Duocel$^{\text{\textregistered}}$ Al}
    \label{pressure-loss-Al}
  \end{subfigure}
\end{figure}
\vspace*{-2.5mm}
  \begin{figure}[H]\ContinuedFloat
  \centering
     \begin{subfigure}[t]{.485\linewidth}
    \centering
    \includegraphics[trim={0 0 0 0},width=0.85\textwidth, clip]{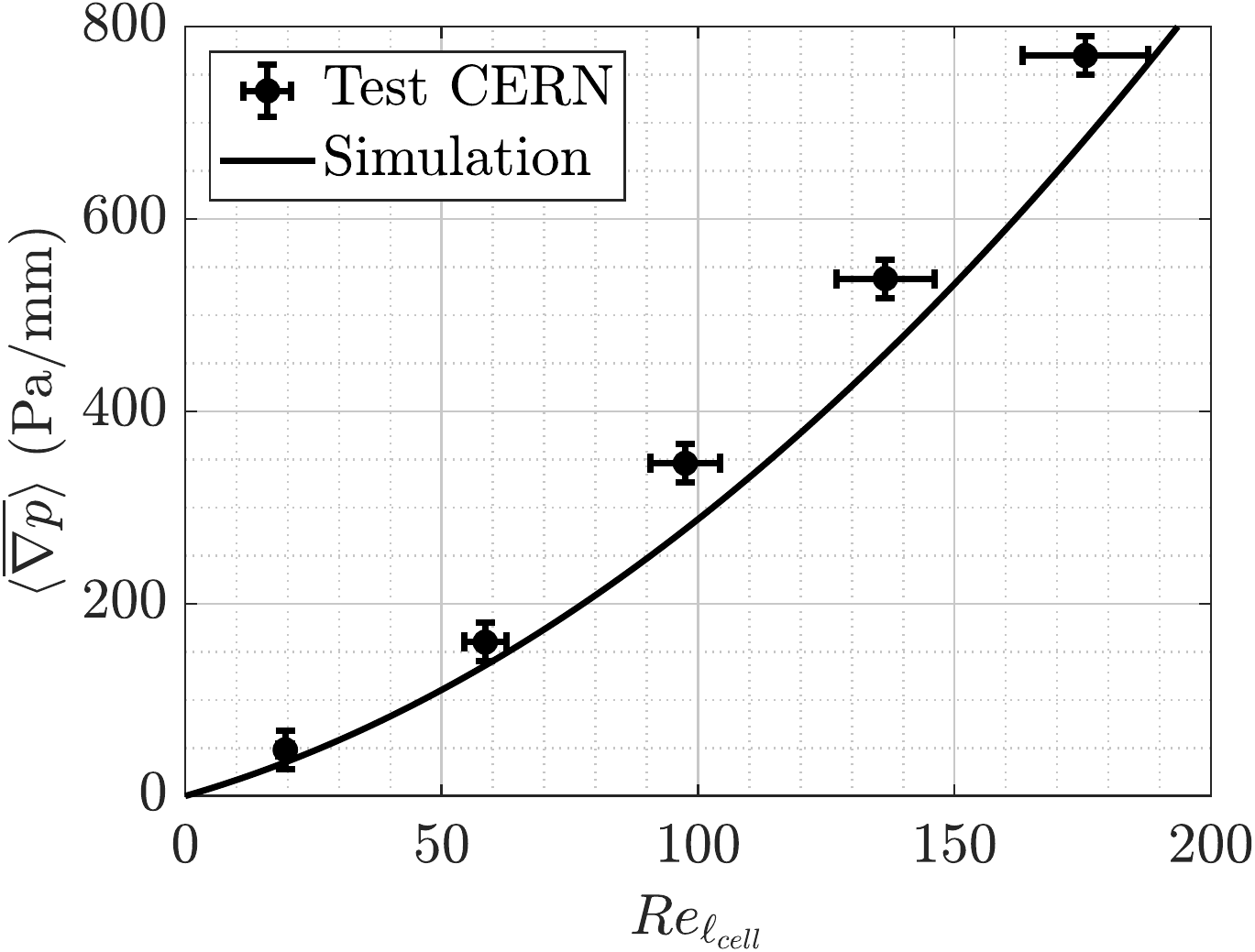}
    \caption{K9}
    \label{k9-deltap}
  \end{subfigure}
       \begin{subfigure}[t]{.485\linewidth}
    \centering
    \includegraphics[trim={0 0 0 0},width=0.85\textwidth, clip]{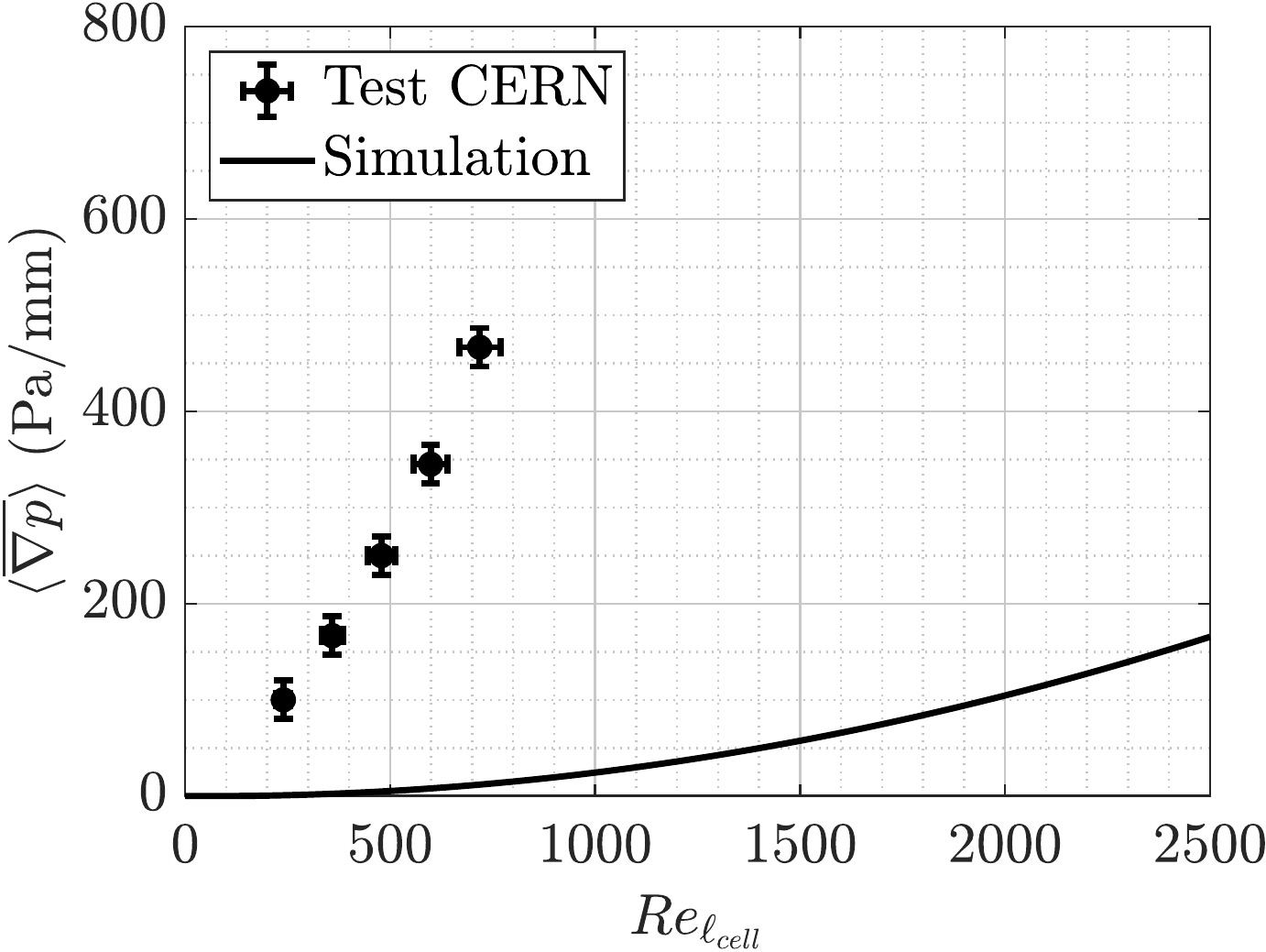}
    \caption{CFOAM$^{\text{\textregistered}}$ 35 HTC }
      \label{res-cfoam}
  \end{subfigure}
  \caption{Comparison between results of the simulations and experimental data of the pressure loss produced by carbon and aluminum foams}
    \label{pressure-loss}
  \end{figure}
  
\cref{res-cfoam} shows that the tetrakaidecahedron model does not predict the pressure loss of the CFOAM$^{\text{\textregistered}}$  correctly. Since the characteristic cell size is 3.5 mm (see \cref{geop-foams}), and the foam length is 6 mm, there are less than two cells in the longitudinal direction. Thus, the periodic geometry assumption is not justified. However, the experimental data gives values that are one order of magnitude higher than the simulation values, which is a large deviation even if the periodicity error is considered. This is attributed to the reduced number of pores and the presence of closed cells in the CFOAM  (see \cref{cfoam-geo}). The CFOAM is not completely an open-cell foam, therefore the microscopic model is not applicable to this foam.

\subsubsection{Thermal conductivity}
\label{s-k}

\begin{figure}[H]
  	\centering
 \begin{subfigure}[t]{.485\linewidth}
 	\centering
\includegraphics[trim={0 0 0 0}, width=0.85\textwidth,clip]{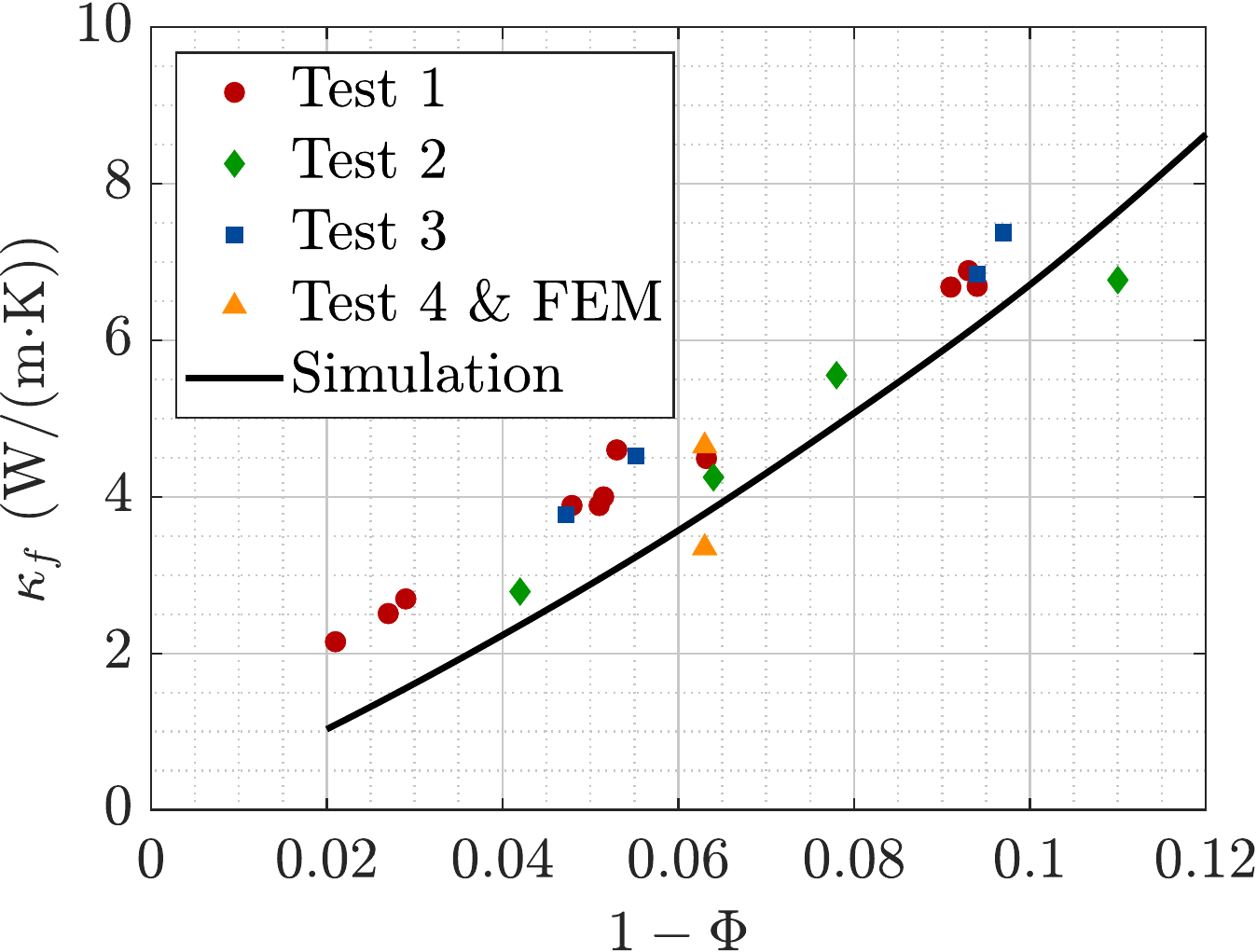}
    \caption{Duocel$^{\text{\textregistered}}$ Al}
    \label{k-foams-comp-al}
  \end{subfigure}
    \begin{subfigure}[t]{.485\linewidth}
    \centering
    \includegraphics[trim={0 0 0 0},width=0.85\textwidth, clip]{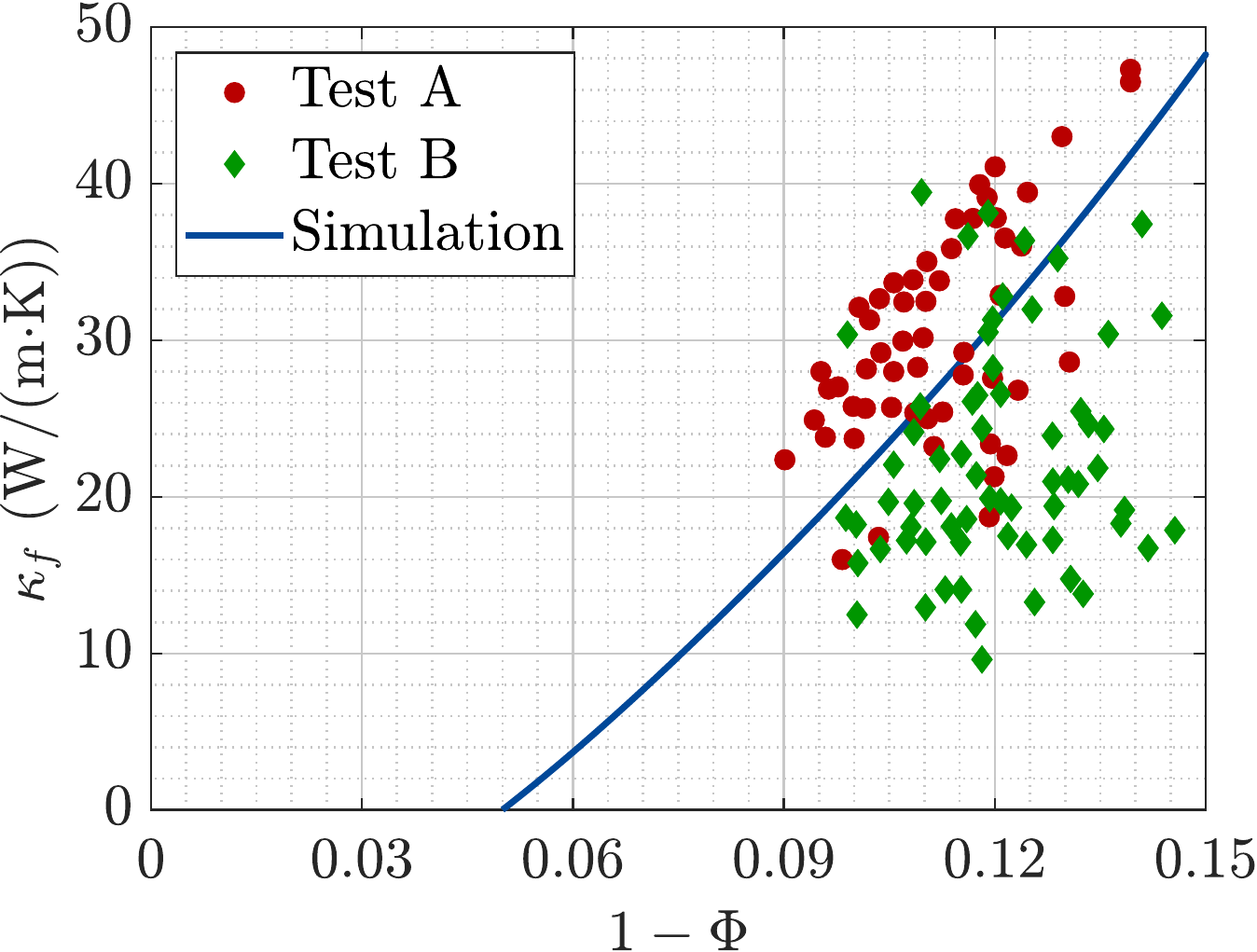}
    \caption{K9}
      \label{k-foams-comp-c}
  \end{subfigure}
  \caption{Thermal conductivity of foams}
  \end{figure}

\cref{k-foams-comp-al} provides the comparison between the thermal conductivity given by the microscopic model and experimental data of aluminum foams, which has been obtained with the direct contact method: \cite{CALMIDI1999} (Test 1), \cite{PAEK2000} (Test 2), \cite{SADEGHI2011} (Test 3), and \cite{AMANI2018} (Test 4). The experiments cited assume that the foams are isotropic, although in Test 4 differences between the planar and vertical directions were claimed, which are represented by the upper and lower values of Test 4 in \cref{k-foams-comp-al}. The simulation results, interpreted as average values in all directions, are correct.

\cref{k-foams-comp-c} illustrates the experimental results obtained from the laser flash method \cite{GILCHRIESE2012} (Test A), from the direct contact method\footnote{Private communication with Lockheed Martin} (Test B), and from the simulations of the K9 foam. The scattering of the experimental data is due to the lack of repeatability of the production process. Assuming that there are no errors in the measurements, it is suspected that the differences between the methods are because the laser flash method is not accurate in foams. In this method the thermal conductivity is obtained from the time derivative of the fundamental solution of the heat equation, which is based on the assumption that Fourier's law of heat conduction is valid. However, recent experiments in metal foams have shown that Fourier's law underpredicts the time derivatives of the temperature distributions \cite{KOVACS2022}. Thus, as a conservative approach, in what follows the values obtained from the direct contact method will be taken as a reference. The simulations predict the mean value of the experimental data, and $\kappa_f \sim (1-\Phi)^2$, which is the case of the experimental results when broader ranges of porosities are considered \cite{GILCHRIESE2012}.

\subsubsection{Nusselt number}
\label{s-h}

To validate the results of the heat transfer coefficient given by the macroscopic model, the correlation obtained from experiments of Duocel$^{\text{\textregistered}}$ Al foams \cite{MANCIN2013} will be taken as a reference:

\begin{equation}
\label{corr-h}
Nu_{\ell_{tr}}=ARe_{\ell_{tr}}^{n} Pr^{1/3},
\end{equation}

\noindent where the constants $A=0.418$ and $n=0.53$, the Nusselt number $Nu_{\ell_{tr}}=h \ell_{tr}/\kappa$, the Reynolds number $Re_{\ell_{tr}}=\rho v_{\infty} \ell_{tr}/(\mu \Phi)$, and the Prandtl number $Pr=\mu c_p/\kappa$.

The measurement of heat (and mass) transfer coefficients is usually more complicated than the measurement of the pressure gradient curve. Thus, the accuracy of the Generalized Lévêque Equation (GLE) will be also considered, which provides a general expression of the heat transfer coefficient as a function of the pressure drop \citep{MARTIN2002}:

\begin{equation}
\label{GLE}
Nu_{\ell_{cell}}= 0.404(\chi f d_h/\ell_{cell})^{1/3}Re_{\ell_{cell}}^{2/3} Pr^{1/3},
\end{equation}

\noindent where the friction factor $f$ is given by \cref{deltap-adj}, $\chi=(f-B)/f$ is the fraction of the pressure drop due to viscous forces with $B$ defined in  \cref{deltap-adj}, and $d_h =4 \Phi /\Sigma_f$ is the hydraulic radius. Given that \cref{corr-h} and \cref{GLE} are referred to different Reynolds numbers, the identity $\ell_{cell}=6.25 \ell_{tr}$ derived from \cref{geop-foams} is used to refer all of the Reynolds numbers to the same length scale.

 \begin{figure}[H]
  	\centering
 \begin{subfigure}[t]{.485\linewidth}
 	\centering
\includegraphics[trim={0 0 0 0}, width=0.85\textwidth,clip]{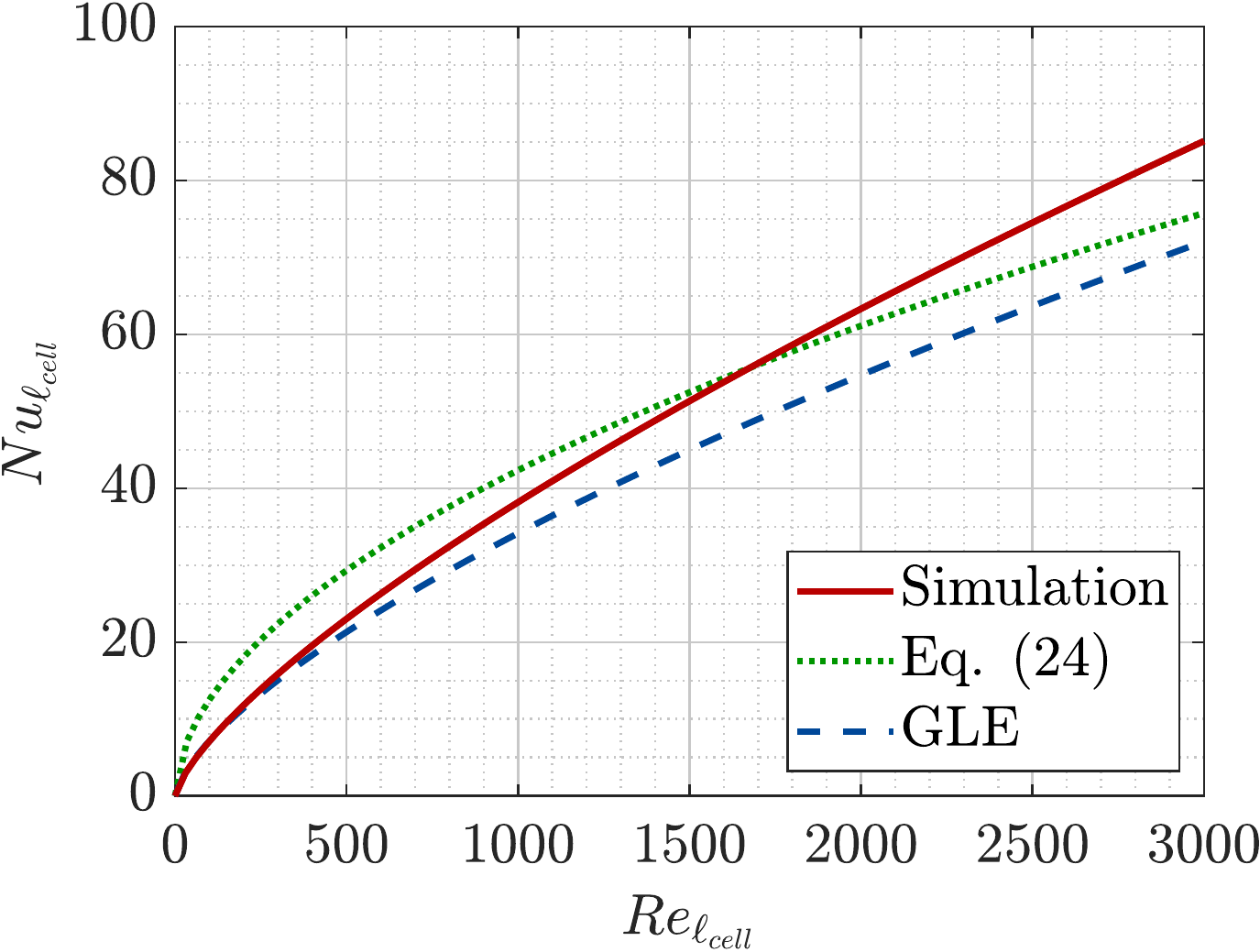}
    \caption{Duocel$^{\text{\textregistered}}$ Al}
    \label{h-foams}
  \end{subfigure}
    \begin{subfigure}[t]{.485\linewidth}
    \centering
    \includegraphics[trim={0 0 0 0},width=0.85\textwidth, clip]{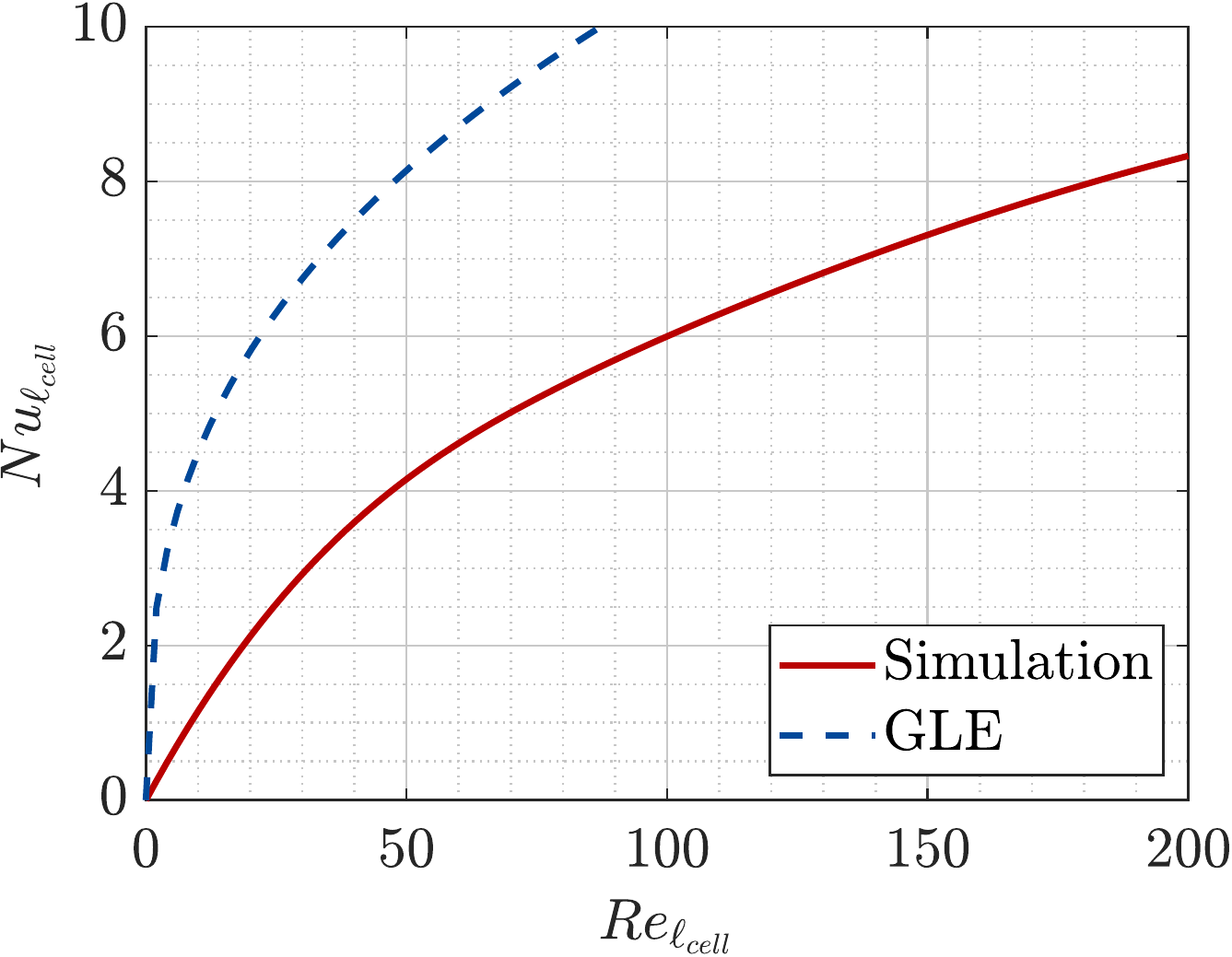}
    \caption{K9}
      \label{h-k9}
  \end{subfigure}
  \caption{Thermal conductivity of foams}
  \end{figure}

The comparison between numerical results and the correlation in \cref{h-foams}  shows that the heat transfer coefficient increases when the cell size decreases, since $Nu \sim Re_{\ell_{tr}}^n/\ell_{tr} \sim \ell_{tr}^{n-1}$ and $n-1 <1$. The numerical values of the Nusselt number of the Al foams are the same in all cases, and they agree well with \cref{corr-h,GLE}. With foams of similar geometric parameters, satisfactory results were also obtained in previous studies \citep{GARRIDO2008}. The opposite conclusion can be drawn from \cref{h-k9}. The reason of the discrepancy can be related to the fact that in \cref{GLE} the product $\chi f$ is approximately constant in the cases considered in \cref{h-foams}, while for the K9 foam $\chi f$ depends strongly on $Re_{\ell_{cell}}$ and alters the $Re_{\ell_{cell}}^{2/3}$-dependency of the GLE equation. 

The correlation of the numerical results gives constants $a \approx 0.12$ and of $n\approx 0.75$ for the Duocel$^{\text{\textregistered}}$ Al foams, while $a = 0.11$ and $b=0.55$ for the K9 foam. These exponents are similar to the values of the laminar flow in a flat plate ($n=0.5$), and the turbulent flow in a flat plate and in a circular tube ($n=0.8$ in both cases) \cite{SCHLICHTING1979}. Since there are no significant variations among the foams studied, what makes the difference in the thermal performance---apart from the foam thermal conductivity---is the specific surface area (see \cref{energy-solid}). This is because the surface areas of foams differ by up to one order of magnitude (see \cref{geop-foams}). It has been verified that the variations of the heat transfer coefficient as a function of the foam porosity are negligible.

\subsection{Macroscopic model}
\label{r-macro}

The values of the pressure loss, the thermal conductivity, and the heat transfer coefficient obtained from the microscopic model are used to validate the results given by the macroscopic model. Two values are analyzed, the overall heat transfer coefficient of Duocel$^{\text{\textregistered}}$ Al foams in \cref{o-htc-r}, and the temperature of the heaters in contact with the K9 foam in \cref{glue-opt}.

\subsubsection{Overall heat transfer coefficient}
\label{o-htc-r}

The overall heat transfer coefficient $U$ of Duocel$^{\text{\textregistered}}$ Al foams has been measured experimentally \cite{MANCIN2013}.  The lengths of the fluid and foam regions are $l_x=100$ mm, $l_y=50$ mm, and $l_z=20$ mm (see \cref{geo-ohtc}). In the plane $z=l_z$, an aluminum plate of $10$ mm is placed in contact with the foam, and a heater made of copper of $\xi=7$ mm of thickness is placed in in contact with the aluminum plate. The heater provides a heat flux $q_2=\num{2.5e4}$ \si{\watt\per\meter\squared} ($q_1=0$).  In \cref{log-T}, the inlet and outlet wall temperatures are defined at $x^{in}=20$ mm and $x^{out}=80$ mm, respectively. In the experiment, the air outlet temperature is measured by six temperature sensors placed after a mixer in the plane $x \approx 200$ mm. In the simulations it is considered to be the area-weighted temperature in that plane. The thermal conductivity and the heat transfer coefficient used in the simulations are given by the microscopic model in  \cref{k-foams-comp-al} and \cref{h-foams}.

\begin{figure}[H]
  	\centering
 \begin{subfigure}[t]{.485\linewidth}
 	\centering
\includegraphics[trim={0 0 0 0}, width=0.85\textwidth,clip]{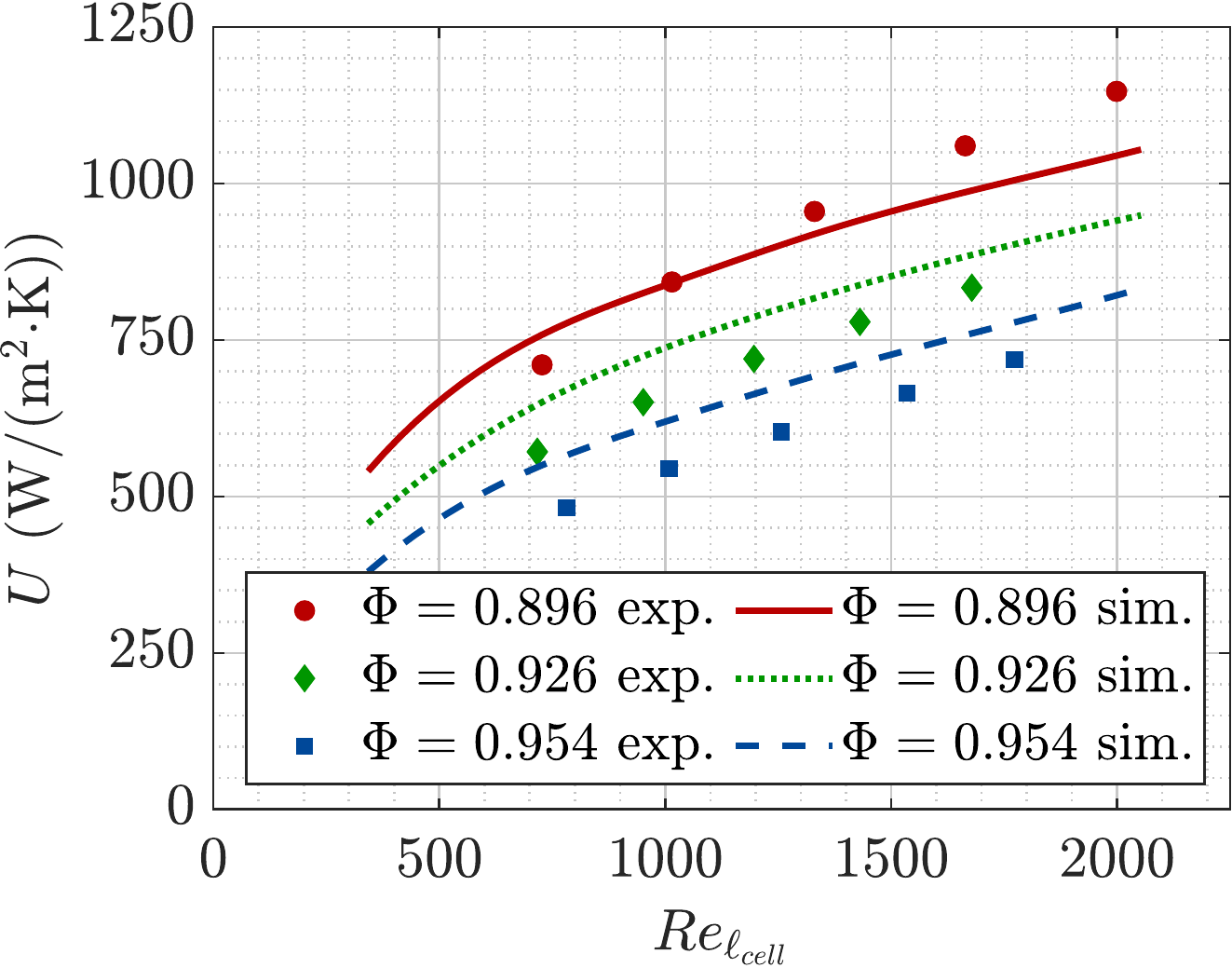}
    \caption{Al-10}
      \label{ohtc}
  \end{subfigure}
    \begin{subfigure}[t]{.485\linewidth}
    \centering
    \includegraphics[trim={0 0 0 0},width=0.85\textwidth, clip]{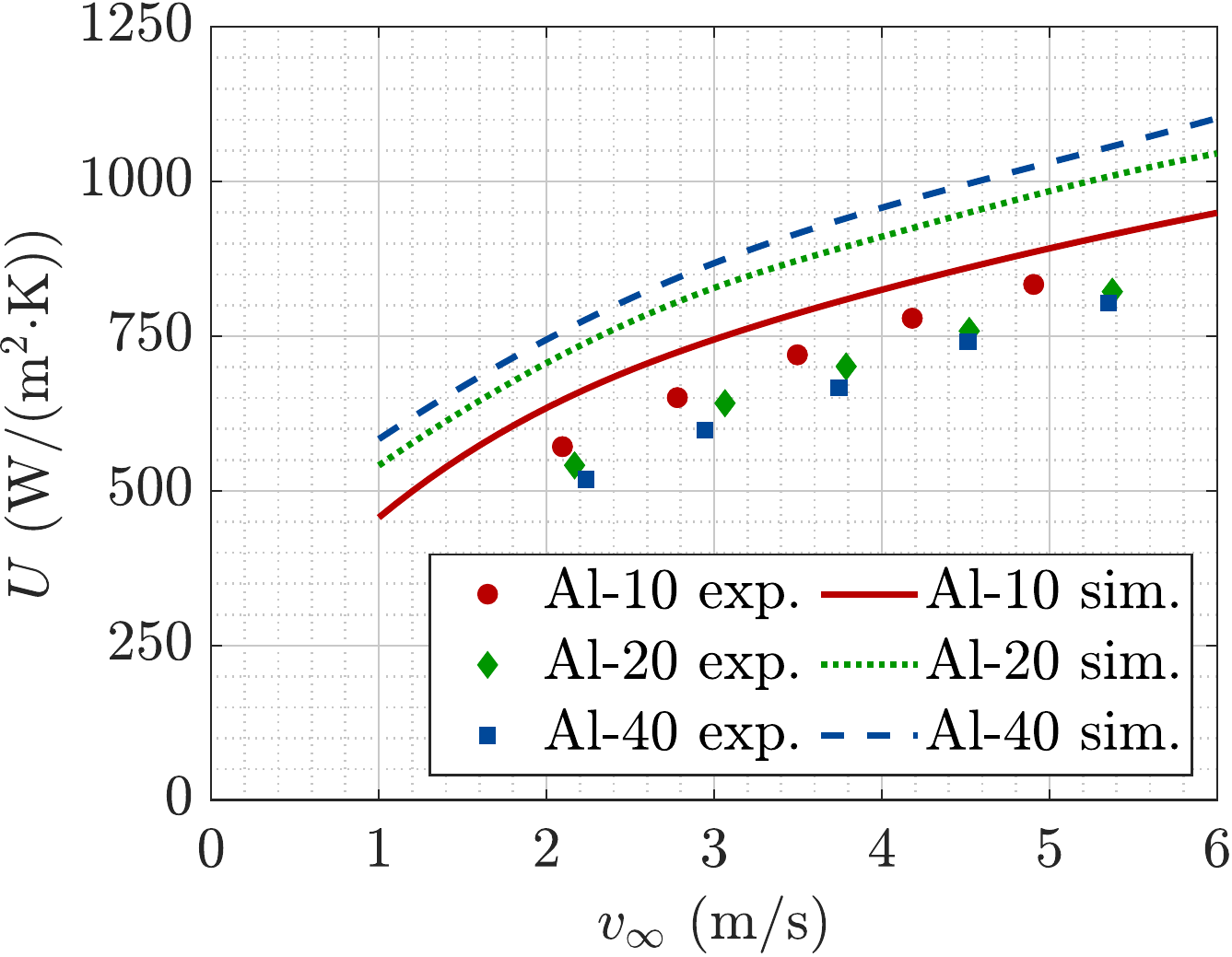}
    \caption{$\Phi \approx 0.93$}
     \label{ohtc-ppi}
  \end{subfigure}
  \caption{Overall heat transfer coefficient of the Duocel$^{\text{\textregistered}}$ Al foams}
  \end{figure}

A comparison between experimental and simulation results of the overall heat transfer coefficient defined by \cref{power-h} is given by \cref{ohtc}.  The maximum error of the simulations is around $15\  \%$, and allows to conclude that the results given by the macroscopic model are accurate. The shape of the curve is better approximated when $\Phi=0.926$ and $\Phi=0.954$, although the increase of the slope of the curve for $Re_{\ell_{cell}}$$ \approx 750$ is captured correctly in all cases. This behavior is closely related to the decrease of the heat transfer coefficient for low velocity values (see \cref{h-foams}). 

\cref{ohtc-ppi} compares the numerical and experimental results of three variants of the Duocel$^{\text{\textregistered}}$ foams for $\Phi \approx 0.93$. The differences increase as the cell size is decreased, since the experimental results predict that the overall heat transfer coefficient decreases as the cell size decreases, while the model developed provides the opposite trends. Since the heat transfer coefficient increases as the cell size decreases (see \cref{s-h}), and the thermal conductivity does not depend on the cell size \citep{AMANI2018}, it is expected that the correct dependency is given by the numerical results. It should be noted that the freestream velocity has been used in the $x$ axis because the discrepancy with the experimental results cannot be deduced when $Re_{\ell_{cell}}$ is used. The anisotropic thermal conductivity (see \cref{s-k}) and the lack of repeatability of the production process may play an important role on the differences numerical and experimental results. In addition, the errors related to the model developed and the experimental data add additional uncertainties to the comparison. The overall heat transfer coefficient defined in \cref{log-T} depends on four temperature measurements, and the $\ln$ function is sensible to small perturbations.

\subsubsection{Heater wall temperature}
\label{glue-opt}

Thanks to its higher specific surface area and thermal conductivity with respect to the Duocel$^{\text{\textregistered}}$ Al foams, the K9 foam is concluded to be the one that provides the best thermal performance among the foams studied. The experimental setup described in \cref{exp-setup} is used to determine the thermal performance of the K9 foam in a case of practical application.

To understand the effect of the glue penetration, six samples of mean glue penetration  $\langle \zeta_g \rangle =30,80,130,180,280$ and $380$ \si{\micro\meter} are tested in the setup. Based on \cref{interface-ts}, in the simulations the foam-heater interface is modeled as a layer of $100+120+\langle \zeta_g \rangle$ of thickness, with the thermal conductivity equal to the glue thermal conductivity. Taking as a reference the experimental data obtained by the CMS Collaboration \cite{MUSSGILLER2016}, the optimum penetration of $\langle \zeta_g \rangle=250$ \si{\micro\meter} is considered in the macroscopic model, thus giving a total glue layer thickness of $100+120+250=470$ \si{\micro\meter} in the simulations. The heat fluxes are $q_1=q_2=q$, and the reference heat flux is defined as $q_0=10^4$ \si{\watt\per\meter\per\kelvin}. Since the pressure loss that the fan can provide is limited, seven equispaced holes of 1.5 mm are drilled in the foam that provide a 20 \% of reduction in the pressure loss. In the simulations the $z=0$ plane is a symmetry plane, so the lengths of the computational domain are $l_x \times l_y \times l_z = 6 \times 30 \times 3 $ mm. The thermal conductivity is given by the simulation curve (see \cref{k-foams-comp-c}) for $\Phi=0.89$ (see \cref{geop-foams}). During the tests, it has been verified that \cref{power-h} holds with a difference of less than 5 \% between the two terms in all of the cases studied.

 \begin{figure}[H]
  	\centering
 \begin{subfigure}[t]{.485\linewidth}
 	\centering
\includegraphics[trim={0 0 0 0}, width=0.85\textwidth,clip]{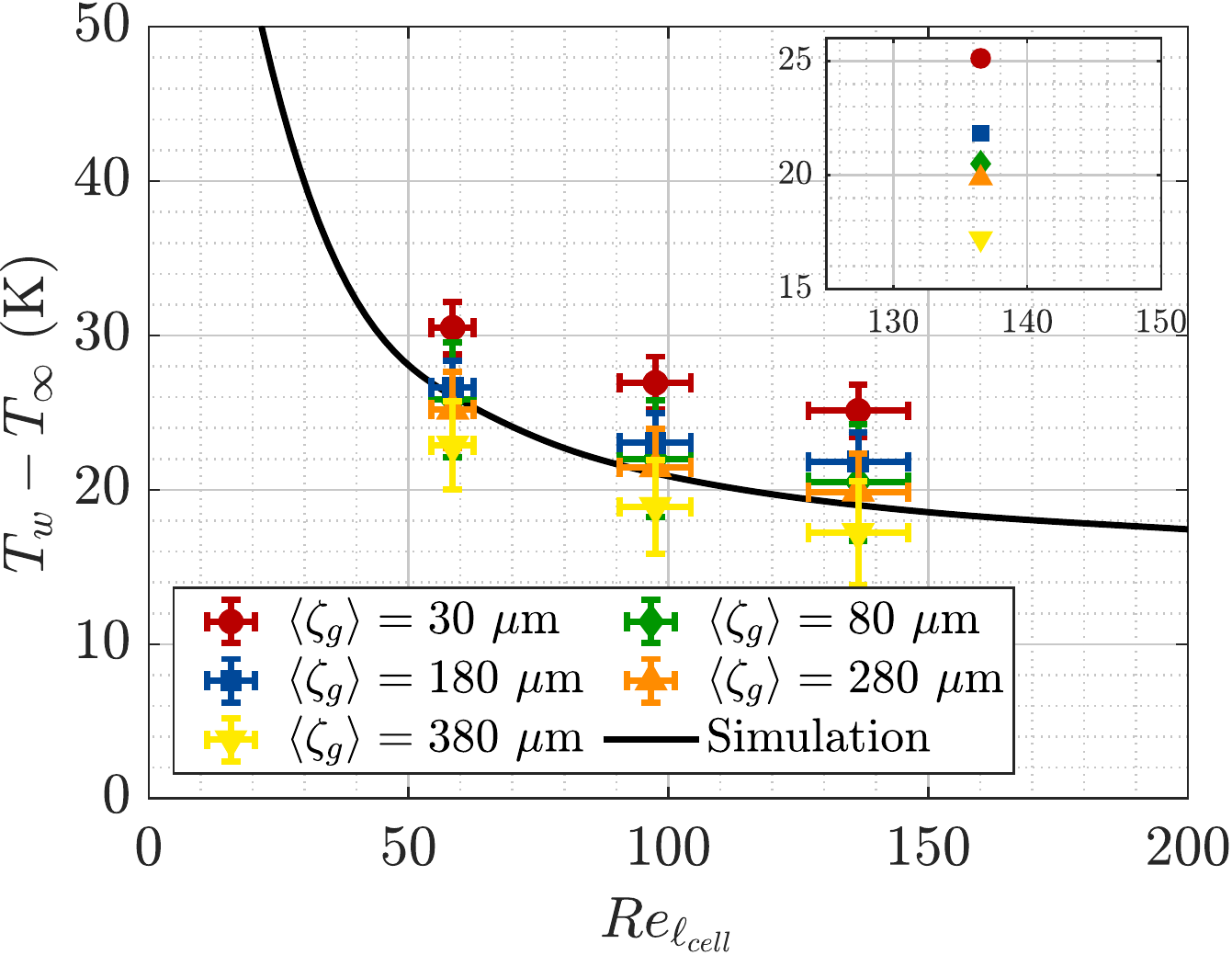}
    \caption{Influence of glue penetration for $q=2 q_0$}
    \label{glue-pen}
  \end{subfigure}
    \begin{subfigure}[t]{.485\linewidth}
    \centering
    \includegraphics[trim={0 0 0 0},width=0.85\textwidth, clip]{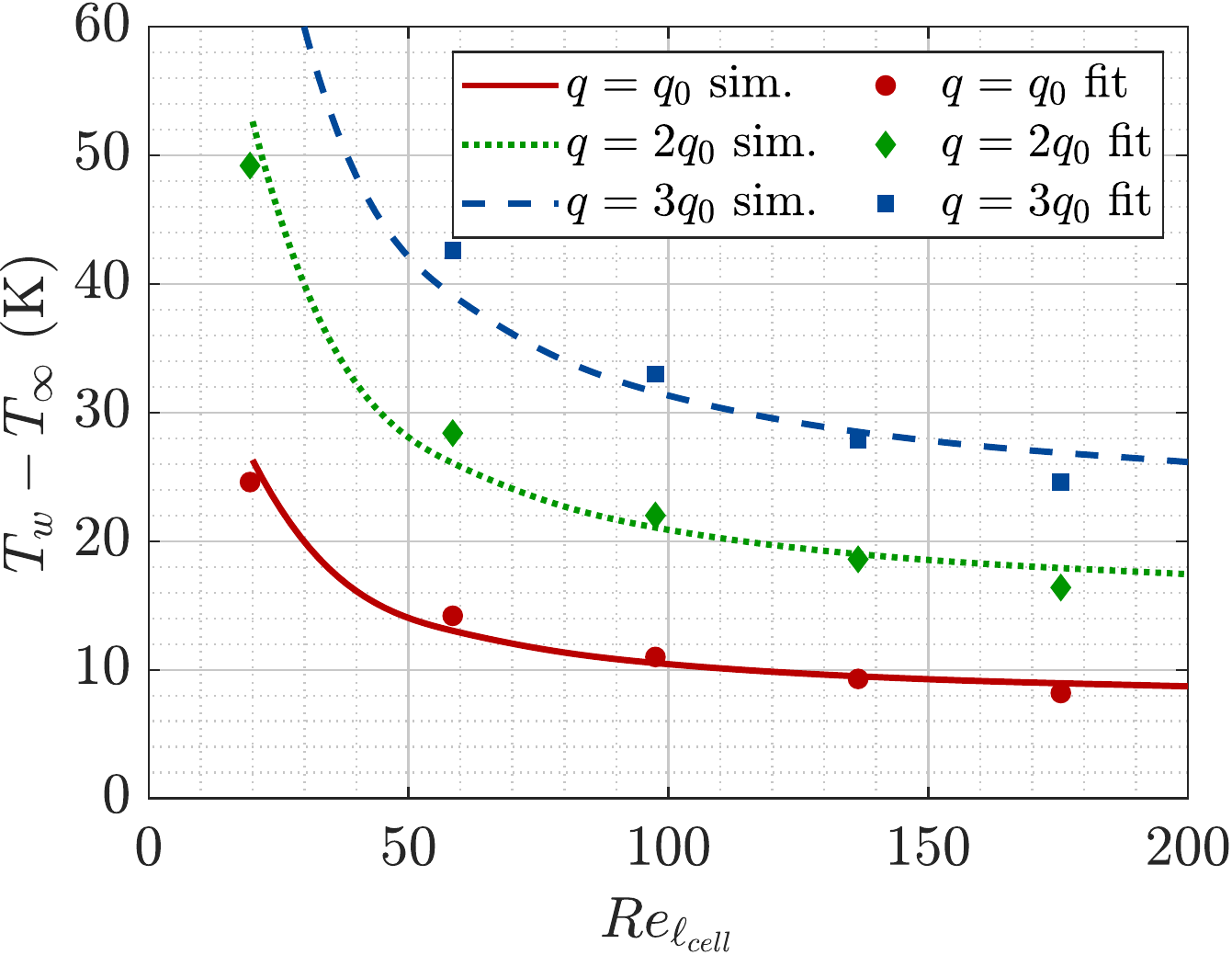}
    \caption{Influence of the heat flux}
      \label{q-influence}
  \end{subfigure}
  \caption{Heater wall temperature difference with the K9 foam}
  \end{figure}

\cref{glue-pen} shows the difference between the heater wall temperature and the freestream temperature for different velocity values. The wall temperature values ($T_w$) are the ones of the temperature sensors located at the center of the heater $T_{center}$ (see \cref{heaters}). The values of $T_{side}$ are in all cases around 20 \% lower than $T_{center}$. This temperature non-uniformity is because the total surface area of the heaters is lower than 6$\times$60 mm because of the presence of soldering points of the power supply. Unlike in the CMS tests \citep{MUSSGILLER2016}, it can be concluded that the temperature variations are reasonably bounded with the proposed assembly procedure explained in \cref{exp-setup}. The case of $\langle \zeta_g \rangle=30$ \si{\micro\meter} gives the highest temperature values, and is expected to be sensitive to variations in the glue thickness, therefore it is discarded for future tests. The decrease of the heater wall temperature when the glue penetration is increased is explained by the combined effect of two aspects: the decrease of the glue uniformity, which translates into the increase of the thermal resistance, and the increase of the thermal conductivity of the foam, since the glue fills the cells of the foam. 
  
The boundary layers of the two heaters interact, and from \cref{bc-inf} it is deduced that $T_c\neq T_{\infty}$. This means that the analytical solution (\cref{analytical}) is not able to provide a good approximation of the heater wall temperature. However, based on the analytical expression, the numerical results have been fitted to a curve as follows:

\begin{equation}
\label{deltat-fit}
    T_w-T_{\infty}=\sqrt{C} \frac{q}{\sqrt{v_{\infty}}}
\end{equation}

\noindent for a constant $C=\num{6.04e-6}$ \si{\kelvin\squared\meter\tothe{5}\per\watt\squared\per\second} that is obtained from the minimization of the differences of the numerical results and \cref{deltat-fit} in a least-squares sense. 

\cref{q-influence} shows the variation of the heater wall temperature for different values of the dimensionless heat dissipation given by the macroscopic model (lines) and the fit of \cref{deltat-fit} (points). It is deduced that the analytical solution provides correct trends of the wall temperature as a function of the heat flux and the freestream velocity. In addition, the analytical solution predicts that $T_w-T_{\infty} \sim 1/\sqrt{\kappa_f}$, which means that the uncertainty in the thermal conductivity of the K9 foam (see \cref{k-foams-comp-c}) is not expected to produce a great impact on the results.

\section{Conclusions}
\label{conclusions}

In this work a multiscale model for open-cell foams has been developed. The microscopic model is based on a periodic unit-cell geometry, and provides as outputs the pressure loss, the thermal conductivity, and the heat transfer coefficient of foams. These parameters are universal and are used as inputs of a macroscopic model, that treats the foams as porous mediums with averaged properties, and gives as a result the thermal performance of a macroscopic system. 

The microscopic approach is referred to characteristic lengths of the order of the foam cell length. The periodic unit cell is modeled with a 14-sided truncated octahedron, which is taken as a reference from previous studies. The model uses as inputs the porosity, which can be easily measured with a precision balance, and the specific area, which can be obtained accurately from experimental techniques such as microscope images or computer tomography scans. It has been shown that the accuracy of the model deteriorates when partially open-cell foams are considered. The model has been shown to be valid for foam porosities higher than 0.82, which is the minimum value under which the overall geometrical structure of the unit cell proposed is maintained. Additional limitations have been noted for very high temperature variations, where the assumption of constant material properties can play an important role. A comparison with experimental data obtained from the literature and from a setup built at CERN has shown that the model provides accurate results in all open-cell foams considered. A correlation to predict the Nusselt number as function of the pressure loss has been confirmed in the cases where both experimental and numerical results are available. It has been concluded that the specific surface area of foams plays an important role in the thermal performance of the systems where convective heat transfer is present.

The macroscopic model is used where the characteristic length is sufficiently large so that the flow is fully developed.  In this model, the foam is considered as a continuum, and the effect of the microscopic geometry in the macroscopic behavior is done with source terms in the governing equations. Since the entrance and exit effects are neglected in the microscopic model, the minimum ratio between the macroscopic and microscopic scales under which the macroscopic model is valid has been deduced. The results given by this model are accurate when compared with experimental data; in particular, in an experimental setup that represents a case of practical application in the ALICE experiment at the Large Hadron Collider. When a foam is glued to a solid material, it has been shown that the thermal performance of the system is proportional to the penetration of the glue in the foam. An optimized assembly procedure developed at CERN has been shown to limit the thermal resistance of the joint in all cases.

The multiscale model developed has been used for characterization of heat exchangers containing open-cell foams, and the methodology can be applied to other geometries such as heat sinks. Since open-cell foams are expected to be used in the inner particle detectors of HEP experiments, the macroscopic model can be used to study the performance of the cooling systems of these detectors, which is expected to be part of the future work. 

\appendix
	
\section{Foams}
\label{foams-appendix}

The key properties of the four foams used to validate the multiscale model are explained.

\setcounter{footnote}{0} 

\subsection*{Duocel$^{\text{\textregistered}}$ Al}  
  
Duocel$^{\text{\textregistered}}$ open-cell aluminum foams are fabricated from 6101 aluminum alloy by ERG Aerospace.  These foams are available in five different porosity ranges from $\Phi=0.88$ to $\Phi=0.96$ with a corresponding range of thermal conductivities of $2-7$ \si{\watt\per\meter\per\kelvin}, and in in four different linear pores per inch (PPI). Although the denominations provided by the company are  ``10 PPI", ``20 PPI", and ``40 PPI", the pore density does not match these values, as the names are classification names versus specific measurements that allow to track the different pore sizes. This keeps the material consistent to each other through the years and across applications\footnote{Private communication with ERG Aerospace}. For example, images extracted from CT scans indicate that the mean cell size of the ``40 PPI" foam is around 2.25 mm \cite{AMANI2018}, which is equivalent to 25.4/2.25 $\approx$ 10 linear pores per inch. In what follows these foams will be named as ``Al-10", ``Al-20", and ``Al-40", respectively. The geometry is anisotropic, with the characteristic cell lengths in the planar directions $\ell_x \approx \ell_y$, while $\ell_z<\ell_x,\ell_y$,  \cite{AMANI2018}. The repeatability of the process is not 100 \%, which is the reason why the foams are offered in a range of porosities and cell sizes that should be treated as approximate. 
\setcounter{footnote}{0} 

 \begin{figure}[H]
 	\centering
  \begin{subfigure}[t]{.485\linewidth}
  \centering
  \scalebox{0.88}{
        	\begin{tikzpicture}[xscale=1, yscale=1]
			\node[anchor=south west,inner sep=0] at (0,0) {  \includegraphics[trim={0 0 0 0},width=0.85\textwidth, clip]{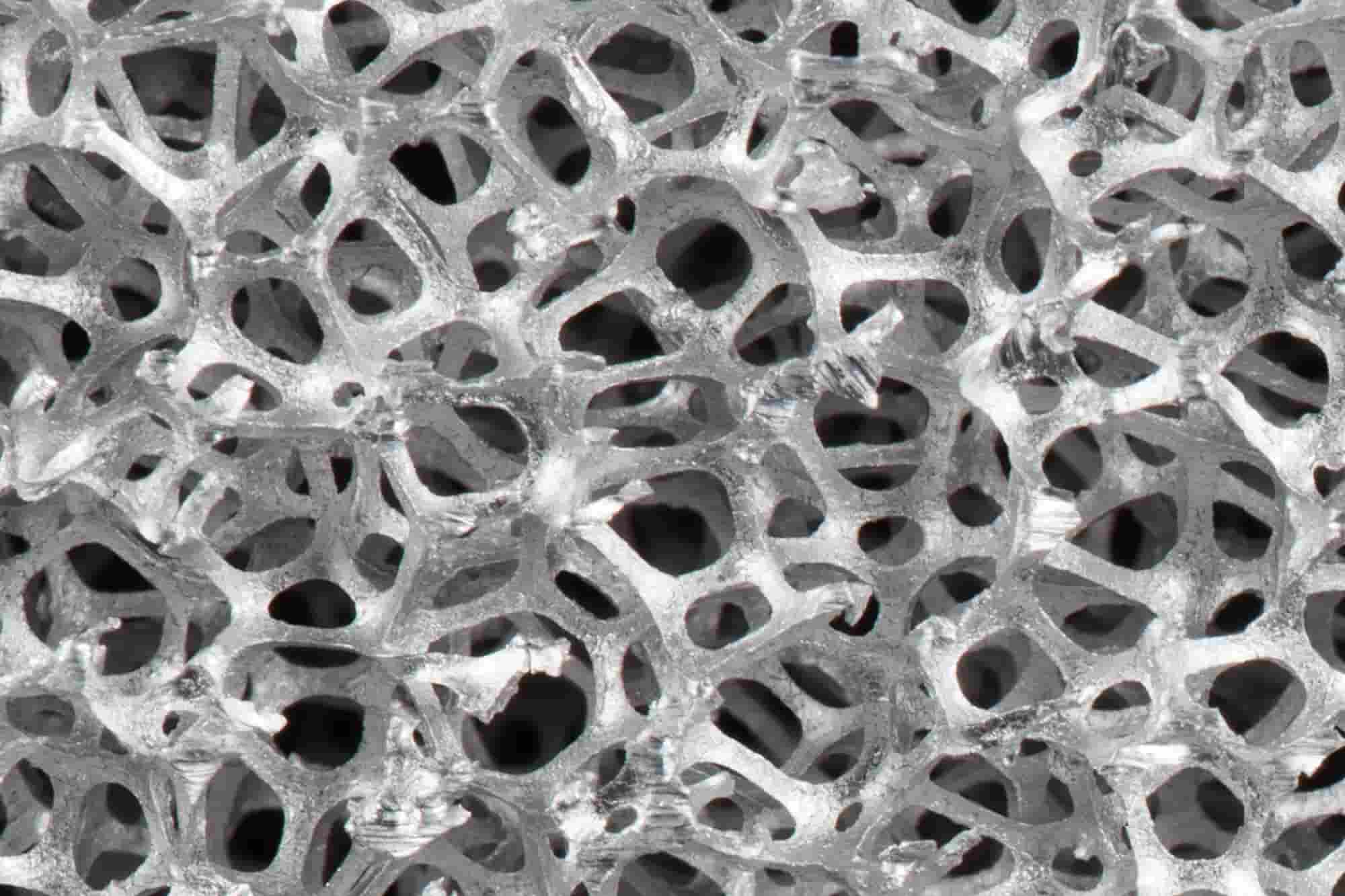}};
			\draw[quote](0.2,-0.25) -- (0.72,-0.25);
			\node  at (0.5, -0.625) {\small \bfseries 1 mm};
	\end{tikzpicture}}
		    \caption{Al-40 ($\Phi=0.92$)\footnotemark}
		    \label{duocel-al}
  \end{subfigure}
  \begin{subfigure}[t]{.485\linewidth}
  \centering
  \scalebox{0.88}{
      	\begin{tikzpicture}[xscale=1, yscale=1]
			\node[anchor=south west,inner sep=0] at (0,0) {\includegraphics[trim={0 85 0 0}, width=0.85\textwidth, clip]{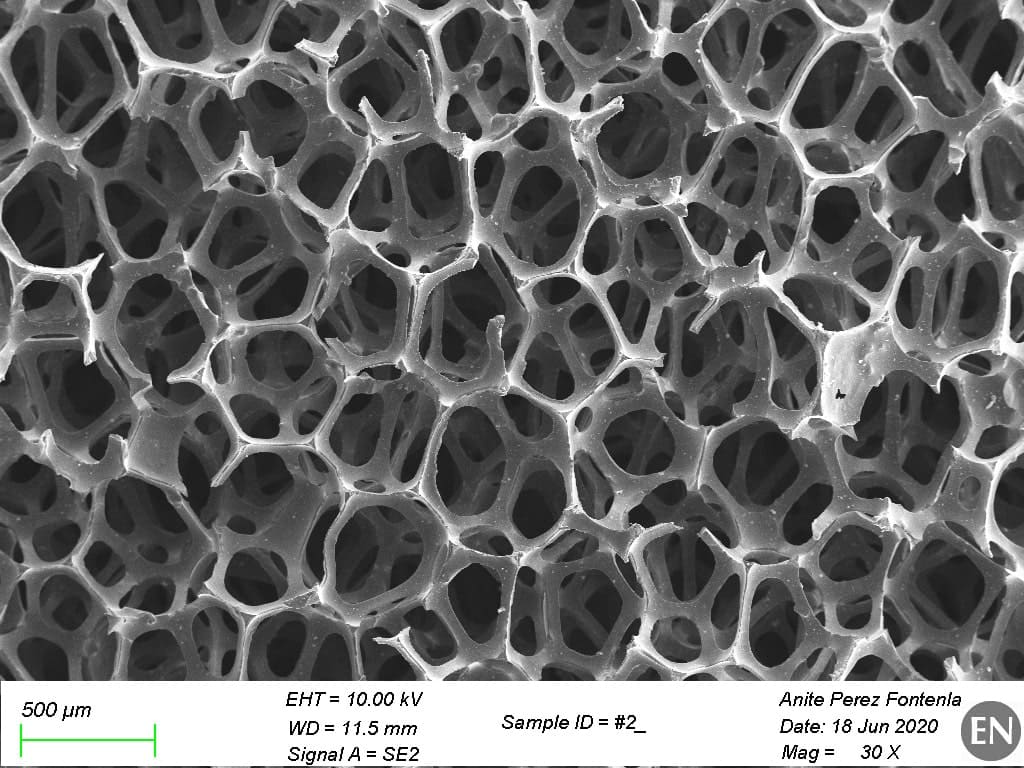}};
			\draw[quote](0.2,-0.25) -- (2.25,-0.25);
			\node  at (1.25, -0.625) {\small \bfseries 1 mm };
	\end{tikzpicture}}
	    \caption{RVC 100 PPI}
	    \label{rvc-sem}
  \end{subfigure}
  \caption[]{Microscopy images of Duocel$^{\text{\textregistered}}$ foams}
  \end{figure}
\footnotetext{Private communication with ERG Aerospace}

\subsection*{Duocel$^{\text{\textregistered}}$ RVC}
  
One of the morphological structures of the vitreous carbon is the reticulated vitreous carbon (RVC). The foam has a density $\rho_f \approx 45$ \si{\kilogram\per\meter\cubed}, and it is available in different PPI values from 5 and 100, which are representative of the actual cell sizes (see \cref{rvc-sem}). The variant chosen is the one of 100 PPI.

\subsection*{Lockheed Martin K9}

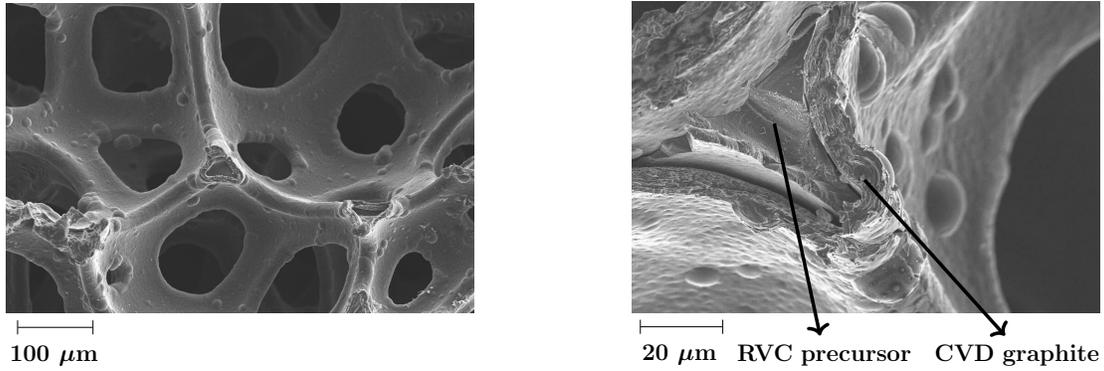
\begin{figure}[H]
 	\centering
  \begin{subfigure}[t]{.485\linewidth}
  	   \centering
        \scalebox{0.75}{
      	\begin{tikzpicture}[xscale=1, yscale=1]
      	   \centering
			\node[anchor=south west,inner sep=0] at (0,0) {\includegraphics[trim={0 0 0 0}, width=\textwidth, clip]{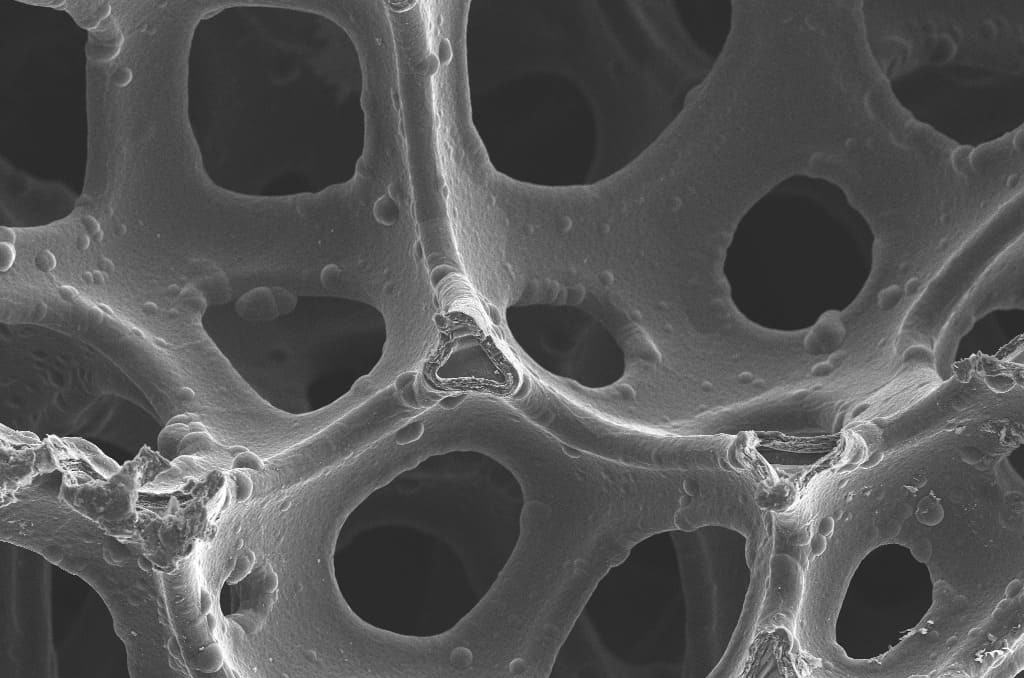}};
			\draw[quote](0.2,-0.25) -- (1.55,-0.25);
			\node  at (0.85, -0.75) { \bfseries 100 $\boldsymbol{\mu}$m};
	\end{tikzpicture}}
  \end{subfigure}
    \begin{subfigure}[t]{.485\linewidth}
    	   \centering
          \scalebox{0.75}{
      	\begin{tikzpicture}[xscale=1, yscale=1]
      	   \centering
			\node[anchor=south west,inner sep=0] at (0,0) {\includegraphics[trim={0 0 0 0}, width=\textwidth, clip]{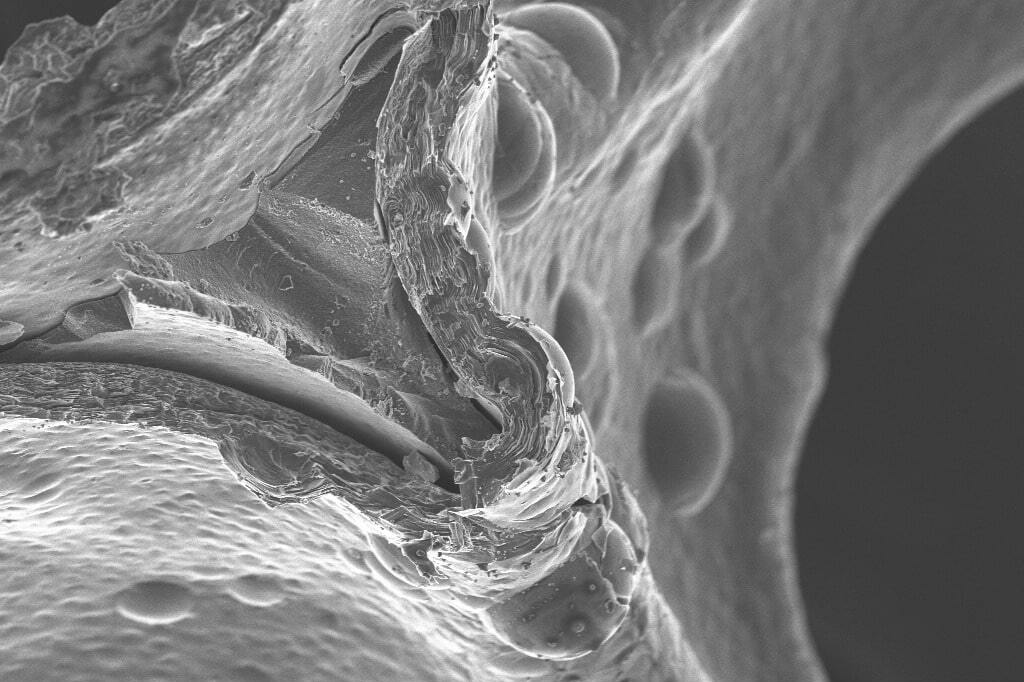}};
		\draw[quote](0.15,-0.25) -- (1.625,-0.25);
			\node  at (0.85, -0.75) { \bfseries 20 $\boldsymbol{\mu}$m};
		\node (A) at (2.5, 3.5) {};
		\node (B) at (3.375, -0.55) {};
		\node (C) at (4, 2.5) {};
		\node (D) at (6.75, -0.55) {};
		\draw [->,line width=0.625mm] (A) -- (B);
		\draw [->,line width=0.625mm] (C) -- (D);
		\node[text width=3.5cm] at (3.625, -0.75) { \bfseries RVC precursor};
		\node[text width=3cm] at (6.875, -0.75) { \bfseries CVD graphite};
	\end{tikzpicture}}
  \end{subfigure}
  \caption{Microscopy images of the K9 low-density carbon foam}
    \label{sem-ld-zoom}
  \end{figure}

RVC foams are thermal insulators with $\kappa_f \approx 0.05$ \si{\watt\per\meter\per\kelvin}, which is attributed to the surface cracks that occur during the heat treatment. Images of the ligaments indicate that the surface cracks are filled when the microstructure is coated by chemical vapor deposition (CVD) \cite{BONAAD2012}. The amount of carbon introduced determines the density of the resulting foam. In this work, the foam with $\rho_f \approx 200$ \si{\kilogram\per\meter\cubed} is studied. Microscopy (SEM) images allow to notice the main microstructural difference between the RVC and the K9 foam: in the latter, the graphite coating is a new structure that increases the thickness of the filaments (see \cref{sem-ld-zoom}) and the thermal conductivity to $\kappa_f \approx 25$ \si{\watt\per\meter\per\kelvin}\footnote{Private communication with Lockheed Martin}.

\setcounter{footnote}{0} 

\subsection*{CFOAM$^{\text{\textregistered}}$ 35 HTC}

CFOAM$^{\text{\textregistered}}$ 35 HTC is made from mesophase pitch feedstock, and has a density $\rho_f \approx 350$ \si{\kilogram\per\meter\cubed}. The thermal conductivity in the vertical direction---20-30 \si{\watt\per\meter\per\kelvin}--is approximately twice as the one in the planar directions\footnote{Private communication with CFOAM LLC}. \cref{cfoam-geo} shows the microscopic structure of the of the low-density variant. The structure is highly anisotropic, and the most homogeneous zone---which coincides with the zone of lower characteristic cell size---has been analyzed in the images. Compared to the K9 foam, the CFOAM$^{\text{\textregistered}}$ 35 HTC has a higher pore size and more irregular surfaces, with multiple microcracks.

     \begin{figure}[H]
      	\centering
  \begin{subfigure}[t]{.485\linewidth}
  \centering
  \scalebox{0.75}{
      	\begin{tikzpicture}[xscale=1, yscale=1]
      	   \centering
			\node[anchor=south west,inner sep=0] at (0,0) {\includegraphics[trim={0 0 0 0}, width=\textwidth, clip]{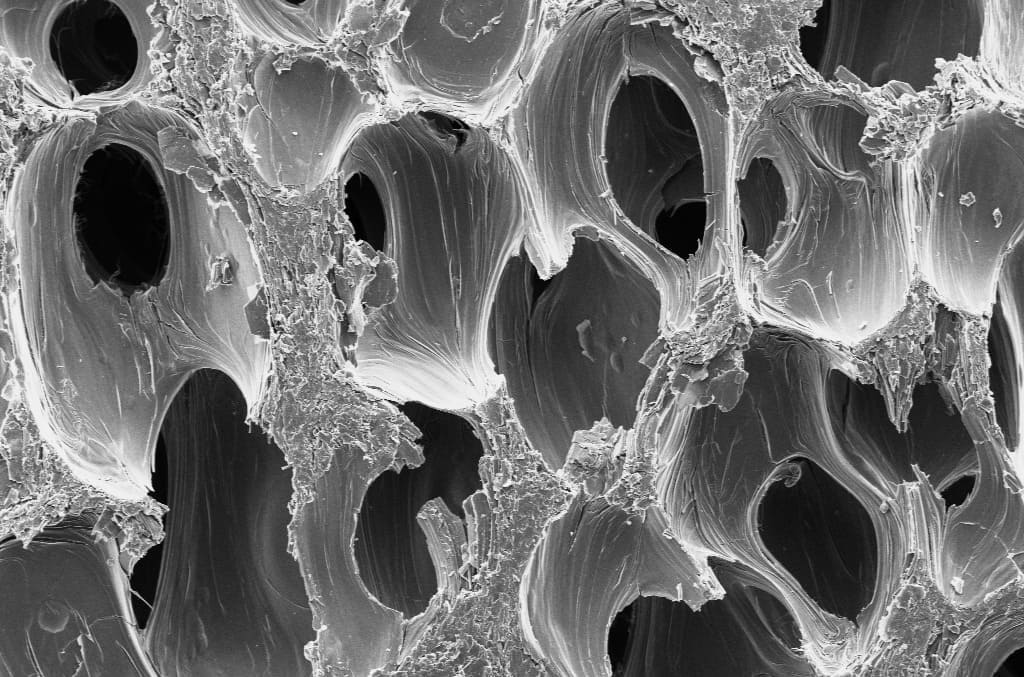}};
			\draw[quote](0.175,-0.25) -- (1.75,-0.25);
			\node  at (0.925, -0.75) { \bfseries 1 mm};
	\end{tikzpicture}}
	  \caption{Microscopy image - Cell structure}
  \end{subfigure}
    \begin{subfigure}[t]{.485\linewidth}
    \centering
      \scalebox{0.75}{
      	\begin{tikzpicture}[xscale=1, yscale=1]
      	   \centering
			\node[anchor=south west,inner sep=0] at (0,0) {\includegraphics[trim={0 0 0 0}, width=\textwidth, clip]{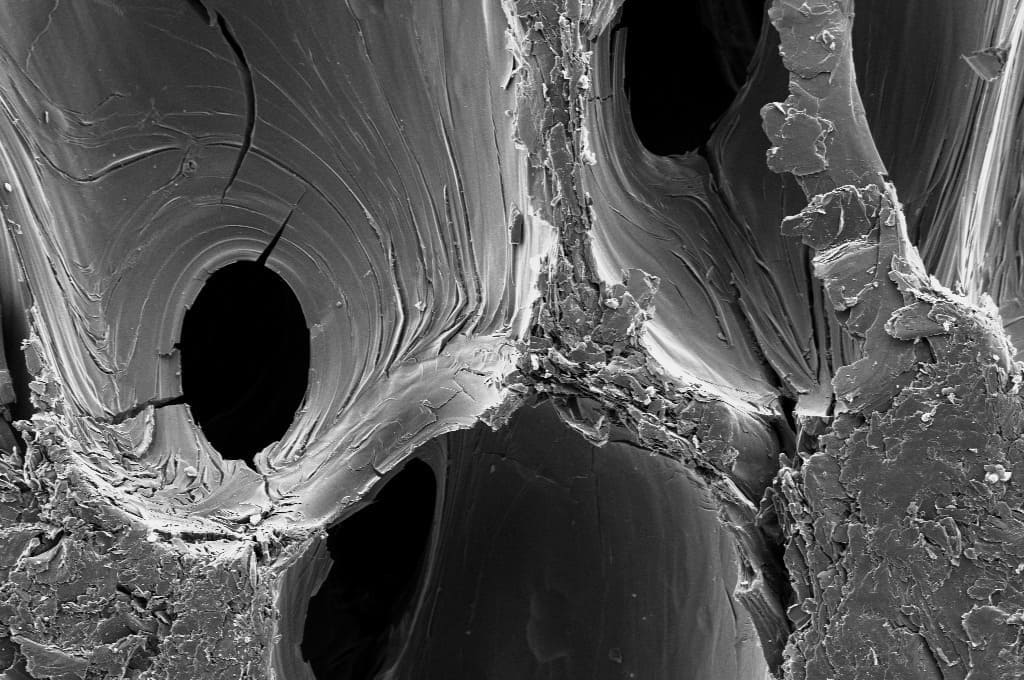}};
			\draw[quote](0.175,-0.25) -- (1.225,-0.25);
			\node  at (0.75, -0.75) { \bfseries 300 $\boldsymbol{\mu}$m};
	\end{tikzpicture}}
	  \caption{Microscopy image - Ligament structure}
  \end{subfigure}
  \end{figure}
     \begin{figure}[H]\ContinuedFloat
      	\centering
  \begin{subfigure}[t]{.485\linewidth}
  \centering
  \scalebox{0.8}{
      	\begin{tikzpicture}[xscale=1, yscale=1, axis/.style={->,thick},line cap=rect]
      	   \centering
			\node[anchor=south west,inner sep=0] at (0,0) {\includegraphics[trim={375 75 375 50}, width=\textwidth, clip]{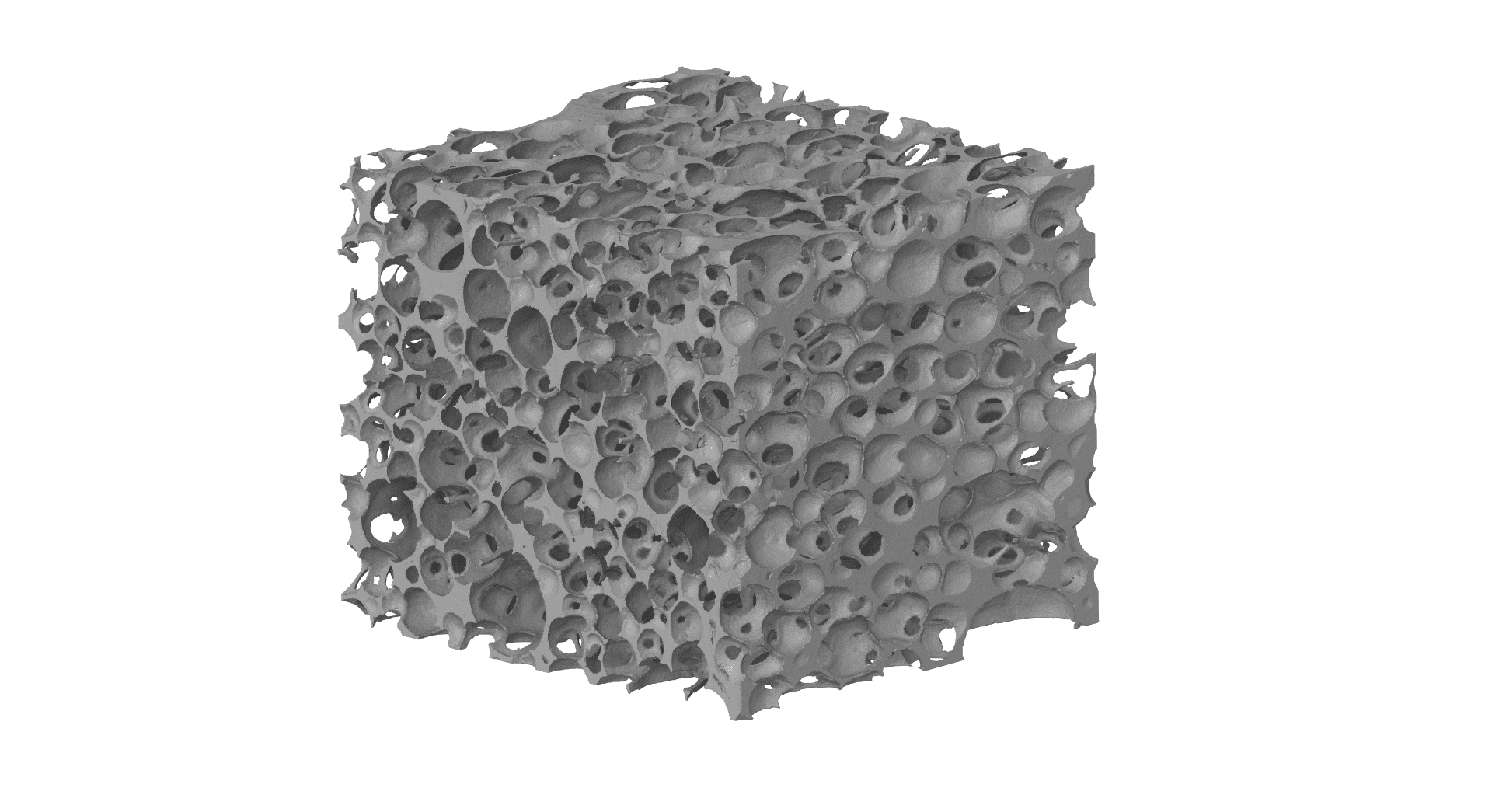}};
			\draw[axis,line width=0.5mm] (4,0.35) -- (2.5,0.7) node[anchor=north]{};
			\draw[axis,line width=0.5mm] (4,0.35) -- (5.5,0.7) node[anchor=north]{};
			\draw[axis,line width=0.5mm] (4,0.35) -- (4,2) node[anchor=north]{};
			\node[]  at (2.5,0.35) {$\boldsymbol{x}$};
			\node[]  at (5.5,0.35) {$\boldsymbol{z}$};
			\node[]  at (4.25,1.75) {$\boldsymbol{y}$};
	\end{tikzpicture}}
	\caption{CT scan - Isometric view}
	\label{ct-isometric}
  \end{subfigure}
    \begin{subfigure}[t]{.485\linewidth}
    \centering
      \scalebox{0.8}{
      	\begin{tikzpicture}[xscale=1, yscale=1, axis/.style={->,thick},line cap=rect]
      	   \centering
			\node[anchor=south west,inner sep=0] at (0,0) {\includegraphics[trim={375 75 375 50}, width=\textwidth, clip]{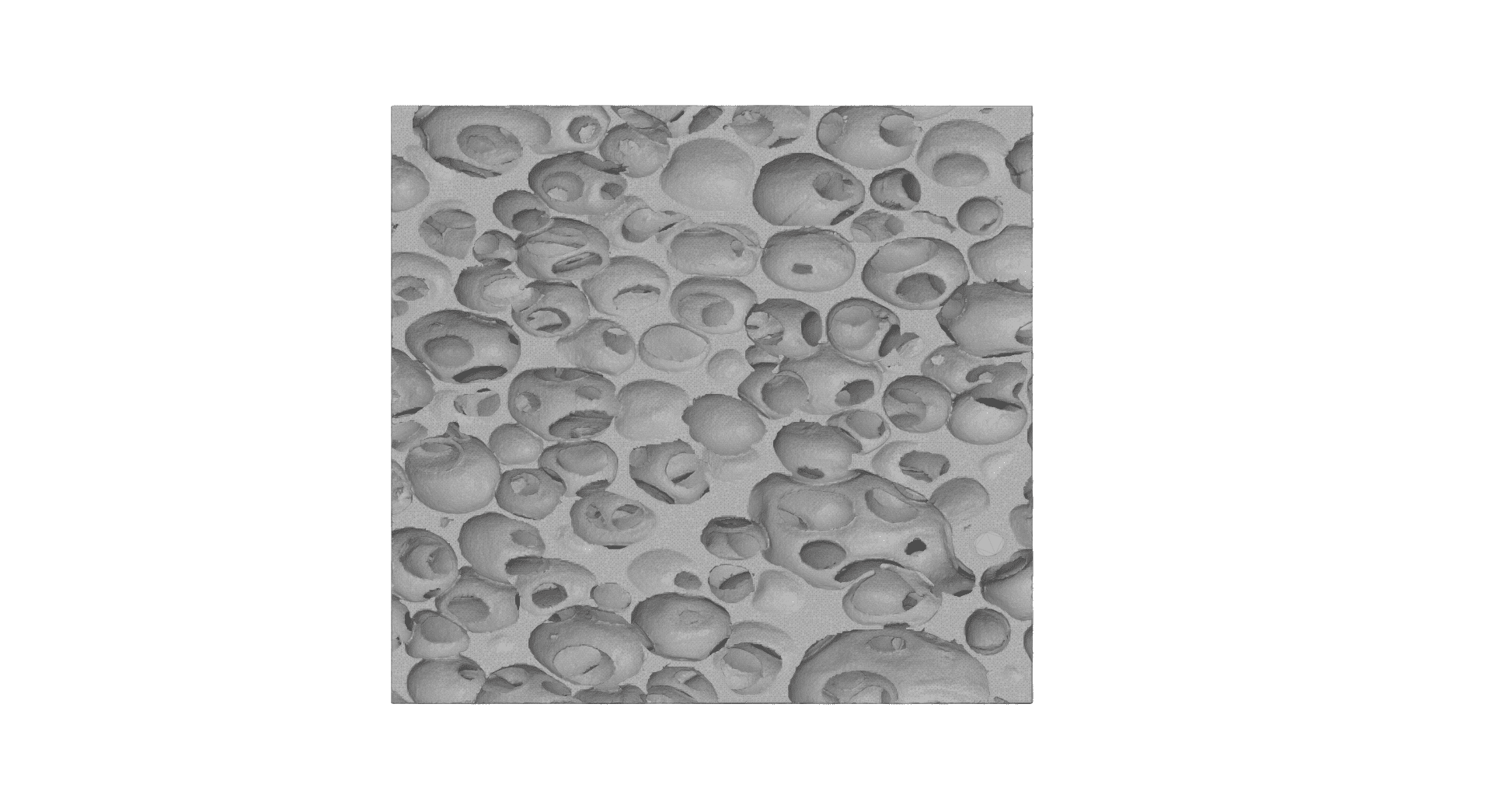}};
			\draw[quote](0.925,0.25) -- (3.25,0.25);
			\node  at (2, -0.1) { \bfseries 5 mm};
	\end{tikzpicture}
	}
	  	\caption{CT scan - Plane $x=0$}
	 \label{ct-plane}
  \end{subfigure}
  \caption{Geometry of the CFOAM$^{\text{\textregistered}}$ 35 HTC carbon foam}
  \label{cfoam-geo}
  \end{figure}
  
A CAD file of the foam has been obtained from a computed tomography scan performed at CERN (see \cref{ct-isometric}). The cut with the $z=0$ plane leads to cells whose two-dimensional projection are circles, while in the $x=0$ plane the projections are ellipses. As \cref{ct-plane} shows, the cells are elongated in the $z$ direction, which leads to anisotropic mechanical and thermal properties. Compared to the number of pores of a teteakaidecahedron (14), the number of pores per cell is much lower in the case of the CFOAM$^{\text{\textregistered}}$ 35 HTC: in particular, some cells are closed.

\bibliographystyle{unsrt}
\bibliography{references.bib}







\end{document}